\documentclass[zpreprint,zbstnp]{zeus_paper}
\usepackage[english]{babel}
\chardef\usc=95
\chardef\til=126
\catcode`\@=11 
\DeclareRobustCommand\xdotspace{\futurelet\@let@token\@xdotspace}
\def\@xdotspace{%
  \ifx\@let@token.\else
  \ifx\@let@token\bgroup.\else
  \ifx\@let@token\egroup.\else
  \ifx\@let@token\/.\else
  \ifx\@let@token\ .\else
  \ifx\@let@token~.\else
  \ifx\@let@token!.\else
  \ifx\@let@token,.\else
  \ifx\@let@token:.\else
  \ifx\@let@token;.\else
  \ifx\@let@token?.\else
  \ifx\@let@token/.\else
  \ifx\@let@token'.\else
  \ifx\@let@token).\else
  \ifx\@let@token-.\else
  \ifx\@let@token\@xobeysp.\else
  \ifx\@let@token\space.\else
  \ifx\@let@token\@sptoken.\else
   .\space
   \fi\fi\fi\fi\fi\fi\fi\fi\fi\fi\fi\fi\fi\fi\fi\fi\fi\fi}
\catcode`\@=12 

\newcommand{\stru}[2]{%
   \relax\ifmmode\hbox{\vrule height#1 depth#2 width0pt}%
   \else\vrule height#1 depth#2 width0pt\fi}

\newcommand{\Ronum}[1]{\uppercase\expandafter{\romannumeral#1}}
\newcommand{\ronum}[1]{\expandafter{\romannumeral#1}}
\DeclareRobustCommand{\LaTeXZ}{%
  \LaTeX\kern-.05em4\kern-.1em
  {\raisebox{-0.2ex}{$\scriptstyle\text{ZEUS}$}}\xspace}

\DeclareMathAlphabet{\mathbf}{OT1}{cmr}{bx}{sl}
\newcommand{\eVdist}{\kern-0.06667em}

\newcommand{\Gev}{{\text{Ge}\eVdist\text{V\/}}}

\newcommand{\slashfrac}[2]{%
  \raisebox{0.5ex}{\ensuremath #1}\kern-0.12em/\kern-0.08em
  \raisebox{-.8ex}{\ensuremath #2}}

\newcommand{\sqr}[3]{%
    {\vcenter{\hrule height.#3ex\hbox{\vrule width.#2ex height#1ex
     \kern#1ex\vrule width.#3ex}\hrule height.#2ex}}}

\catcode`\@=11 
\newcommand{\parenbar}{\mathpalette\p@renb@r}
\def\p@renb@r#1#2{\vbox{%
  \ifx#1\scriptscriptstyle \dimen@.7em\dimen@ii.2em\else
  \ifx#1\scriptstyle \dimen@.8em\dimen@ii.25em\else
  \dimen@1em\dimen@ii.4em\fi\fi \offinterlineskip
  \ialign{\hfill##\hfill\cr
    \vbox{\hrule width\dimen@ii}\cr
    \noalign{\vskip-.3ex}%
    \hbox to\dimen@{$\mathchar300\hfil\mathchar301$}\cr
    \noalign{\vskip-.3ex}%
    $#1#2$\cr}}}
\catcode`\@=12 

\newcommand{\IP}{{\rm I$\kern-0.01667em$P}\xspace}

\mathchardef\qsm=63
\mathchardef\pls=43
\mathchardef\mns=512
\mathchardef\plm=518
\mathchardef\eql=61
\mathchardef\smallleft=300
\mathchardef\smallright=301
\mathchardef\les=316
\mathchardef\gre=318
\mathchardef\leq=532
\mathchardef\grq=533

\catcode`\@=11 
\newcounter{pict@width}
\newcounter{pict@height}
\newlength{\pict@scale}
\newcommand{\psfigadd}[4]{%
\setcounter{pict@width}{1*\ratio{#2+\pict@scale/2}{\pict@scale}}
\setcounter{pict@height}{1*\ratio{#3+\pict@scale/2}{\pict@scale}}
\setlength{\unitlength}{\pict@scale}
\hbox to #2{\hspace{-\fill}\begin{picture}(\thepict@width,\thepict@height)
\put(0,0){\psfig{figure=#1,width=#2,height=#3,clip=}}
\SetScale{0.283466457}
\SetWidth{1.763889}
{#4}
\end{picture}}
}
\newcounter{pict@widthfst}
\newcounter{pict@widthscd}
\newcounter{pict@widthtot}
\newcommand{\psfigaddtwo}[7]{%
\setcounter{pict@widthfst}{1*\ratio{#2+\pict@scale/2}{\pict@scale}}
\setcounter{pict@widthscd}{1*\ratio{#2+#4+\pict@scale/2}{\pict@scale}}
\setcounter{pict@widthtot}{1*\ratio{#2+#4+#6+\pict@scale/2}{\pict@scale}}
\setcounter{pict@height}{1*\ratio{#3+\pict@scale/2}{\pict@scale}}
\setlength{\unitlength}{\pict@scale}
\hbox{\hspace{-\fill}\begin{picture}(\thepict@widthtot,\thepict@height)
\put(0,0){\psfig{figure=#1,width=#2,height=#3,clip=}}
\put(\thepict@widthscd,0){\psfig{figure=#5,width=#6,height=#3,clip=}}
\SetScale{0.283466457}
\SetWidth{1.763889}
{#7}
\end{picture}}
}
\newcommand{\psfigror}[4]{%
\setcounter{pict@width}{1*\ratio{#2+\pict@scale/2}{\pict@scale}}
\setcounter{pict@height}{1*\ratio{#3+\pict@scale/2}{\pict@scale}}
\setlength{\unitlength}{\pict@scale}
\hbox{\begin{picture}(\thepict@width,\thepict@height)
\put(0,\thepict@height){\psfig{figure=#1,width=#3,height=#2,clip=,angle=270}}
\SetScale{0.283466457}
\SetWidth{1.763889}
{#4}
\end{picture}}
}
\newcommand{\psfigrol}[4]{%
\setcounter{pict@width}{1*\ratio{#2+\pict@scale/2}{\pict@scale}}
\setcounter{pict@height}{1*\ratio{#3+\pict@scale/2}{\pict@scale}}
\setlength{\unitlength}{\pict@scale}
\hbox{\begin{picture}(\thepict@width,\thepict@height)
\put(0,0){\psfig{figure=#1,width=#3,height=#2,clip=,angle=90}}
\SetScale{0.283466457}
\SetWidth{1.763889}
{#4}
\end{picture}}
}
\catcode`\@=12 
\newlength\listtextwidth

\catcode`\@=11 
\newlength{\@tabfninsert}
\newlength{\@tabfnwidth}
\newcommand{\tabfootnote}[2]{%
  \setlength{\@tabfninsert}{0.8em}
  \setlength{\@tabfnwidth}{\textwidth}
  \addtolength{\@tabfnwidth}{-\@tabfninsert}
  \addtolength{\@tabfnwidth}{-0.4em}
  \noindent\makebox[\@tabfninsert][r]{\footnotesize$^{#1}$\hfil}\hfill%
  \parbox[t]{\@tabfnwidth}{\footnotesize #2\hfill}}
\catcode`\@=12 

\begin{document}
\prepnum{{DESY--07--118}}
\date{August 2007}


\title{
Exclusive $\rho^0$ production \\
in deep inelastic scattering\\
at HERA
}                                                       
                    
\author{ZEUS Collaboration}

\date{August 2007}

\abstract{Exclusive $\rho^0$ electroproduction at HERA has been studied
with the ZEUS detector using 120 pb$^{-1}$ of integrated luminosity
collected during 1996-2000. The analysis was carried out in the
kinematic range of photon virtuality $2 < Q^2 < 160$ GeV$^2$, and
$\gamma^* p$ centre-of-mass energy $32 < W < 180$ GeV.  The
results include the $Q^2$ and $W$ dependence of the $\gamma^* p
\to \rho^0 p$ cross section and the distribution of the
squared-four-momentum transfer to the proton. The helicity analysis of
the decay-matrix elements of the $\rho^0$ was used to study the ratio
of the $\gamma^* p$ cross section for longitudinal and transverse
photon as a function of $Q^2$ and $W$. Finally, an effective Pomeron
trajectory was extracted.  The results are compared to various
theoretical predictions.  }

\makezeustitle

\newpage

%
%
%
%
                                                   %
\begin{center}                                                                                     
{                      \Large  The ZEUS Collaboration              }                               
\end{center}                                                                                       
  S.~Chekanov$^{   1}$,                                                                            
  M.~Derrick,                                                                                      
  S.~Magill,                                                                                       
  B.~Musgrave,                                                                                     
  D.~Nicholass$^{   2}$,                                                                           
  \mbox{J.~Repond},                                                                                
  R.~Yoshida\\                                                                                     
 {\it Argonne National Laboratory, Argonne, Illinois 60439-4815, USA}~$^{n}$                       
\par \filbreak                                                                                     
  M.C.K.~Mattingly \\                                                                              
 {\it Andrews University, Berrien Springs, Michigan 49104-0380, USA}                               
\par \filbreak                                                                                     
  M.~Jechow, N.~Pavel~$^{\dagger}$, A.G.~Yag\"ues Molina \\                                        
  {\it Institut f\"ur Physik der Humboldt-Universit\"at zu Berlin,                                 
           Berlin, Germany}~$^{b}$
% ...
\newpage                                                                                           
$^{\    1}$ supported by DESY, Germany \\                                                          
$^{\    2}$ also affiliated with University College London, UK \\                                  
% ...
$^{\dagger}$ deceased \\                                                                           
%
                                                           %
                                                           %
\begin{tabular}[h]{rp{14cm}}                                                                       
$^{a}$ &  supported by the Natural Sciences and Engineering Research Council of Canada (NSERC) \\  
$^{b}$ &  supported by the German Federal Ministry for Education and Research (BMBF), under        
          contract numbers 05 HZ6PDA, 05 HZ6GUA, 05 HZ6VFA and 05 HZ4KHA\\                         
$^{c}$ &  supported in part by the MINERVA Gesellschaft f\"ur Forschung GmbH, the Israel Science   
          Foundation (grant no. 293/02-11.2) and the U.S.-Israel Binational Science Foundation \\  
$^{d}$ &  supported by the German-Israeli Foundation and the Israel Science Foundation\\           
$^{e}$ &  supported by the Italian National Institute for Nuclear Physics (INFN) \\                
$^{f}$ &  supported by the Japanese Ministry of Education, Culture, Sports, Science and Technology 
          (MEXT) and its grants for Scientific Research\\                                          
$^{g}$ &  supported by the Korean Ministry of Education and Korea Science and Engineering          
          Foundation\\                                                                             
$^{h}$ &  supported by the Netherlands Foundation for Research on Matter (FOM)\\                   
$^{i}$ &  supported by the Polish State Committee for Scientific Research, grant no.               
          620/E-77/SPB/DESY/P-03/DZ 117/2003-2005 and grant no. 1P03B07427/2004-2006\\             
$^{j}$ &  partially supported by the German Federal Ministry for Education and Research (BMBF)\\   
$^{k}$ &  supported by RF Presidential grant N 8122.2006.2 for the leading                         
          scientific schools and by the Russian Ministry of Education and Science through its grant
          Research on High Energy Physics\\                                                        
$^{l}$ &  supported by the Spanish Ministry of Education and Science through funds provided by     
          CICYT\\                                                                                  
$^{m}$ &  supported by the Particle Physics and Astronomy Research Council, UK\\                   
$^{n}$ &  supported by the US Department of Energy\\                                               
$^{o}$ &  supported by the US National Science Foundation. Any opinion,                            
findings and conclusions or recommendations expressed in this material                             
are those of the authors and do not necessarily reflect the views of the                           
National Science Foundation.\\                                                                     
$^{p}$ &  supported by the Polish Ministry of Science and Higher Education                         
as a scientific project (2006-2008)\\                                                              
$^{q}$ &  supported by FNRS and its associated funds (IISN and FRIA) and by an Inter-University    
          Attraction Poles Programme subsidised by the Belgian Federal Science Policy Office\\     
$^{r}$ &  supported by the Malaysian Ministry of Science, Technology and                           
Innovation/Akademi Sains Malaysia grant SAGA 66-02-03-0048\\                                       
\end{tabular}                                                                                      
                                                           %

\pagenumbering{arabic} 
\pagestyle{plain}

\newcommand {\pom} {I\!\!P}
\newcommand {\pomsub} {{\scriptscriptstyle \pom}}
\newcommand {\reg} {I\!\!R}
\newcommand {\regsub} {{\scriptscriptstyle \reg}}
\newcommand {\ppp} {\pom\pom\pom}
\newcommand {\ppr} {\pom\pom\reg}
\newcommand {\xpom} {x_{\pomsub}}
\newcommand {\apom} {\alpha_{\pomsub}}
\newcommand {\areg} {\alpha_{\regsub}}
\newcommand {\aprime} {\alpha^\prime_\pomsub}
\newcommand {\deta} {\Delta\eta}
\newcommand {\aveapom} {\bar{\alpha}_\pomsub}
\newcommand{\mxtwo}{M^2_X}
\newcommand{\xbj}{x_{\mathrm{Bj}}}
\newcommand {\xl} {x_L}
\newcommand{\yjb}{y_{\scriptscriptstyle JB}}
\newcommand {\mxlps} {M_{X}^{\mathrm{LPS}}}
\newcommand{\gp}{\gamma p}
\newcommand{\gvp}{\gamma^* p}

\section{Introduction}
\label{sec-int}

Two of the most surprising aspects of high-energy deep inelastic
scattering (DIS) observed at the HERA $ep$ collider have been the
sharp rise of the proton structure function, $F_2$, with decreasing
value of Bjorken $x$ and the abundance of events with a large rapidity
gap in the hadronic final state~\cite{rmp}. The latter are identified
as due to diffraction in the deep inelastic regime.  A contribution to
the diffractive cross section arises from the exclusive production of
vector mesons (VM).

High-energy exclusive VM production in DIS has been postulated to
proceed through two-gluon exchange~\cite{strikfurtHA,brodsky}, once
the scale, usually taken as the virtuality $Q^2$ of the exchanged
photon, is large enough for perturbative Quantum Chromodynamics (pQCD)
to be applicable.  The gluons in the proton, which lie at the origin
of the sharp increase of $F_2$, are also expected to cause the VM
cross section to increase with increasing photon proton centre-of-mass
energy, $W$, with the rate of increase growing with $Q^2$. Moreover,
the effective size of the virtual photon decreases with increasing
$Q^2$, leading to a flatter distribution in $t$, the
four-momentum-transfer squared at the proton vertex.  All these
features, with varying levels of significance, have been observed at
HERA~\cite{review-data1,review-data2,review-data3,review-data4,h1apom,
h1-disrho,z-disrho} in the exclusive production of $\rho^0$, $\omega$,
$\phi$, and $J/\psi$ mesons.

This paper reports on an extensive study of the properties of
exclusive $\rho^0$-meson production,
\begin{equation}
\gamma^* p \to \rho^0 p,  \nonumber
\label{eq:g*prp}
\end{equation}
based on a high statistics data sample collected with the
ZEUS detector during the period 1996-2000, corresponding to an
integrated luminosity of about 120 pb$^{-1}$.


\section{Theoretical background}
\label{sec:theo}

Calculations of the VM production cross section in DIS require
knowledge of the $q\bar{q}$ wave-function of the virtual photon,
specified by QED and which depends on the polarisation of the virtual
photon. For longitudinally polarised photons, $\gamma^*_L$, $q\bar{q}$
pairs of small transverse size dominate~\cite{brodsky}. The opposite
holds for transversely polarised photons, $\gamma^*_T$, where
$q\bar{q}$ configurations with large transverse size dominate. The
favourable feature of exclusive VM production is that, at high $Q^2$,
the longitudinal component of the virtual photon is dominant. The
interaction cross section in this case can be fully calculated in
pQCD~\cite{col-fran-stri}, with two-gluon exchange as the leading
process in the high-energy regime.  For heavy vector mesons, such as
the $J/\psi$ or the $\Upsilon$, perturbative calculations apply even
at $Q^2=0$, as the smallness of the $q\bar{q}$ dipole originating from
the photon is guaranteed by the mass of the quarks.

Irrespective of particular calculations~\cite{reviewVM}, in the
region dominated by perturbative QCD the following features are
predicted:
\begin{itemize}
\item the total $\gamma^* p \rightarrow V p$ cross section,
  $\sigma_{\gamma^* p}$, exhibits a steep rise with $W$, which can be
  parameterised as $\sigma \sim W^{\delta}$, with $\delta$ increasing
  with $Q^2$;
\item the $Q^2$ dependence of the cross-section, which for a longitudinally polarised
  photon is expected to behave as $Q^{-6}$, is moderated to become
  $Q^{-4}$ by the rapid increase of the gluon density with $Q^2$;
\item the distribution of $t$ becomes universal, 
    with little or no dependence on $W$ or $Q^2$;
\item breaking of the $s$-channel helicity conservation (SCHC) is expected.
\end{itemize}
In the region where perturbative calculations are applicable,
exclusive vector-meson production could become a complementary source
of information on the gluon content of the proton. At present, the
following theoretical uncertainties have been identified:
\begin{itemize}
\item the calculation of $\sigma({\gamma^* p \rightarrow Vp})$
  involves the generalised parton distributions~\cite{skewed1,skewed2}, which
  are not well tested; in addition~\cite{jhep:103:45}, it involves
  gluon densities outside the range constrained by global QCD analyses
  of parton densities; 
\item higher-order corrections have not been fully calculated~\cite{ivanov};
  therefore the overall normalisation is uncertain and the scale at
  which the gluons are probed is not known;
\item the rapid rise of $\sigma_{\gamma^* p}$ with $W$ implies a non-zero real part 
    of the scattering amplitude, which is not known;
\item the wave-functions of the vector mesons are not fully known.
\end{itemize}
In spite of all these problems, precise measurements of differential
cross sections separated into longitudinal and transverse
components~\cite{amirim}, should help to resolve the above
theoretical uncertainties.

It is important in these studies to establish a region of phase space
where hard interactions dominate over the non-perturbative soft
component.  If the relative transverse momentum of the $q\bar{q}$ pair
is small, the colour dipole is large and perturbative calculations do
not apply.  In this case the interaction looks similar to
hadron-hadron elastic scattering, described by soft Pomeron exchange
as in Regge phenomenology~\cite{collins}.

The parameters of the soft Pomeron are known from measurements of total
cross sections for hadron-hadron interactions and elastic
proton-proton measurements.  It is usually assumed that the Pomeron
trajectory is linear in $t$:
\begin{equation}
\apom (t) = \apom(0) + \aprime\ t \, .
\label{eq:trajectory}
\end{equation}
The parameter $\apom(0)$ determines the energy behaviour of the
total cross section,
\begin{equation}
\sigma_{\rm tot} \sim (W^2)^{\apom(0)-1} \nonumber
\end{equation}
and $\aprime$ describes the increase of the slope $b$ of the $t$
distribution with increasing $W$. The value of $\aprime$ is inversely
proportional to the square of the typical transverse momenta
participating in the exchanged trajectory.  A large value of $\aprime$
suggests the presence of low transverse momenta typical of soft
interactions. The accepted values of $\apom(0)$~\cite{cudell} and $
\aprime$~\cite{landshoff-aprime} are
\begin{eqnarray}
\apom(0) &=& 1.096 \pm 0.003 \ \, \nonumber \\
\aprime &=& 0.25\ \ {\rm GeV^{-2}} . \nonumber
\end{eqnarray}
The non-universality of $\apom(0)$ has been established in inclusive
DIS, where the slope of the $\gamma^* p$ total cross
section with $W$ has a pronounced $Q^2$
dependence~\cite{h1lambda}. 
The value of $\aprime$ can be determined from exclusive VM production
at HERA via the $W$ dependence of the exponential $b$ slope of the $t$
distribution for fixed values of $W$, where $b$ is expected to behave
as
\begin{equation}
b(W) = b_0 +4 \aprime \ln\frac{W}{W_0}  \, , \nonumber
\end{equation}
where $b_0$ and $W_0$ are free parameters.  The value of $\aprime$ can
also be derived from the $W$ dependence of $d\sigma /dt$ at fixed $t$,
\begin{equation}
\frac{d\sigma}{dt}(W) = F(t) W^{2[2\apom(t)-2]} \, ,
\label{eq:apom}
\end{equation}
where $F(t)$ is an arbitrary function.  This approach has
the advantage that no assumption needs to be made about the $t$
dependence. The first indications from measurements of $\apom(t)$ in
exclusive $J/\psi$ photoproduction~\cite{h1apom,zeusapom} are that
$\apom(0)$ is larger and $\aprime$ is smaller than those of the above
soft Pomeron trajectory.


\section{Experimental set-up}
\label{sec:exp}

The present measurement is based on data taken with the ZEUS detector
during two running periods of the HERA $ep$ collider. During
1996-1997, protons with energy 820 GeV collided with 27.5 GeV
positrons, while during 1998-2000, 920 GeV protons collided with 27.5
GeV electrons or positrons. The sample used for this study corresponds
to an integrated luminosity of 118.9 pb$^{-1}$, consisting of 37.2
pb$^{-1}$ $e^+p$ sample from 1996-1997 and 81.7 pb$^{-1}$ from the
1998-2000 sample (16.7 pb$^{-1}$ $e^-$ and 65.0 pb$^{-1}$
$e^+$)\footnote{From now on, the word ``electron'' will be used as a
generic term for both electrons and positrons.}.

A detailed description of the ZEUS detector can be found
elsewhere~\cite{bluebook,pl:b293:465}. A brief outline of the components
that are most relevant for this analysis is given below.

Charged particles are tracked in the central tracking detector
(CTD)~\cite{nim:a279:290,npps:b32:181,nim:a338:254}. The CTD
consists of 72 cylindrical drift chamber layers, organised in nine
superlayers covering the polar-angle\footnote{The ZEUS coordinate
system is a right-handed Cartesian system, with the $Z$ axis pointing
in the proton direction, referred to as the ``forward direction'', and
the $X$ axis pointing left towards the centre of HERA. The coordinate
origin is at the nominal interaction point. }
region $15^\circ <\theta < 164^\circ$. The CTD operates in a magnetic
field of 1.43 T provided by a thin solenoid. The transverse-momentum
resolution for full-length tracks is $\sigma(p_T)/p_T = 0.0058p_T
\oplus 0.0065 \oplus 0.0014/p_T$, with $p_T$ in GeV.

The high-resolution uranium-scintillator calorimeter
(CAL)~\cite{nim:a309:77,nim:a309:101,nim:a321:356,nim:a336:23}
covers 99.7\% of the total solid angle and consists of three parts:
the forward (FCAL), the barrel (BCAL) and the rear (RCAL)
calorimeters. Each part is subdivided transversely into towers and
longitudinally into one electromagnetic section (EMC) and either one
(in RCAL) or two (in BCAL and FCAL) hadronic sections.  The CAL
energy resolutions, as measured under test-beam conditions, are
$\sigma(E)/E=0.18/\sqrt{E}$ for electrons and
$\sigma(E)/E=0.35/\sqrt{E}$ for hadrons, with $E$ in $\Gev$.

The position of the scattered electron was determined by combining
information from the CAL, the small-angle rear tracking
detector~\cite{nim:a401:63} and the hadron-electron
separator~\cite{nim:a277:176}.

In 1998, the forward plug calorimeter (FPC)~\cite{fpc} was
installed in the 20$\times$20 cm$^2$ beam hole of the FCAL with a
small hole of radius 3.15 cm in the centre to accommodate the beam
pipe. The FPC increased the forward calorimeter coverage by about one unit
in pseudorapidity to $\eta \leq 5$.

The leading-proton spectrometer (LPS)~\cite{lps} detected positively
charged particles scattered at small angles and carrying a substantial
fraction, $x_L$, of the incoming proton momentum; these particles
remained in the beam-pipe and their trajectories were measured by a
system of silicon microstrip detectors, located between 23.8 m and
90.0 m from the interaction point. The particle deflections induced by
the magnets of the proton beam-line allowed a momentum analysis of the
scattered proton.

During the 1996-1997 data taking, a proton-remnant tagger (PRT1) was
used to tag events in which the proton dissociates. It consisted of
two layers of scintillation counters perpendicular to the beam at $Z$
= 5.15 m. The two layers were separated by a 2 mm-thick lead absorber.
The pseudorapidity range covered by the PRT1 was $4.3 < \eta < 5.8$.

The luminosity was measured from the rate of the bremsstrahlung process 
$ep \rightarrow e\gamma p$. The photon was measured in a 
lead-scintillator calorimeter~\cite{lumi1,lumi2,lumi3} placed in the 
HERA tunnel at $Z=-107$~m.


\section{Data selection and reconstruction}
\label{sec:data}

The following kinematic variables
are used to describe exclusive $\rho^0$ production and its subsequent decay into a $\pi^+\pi^-$ pair:
\begin{itemize}
\item the four-momenta of the incident electron ($k$), scattered electron 
($k^{\prime}$), incident proton ($P$), 
scattered proton ($P^{\prime}$) and virtual 
photon ($q$);
\item $Q^2=-q^2=-(k-k^{\prime})^2$, the negative squared four-momentum of 
the virtual photon; 
\item $W^2 = (q+P)^2$, the squared centre-of-mass energy of the 
photon-proton system; 
\item $y = (P\cdot q)/(P\cdot k)$, the 
fraction of the electron energy transferred to the proton 
in its rest frame; 
\item $M_{\pi\pi}$, the invariant mass of the two decay pions;
\item $t = (P-P^{\prime})^2$,  the squared four-momentum transfer
at the proton vertex; 
\item three helicity angles, $\Phi_h$, $\theta_{h}$ and $\phi_{h}$ 
(see Section~\ref{sec:decay}).
\end{itemize}
The kinematic variables were reconstructed using the so-called
``constrained'' method~\cite{z-disrho,rhodis94}, which uses the
momenta of the decay particles measured in the CTD and the
reconstructed polar and azimuthal angles of the scattered electron.

The online event selection required an electron candidate in the CAL,
along with the detection of at least one and not more than six tracks
in the CTD.  

In the offline selection, the following further
requirements were imposed:
\begin{itemize}
\item the presence of a scattered electron, with energy in the 
CAL greater than 10 GeV and with an impact point on the face of the
RCAL outside a rectangular area of $26.4\times 16\ {\rm cm^2}$;
\item $E-P_Z > 45$~GeV, where $E-P_Z = \sum_i (E_i -
  p_{Z_i})$ and the summation is over the energies and longitudinal
  momenta of the final-state electron and pions, was imposed. This cut
  excludes events with high energy photons radiated in the initial
  state;
\item the $Z$ coordinate of the interaction vertex
  within $\pm50$ cm of the nominal interaction point;
\item in addition to the scattered electron, exactly
  two oppositely charged tracks, each associated with the
  reconstructed vertex, and each having pseudo\-rap\-i\-di\-ty $\left|
  \eta \right|$ less than 1.75 and transverse momentum greater than
  150~MeV; this excluded regions of low reconstruction efficiency and
  poor momentum resolution in the CTD. These tracks were treated in
  the following analysis as a $\pi^+\pi^-$ pair;
\item events with any energy deposit larger than 300~MeV in the CAL
  and not associated with the pion tracks (so-called `unmatched
  islands') were rejected~\cite{Heiko,Teresa,Arik}.
\end{itemize}

In addition, the following requirements were applied to select 
kinematic  regions of high acceptance:
\begin{itemize}
\item the analysis was restricted to the kinematic regions 
$2<Q^2<80$~GeV$^2$ and $32 < W < 160$~GeV in the 1996-1997 data and
$2<Q^2<160$~GeV$^2$ and $32 < W < 180$~GeV in the 1998-2000 sample;

\item only events in the $\pi^+\pi^-$ mass interval
  $0.65<M_{\pi\pi}<1.1$~GeV and with $|t|<1$~GeV$^2$ were taken.  The
  mass interval is slightly narrower than that used
  previously~\cite{z-disrho}, in order to reduce the effect of the
  background from non-resonant $\pi^+ \pi^-$ production.  In the
  selected $M_{\pi \pi}$ range, the resonant contribution is $\approx
  100\%$ (see Section~\ref{sec:mass}).
\end{itemize}

The above selection yielded 22,400 events in the 1996-1997
sample and 49,300 events in the 1998-2000 sample, giving a total
of 71,700 events for this analysis. 


\section{Monte Carlo simulation}
\label{sec:MC}

The relevant Monte Carlo (MC) generators have been described in detail 
previously~\cite{z-disrho}. Here their main features are summarised.

The program {\sc{Zeusvm}}~\cite{ref:muchor} interfaced to
{\sc{Heracles4.4}}~\cite{ref:heracles} was used.  The effective
$Q^2$, $W$ and $t$ dependences of the cross section were parameterised
to reproduce the data~\cite{Arik}.

The decay angular distributions were generated uniformly and the MC
events were then iteratively reweighted using the results of the
present analysis for the 15 combinations of matrix elements
$r^{04}_{ik}$, $r^{\alpha}_{ik}$ (see Section~\ref{sec:decay}).

The contribution of the proton-dissociative process was studied with
the {\sc{Epsoft}}~\cite{epsoft} generator for the 1996-1997 data and
with {\sc{Pythia}}~\cite{phytia} for the 1998-2000 data. The $Q^2$,
$W$ and $t$ dependences were parameterised to reproduce the control
samples in the data.  The decay angular distributions were generated
as in the {\sc{Zeusvm}} sample.

The generated events were processed through the same chain of
selection and reconstruction procedures as the data, thus accounting
for trigger as well as detector acceptance and smearing effects.  For
both MC sets, the number of simulated events after reconstruction was
about a factor of seven greater than the number of reconstructed data
events.

\begin{figure}[h]
\vfill
\begin{center}
\includegraphics[width=\hsize]{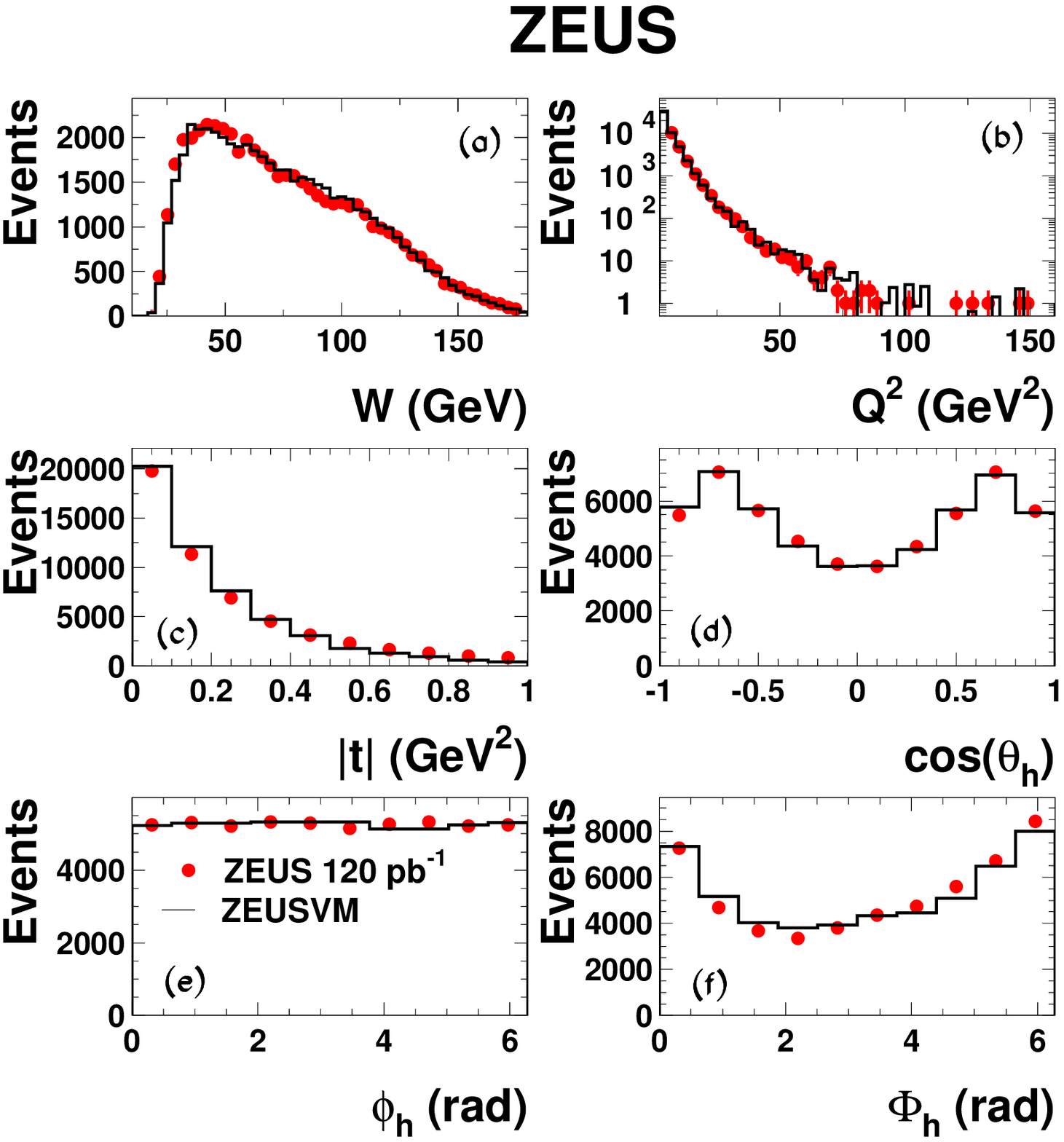}
\end{center}
\caption{
Comparison between the data and the {\sc{Zeusvm}} MC distributions for (a)
$W$, (b) $Q^2$, (c) $|t|$, (d) $\cos\theta_h$, (e) $\phi_h$ and (f)
$\Phi_h$  for events with $0.65<M_{\pi\pi}<1.1$ {\rm GeV} and
$|t|<1.0$ {\rm GeV$^2$}. The MC distributions are normalised to the data. }
\label{fig:control}
\vfill
\end{figure}

\begin{figure}[h]
\vfill
\begin{center}
\includegraphics[width=\hsize]{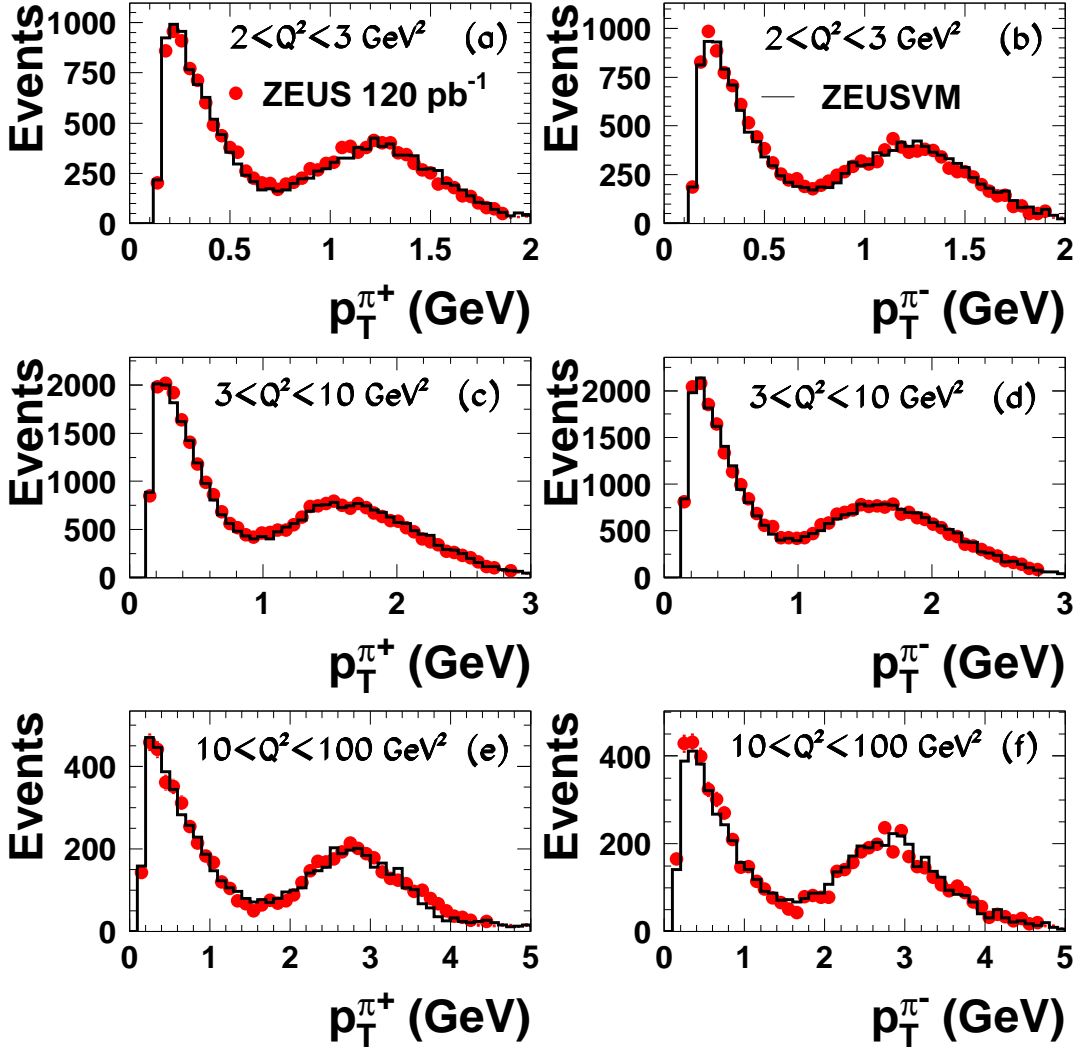}
\end{center}
\caption{
Comparison between the data and the {\sc{Zeusvm}} MC distributions for
the transverse momentum, $p_T$, of $\pi^+$ and $\pi^-$ particles, for
different ranges of $Q^2$, as indicated in the figure.  The events are
selected to be within $0.65<M_{\pi\pi}<1.1$ {\rm GeV} and $|t|<1.0$
{\rm GeV$^2$}. The MC distributions are normalised to the data. }
\label{fig:control-pt}
\vfill
\end{figure}

All measured distributions are well described by the MC
simulations. Some examples are shown in Fig.~\ref{fig:control}, for
the $W$, $Q^2$, $t$ variables, and the three helicity angles,
$\theta_h$, $\phi_h$, and $\Phi_h$, and in Fig.~\ref{fig:control-pt}
for the transverse momentum $p_T$ of the pions, for different $Q^2$
bins.


\section{Systematics}
\label{sec:syst}

The systematic uncertainties of the cross section were evaluated by varying
the selection cuts and the MC simulation parameters.  The 
following selection cuts were varied:
\begin{itemize}
\item the $E-P_Z$ cut was changed within the appropriate resolution of $\pm3$ GeV;
\item the $p_T$ of the pion tracks (default 0.15 GeV) was increased to 0.2 GeV;
\item the distance of closest approach of the extrapolated track to the matched
 island in the CAL was changed from 30 cm to 20 cm;
\item the $\pi^+\pi^-$-mass window was changed to 0.65--1.2 GeV;
\item the $Z$ vertex cut was varied by $\pm 10$ cm;
\item the rectangular area of the electron impact point on the CAL
 was increased by 0.5 cm in $X$ and $Y$;
\item the energy of an unmatched island was lowered to 0.25 GeV and then
  raised to 0.35 GeV.
\end{itemize}

The dependence of the results on the precision with which the MC
reproduces the performance of the detector and the data was checked by
varying the following inputs within their estimated uncertainty:
\begin{itemize}
\item the reconstructed position of the electron was shifted with
  respect to the MC by $\pm 1\ {\rm mm}$; 
\item the electron-position resolution was varied by $\pm 10\%$ in the MC;
\item the $W^\delta$-dependence in the MC was changed by varying
  $\delta$ by $\pm 0.03$;
\item the exponential $t$-distribution in the MC was reweighted by
  changing the nominal slope parameter $b$ by $\pm0.5$ GeV$^{-2}$;
\item the angular distributions in the MC were reweighted assuming SCHC;
\item the $Q^2$-distribution in the MC was reweighted by $(Q^2+M_\rho^2)^k$,
where $k = \pm0.05$.
\end{itemize}

The largest uncertainty of about $\pm$4\% originated from the
variation of the energy of the unmatched islands. All the other checks
resulted on average in a 0.5\% change in the measured cross
sections. All the systematic uncertainties were added in
quadrature. In addition, the cross-section measurements have an
overall normalisation uncertainty of~$\pm$2\% due to the luminosity
measurement.


\section{Proton dissociation}
\label{sec:pdis}

The production of $\rho^0$ mesons may be accompanied by the
proton-dissociation process, $\gamma^* p \to \rho^0 N$. For low masses
$M_N$ of the dissociative system $N$, the hadronisation products may
remain inside the beam-pipe, leaving no signals in the main detector.
The contribution of these events to the exclusive $\rho^0$ cross
section was estimated from MC generators for proton-dissociative
processes.

\begin{figure}[h]
\vfill
\begin{center}
\includegraphics[width=\hsize]{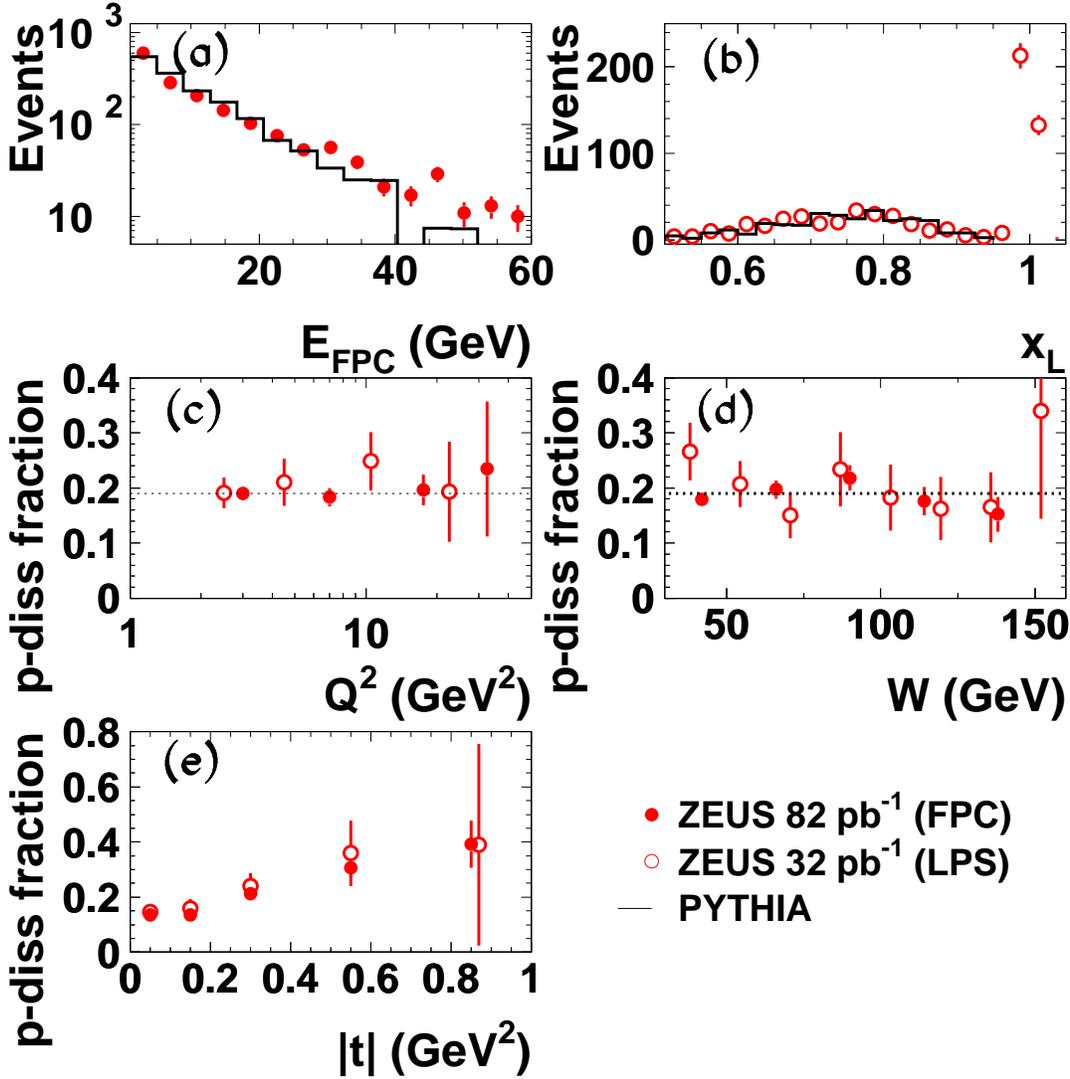}
\end{center}
\caption{\it
(a) The energy distribution in the FPC. The data (full dots) are
compared to the expectations from the {\sc{Pythia}} MC, normalised to
the data. (b) The $x_L$ distribution in the LPS. The data (open
circles) are compared to the expectations from the {\sc{Pythia}} MC,
normalised to the data for $x_L <$ 0.95.  The extracted fraction of
proton-dissociation events, from the FPC data (dots) and from the LPS
data (open circles), as a function of (c) $Q^2$, (d) $W$ and (e)
$|t|$. All events were selected in the $\rho^0$ mass window (0.65-1.1
GeV). The dotted line in (c) and (d) represents a fit of a constant to
the proton-dissociation fraction.}
\label{fig:pdiss}
\vfill
\end{figure}

A class of proton dissociative events for which the final-state
particles leave observed signals in the surrounding detectors was used
to tune the $M_N$ and the $t$ distribution in the MC. In the 1998-2000
running period, these events were selected by requiring a signal in
the FPC detector with energy above 1 GeV.  The comparison of the data
with {\sc{Pythia}} expectations for the energy distribution in the FPC
is shown in Fig.~\ref{fig:pdiss}(a).  The same procedure was repeated
with a sample of $\rho^0$ events for which the FPC energy was less
than 1 GeV and a leading proton was measured in the LPS detector, with
the fraction of the incoming proton momentum $x_L<0.95$. The
comparison between the $x_L$ distribution measured in the data and
that expected from {\sc{Pythia}} is shown in Fig.~\ref{fig:pdiss}(b),
where the elastic peak in the data ($x_L>0.95$) is also observed.
Also shown in Fig.~\ref{fig:pdiss}(c-e) is the fraction of
proton-dissociative events expected in the selected $\rho^0$ sample as
a function of $Q^2$, $W$ and $t$. The fraction is at the level of
$19\%$, independent of $Q^2$ and $W$, but increasing with increasing
$|t|$.  The combined use of the FPC and LPS methods leads to an
estimate of the proton dissociative contribution for $|t| <$ 1 GeV$^2$
of 0.19~$\pm$~0.02(stat.)~$\pm$~0.03(syst.). The systematic uncertainty was
estimated by varying the parameters of the $M_N$ distribution and by
changing the FPC cut.

In the 1996-1997 data-taking period, a similar procedure was applied,
after tuning the {\sc{Epsoft}} MC to reproduce events with hits in the
PRT1 or energy deposits in the FCAL. The proton-dissociative
contribution for $|t| <$ 1 GeV$^2$ was determined to be $0.07 \pm
0.02$ after rejecting events with hits in the PRT1 or energy deposits
in the FCAL. This number is consistent with that determined from the
LPS and FPC because of the different angular coverage of the PRT1.

After subtraction of the proton-dissociative contribution, a good
agreement between the cross sections derived from the two data-taking
periods was found. For all the quoted cross sections integrated over
$t$, the overall normalisation uncertainty due to the subtraction of
the proton-dissociative contributions was estimated to be $\pm$4\% and
was not included in the systematic uncertainty.  The
proton-dissociative contribution was statistically subtracted in each
analysed bin, unless stated otherwise.


\section{Mass distributions}
\label{sec:mass}

\begin{figure}[h]
\vfill
\begin{center}
\includegraphics[width=\hsize]{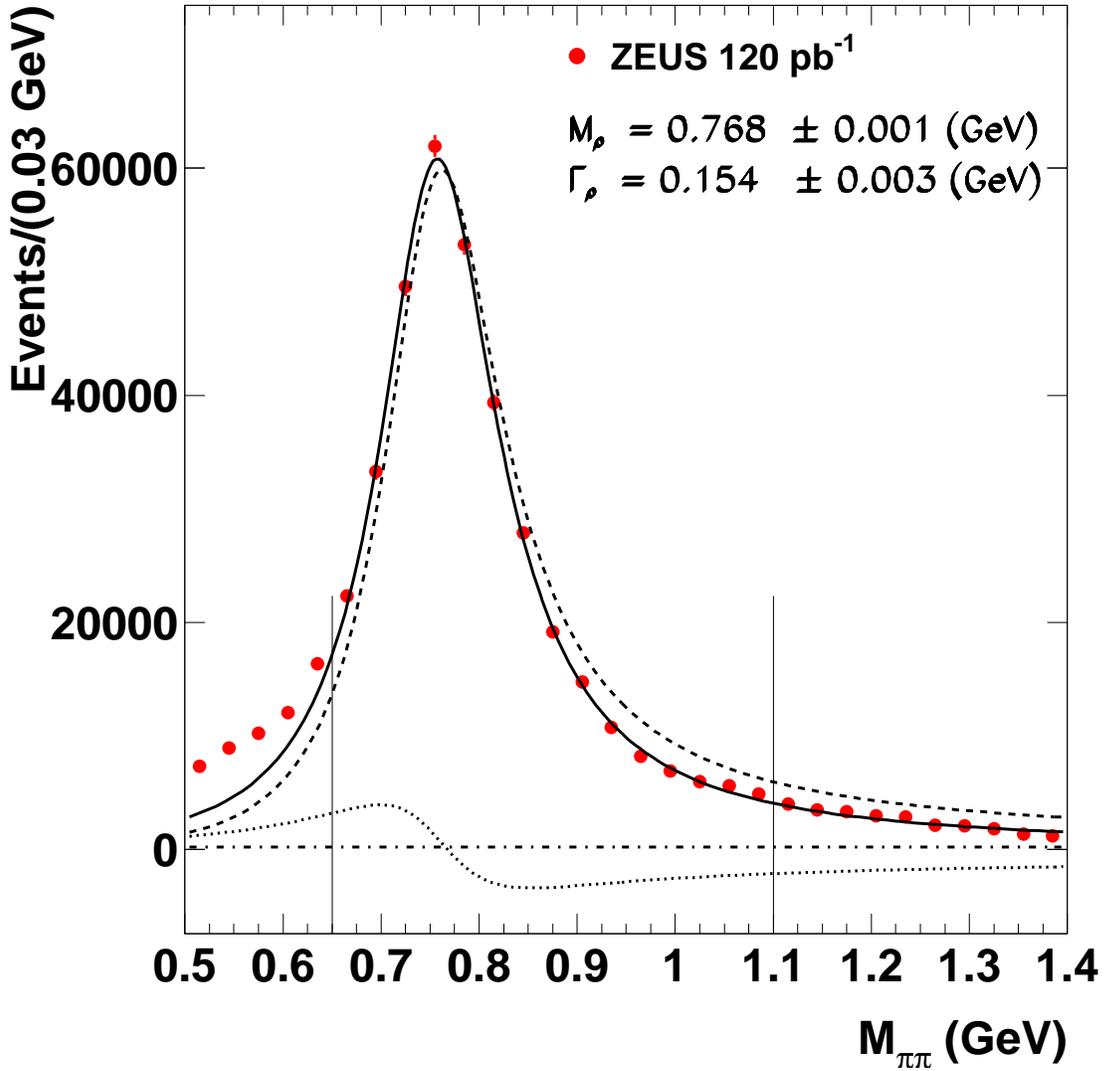}
\end{center}
\caption{\it
The $\pi^+\pi^-$ acceptance-corrected invariant-mass distribution. The
line represent the best fit of the S\"oding form to the data in the
range $0.65< M_{\pi\pi}<1.1$ GeV. The vertical lines indicate the
range of masses used for the analysis. The dashed line is the shape
of a relativistic Breit-Wigner with the fitted parameters given in the
figure. The dotted line is the interference term between the
non-resonant background (dash-dotted line) and the $\rho^0$ signal.
}
\label{fig:mpipi-all}
\vfill
\end{figure}

The $\pi^+\pi^-$-invariant-mass distribution is presented in
Fig.~\ref{fig:mpipi-all}. A clear enhancement in the $\rho^0$ region
is observed. Background coming from the decay $\phi \to K^+ K^-$,
where the kaons are misidentified as pions, is expected~\cite{Arik} in
the region $M_{\pi\pi} < 0.55$ GeV. That coming from $\omega$ events
in the decay channel $\omega \to
\pi^+ \pi^- \pi^0$, where the $\pi^0$ remains undetected, contributes~\cite{Arik} in the
region $M_{\pi\pi} < 0.65$ GeV. Therefore defining the selected
$\rho^0$ events to be in the window $0.65 < M_{\pi\pi} < 1.1$ GeV
ensures no background from these two channels.

In order to estimate the non-resonant $\pi^+\pi^-$ background under
the $\rho^0$, the S\"oding parameterisation~\cite{Soeding} was
fitted to the data, with results shown in the figure. The resulting
mass and width values are in agreement with those given in the
Particle Data Group~\cite{pdg} compilation. The integrated
non-resonant background is of the order of $1\%$ and is thus
neglected.

\begin{figure}[h]
\vfill
\begin{center}
\includegraphics[width=\hsize]{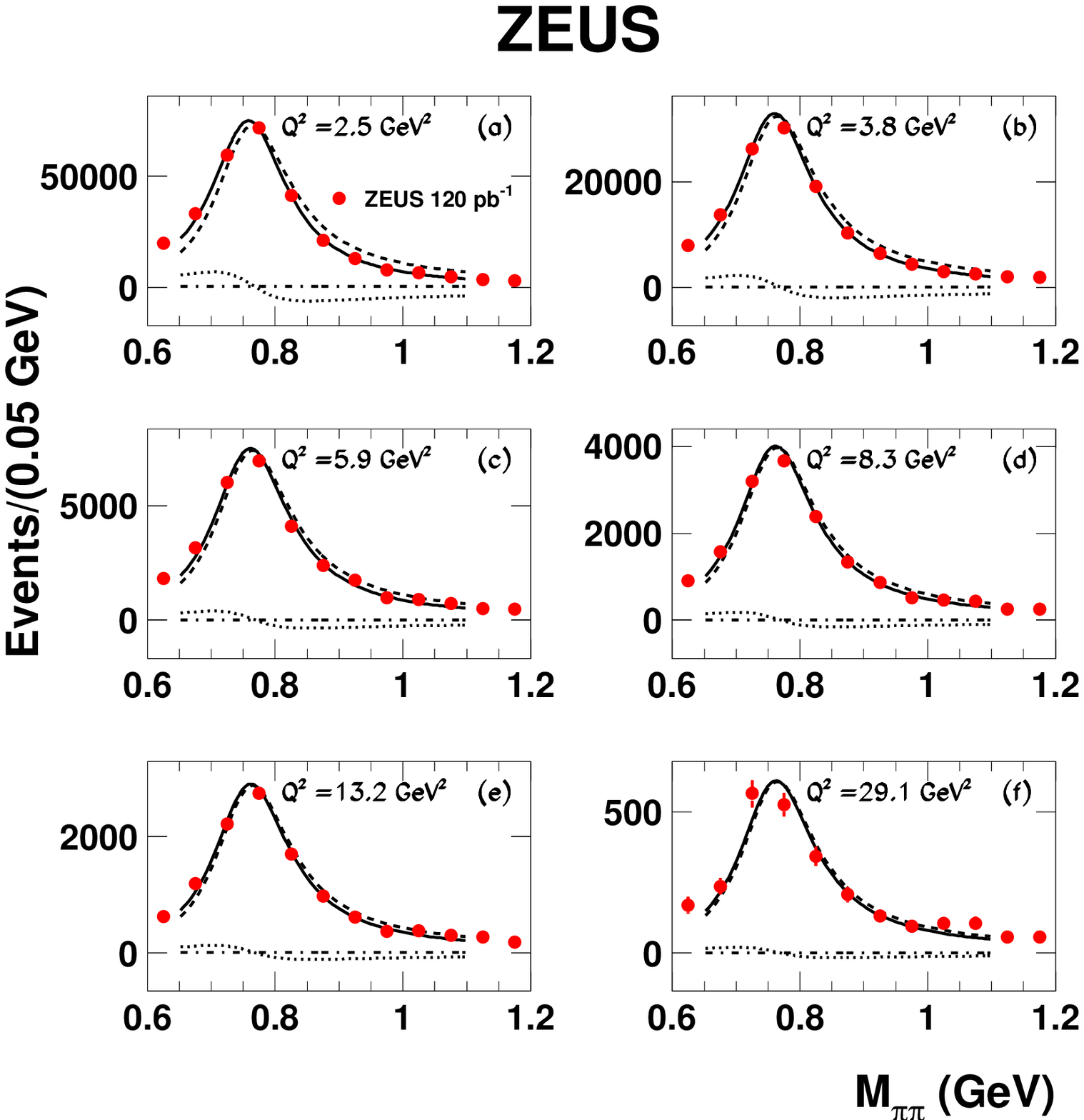}
\end{center}
\caption{\it
The $\pi^+\pi^-$ acceptance-corrected invariant-mass distribution, for
different $Q^2$ intervals, with mean values as indicated in the
figure.  The lines are defined in the caption of
Fig.~\ref{fig:mpipi-all}.  }
\label{fig:mpipi-q2}
\vfill
\end{figure}

\begin{figure}[h]
\vfill
\begin{center}
\includegraphics[width=\hsize]{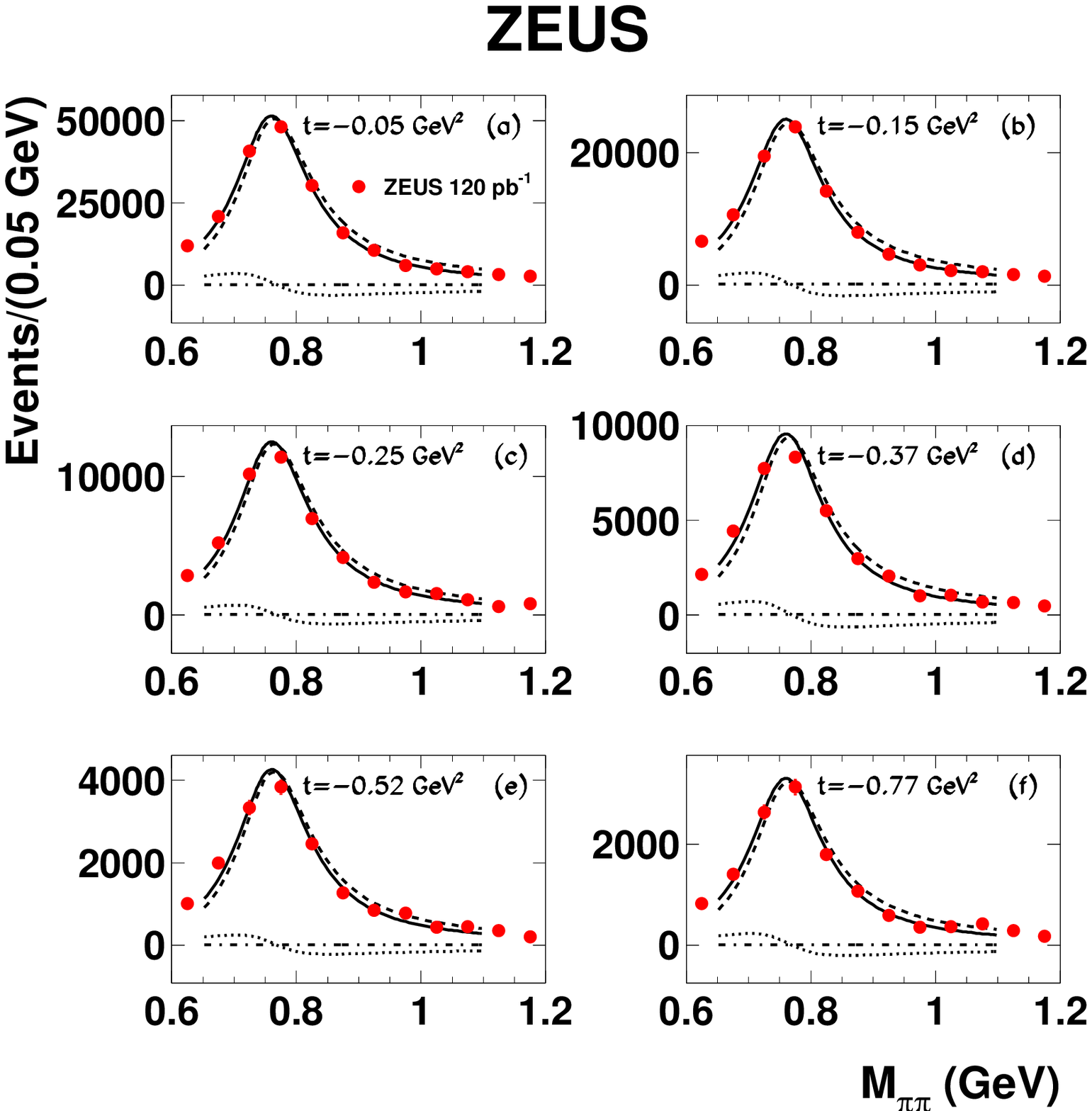}
\end{center}
\caption{\it
The $\pi^+\pi^-$ acceptance-corrected invariant-mass distribution, for
different $t$ intervals, with mean values as indicated in the figure.
The lines are defined in the caption of Fig.~\ref{fig:mpipi-all}.  }
\label{fig:mpipi-t}
\vfill
\end{figure}

The $\pi^+\pi^-$ mass distributions in different regions of $Q^2$ and
$t$ are shown in Fig.~\ref{fig:mpipi-q2} and Fig.~\ref{fig:mpipi-t},
respectively. The shape of the mass distribution changes neither with
$Q^2$ nor with $t$.  The results of the fit to the S\"oding
parameterisation are also shown. Note that the interference term
decreases with $Q^2$ as expected but is independent of $t$, indicating
that the non-exclusive background is negligible.


\section{Angular distributions and decay-matrix density}
\label{sec:decay}

The exclusive electroproduction and decay of $\rho^0$ mesons is
described, at fixed $W$, $Q^2$, $M_{\pi\pi}$ and $t$, by three
helicity angles: $\Phi_h$ is the angle between the $\rho^0$ production
plane and the electron scattering plane in the $\gamma^* p$
centre-of-mass frame; $\theta_{h}$ and $\phi_{h}$ are the polar and
azimuthal angles of the positively charged decay pion in the
$s$-channel helicity frame. In this frame, the spin-quantisation axis
is defined as the direction opposite to the momentum of the
final-state proton in the $\rho^0$ rest frame.  In the $\gamma^* p$
centre-of-mass system, $\phi_{h}$ is the angle between the decay plane
and the $\rho^0$ production plane.  The angular distribution as a
function of these three angles, $W(\cos\theta_{h},\phi_{h},\Phi_{h})$,
is parameterised by the $\rho^0$ spin-density matrix elements,
$\rho_{ik}^{\alpha}$, where $i,k=-1,0,1$ and by convention
$\alpha$=0,1,2,4,5,6 for an unpolarised charged-lepton
beam~\cite{angle}.  The superscript denotes the decomposition of the
spin-density matrix into contributions from the following
photon-polarisation states: unpolarised transverse photons (0);
linearly polarised transverse photons (1,2); longitudinally polarised
photons (4); and from the interference of the longitudinal and
transverse amplitudes (5,6).

The decay angular distribution can be expressed in terms of
combinations, $r^{04}_{ik}$
and $r^{\alpha}_{ik}$, of the density matrix elements
\begin{eqnarray}
r^{04}_{ik} &=& {\rho^0_{ik} \, + \, \epsilon R \rho^{4}_{ik}\over 1 \,
+ \, \epsilon R}, \label{r04}    \nonumber       \\
r^{\alpha}_{ik} &=& \,
\left\{
\begin{array}{ll}
{\displaystyle {\rho^{\alpha}_{ik}\over 1 \, + \, \epsilon R}}, &
{\alpha}=1,2\\*[5mm]
{\displaystyle {\sqrt{R} \; \rho^{\alpha}_{ik}\over 1 \, + \, \epsilon
R}}, & {\alpha}=5,6,
\end{array} \right. \nonumber
\end{eqnarray}

\noindent
where $\epsilon$ is the ratio of the longitudinal- to transverse-photon
fluxes and $R=\sigma_{L}/\sigma_{T}$, with $\sigma_{L}$ and
$\sigma_{T}$ the cross sections for exclusive $\rho^0$ production from
longitudinal and transverse virtual photons, respectively.  In the
kinematic range of this analysis, the value of $\epsilon$ varies
between 0.96 and 1 with an average value of 0.996; hence
$\rho^{0}_{ik}$ and $\rho^{4}_{ik}$ cannot be distinguished.

\begin{figure}[h]
\vfill
\begin{center}
\includegraphics[width=\hsize]{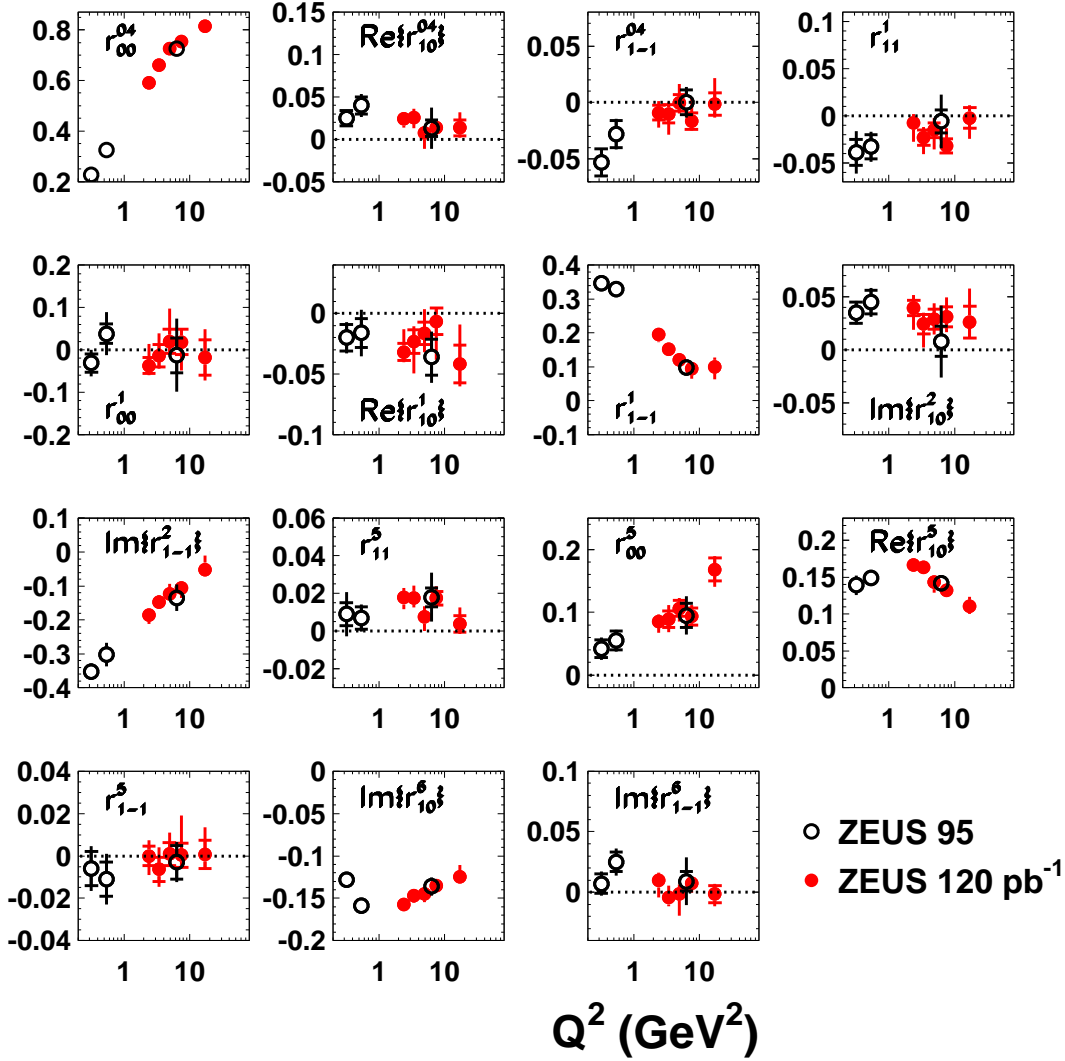}
\end{center}
\protect\caption{\it
The 15 density-matrix elements obtained from a fit to the data (dots),
as a function of $Q^2$. Also shown in the figure are results from an
earlier measurement~\protect\cite{zeusschc} (open circles).  The inner
error bars indicate the statistical uncertainty, the outer error bars
represent the statistical and systematic uncertainty added in
quadrature.  The dotted line at zero is the expectation from SCHC when
relevant.  }
\label{fig:me15}
\vfill
\end{figure}

The Hermitian nature of the spin-density matrix and the requirement of
parity conservation reduces the number of independent parameters to
15~\cite{angle}.  A 15-parameter fit was performed to the data and the
obtained results are listed in Table~\ref{tab:15dme} and shown in
Fig.~\ref{fig:me15} as a function of $Q^2$. The published ZEUS
results~\cite{zeusschc} at lower $Q^2$ values and the expectations of
SCHC, when relevant, are also included. The observed $Q^2$
dependence, expected in some calculations~\cite{kirschner} and
previously reported by H1~\cite{h1schc}, is driven by the $R$
dependence on $Q^2$ under the assumption of helicity conservation and
natural parity exchange. The significant deviation of $r_{00}^5$ from
zero shows that SCHC does not hold~\cite{kirschner} as was observed
previously~\cite{zeusschc,h1schc}.

\begin{figure}[h]
\vfill
\begin{center}
\includegraphics[width=\hsize]{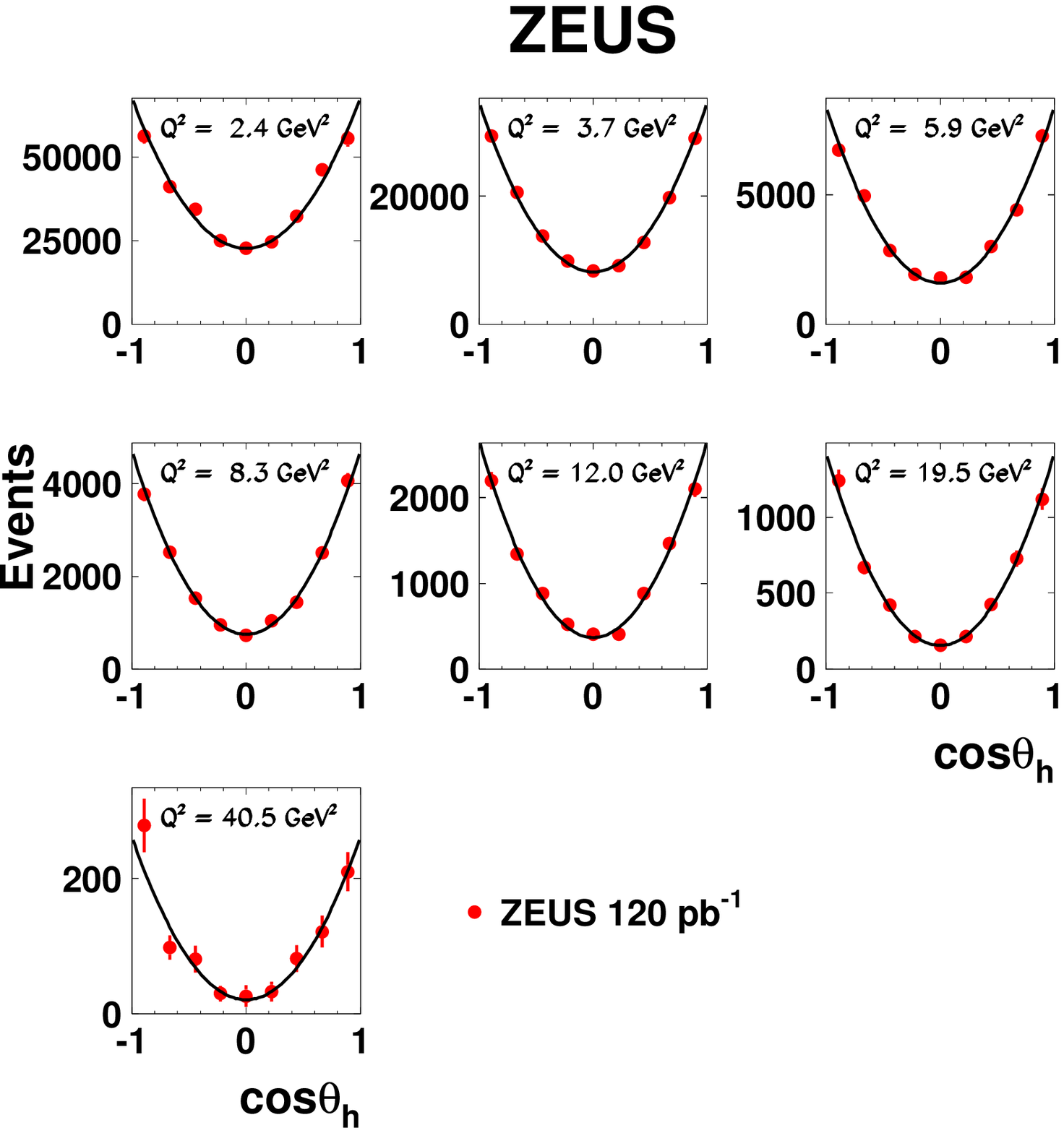}
\end{center}
\caption{\it
The acceptance-corrected $\cos\theta_{h}$ distribution, for different
$Q^2$ intervals, with mean values indicated in the figure. The line
represent the fit to the data of Eq.~(\ref{eq:angular}).}
\label{fig:theta-q2-comb}
\vfill
\end{figure}

The angular distribution for the decay of the $\rho^0$ meson,
integrated over $\phi_h$ and $\Phi_h$, reduces to
\begin{equation}\label{eq:angular}
W(\cos{\theta_{h}}) \propto \left[(1-r^{04}_{00})+(3r^{04}_{00}-1)
\cos^2{\theta_{h}}\right].
\end{equation}
The element $r_{00}^{04}$ may be extracted from a one-dimensional fit
to the $\cos{\theta_h}$ distribution.  The $\cos{\theta_h}$
distributions, for different $Q^2$ intervals, are shown in
Fig.~\ref{fig:theta-q2-comb}, together with the results of a
one-dimensional fit of the form~(\ref{eq:angular}).  The data are well
described by the fitted parameter $r^{04}_{00}$ at each value of
$Q^2$.


\section{Cross section}
\label{sec:xsec}

The measured $\gamma^* p$ cross sections are averaged over intervals
listed in the appropriate tables and are quoted at fixed values of
$Q^2$ and $W$. The cross sections are corrected for the mass range
$0.28 < M_{\pi\pi} < 1.5$ GeV and integrated over the full $t$-range,
where applicable.

\subsection{$t$ dependence of  $\sigma(\gamma^* p \to \rho^0 p)$}
\label{sec:dsdt}

\begin{figure}[h]
\vfill
\begin{center}
\includegraphics[width=\hsize]{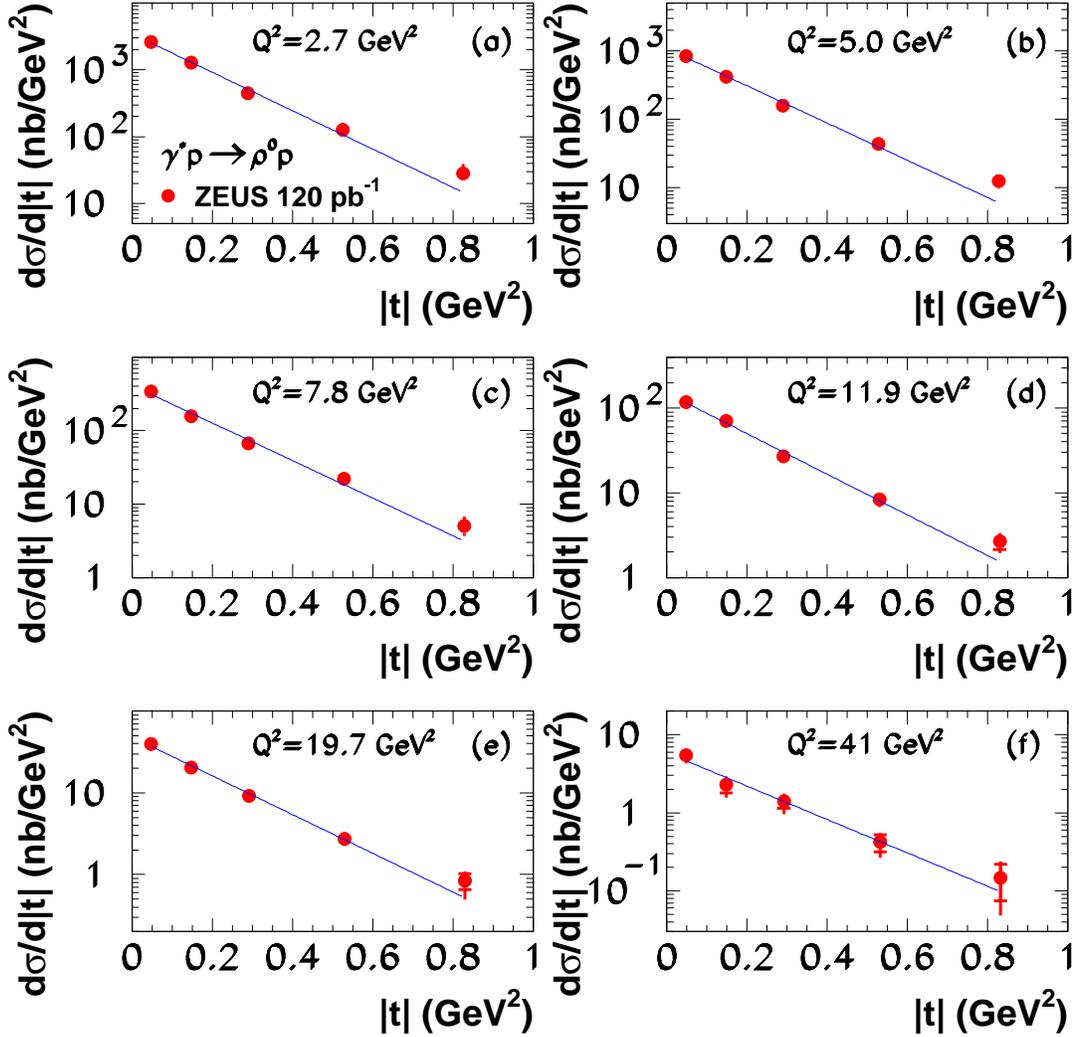}
\end{center}
\caption{\it
The differential cross-section $d\sigma/d|t|$ as a function of $|t|$ for
$\gamma^* p \to \rho^0 p$, for fixed values of $Q^2$, as indicated in
the figure. The line represents an exponential fit to the data. The
inner error bars indicate the statistical uncertainty, the outer error
bars represent the statistical and systematic uncertainty added in
quadrature.}
\label{fig:tdep}
\vfill
\end{figure}

\begin{figure}[h]
\vfill
\begin{center}
\includegraphics[width=\hsize]{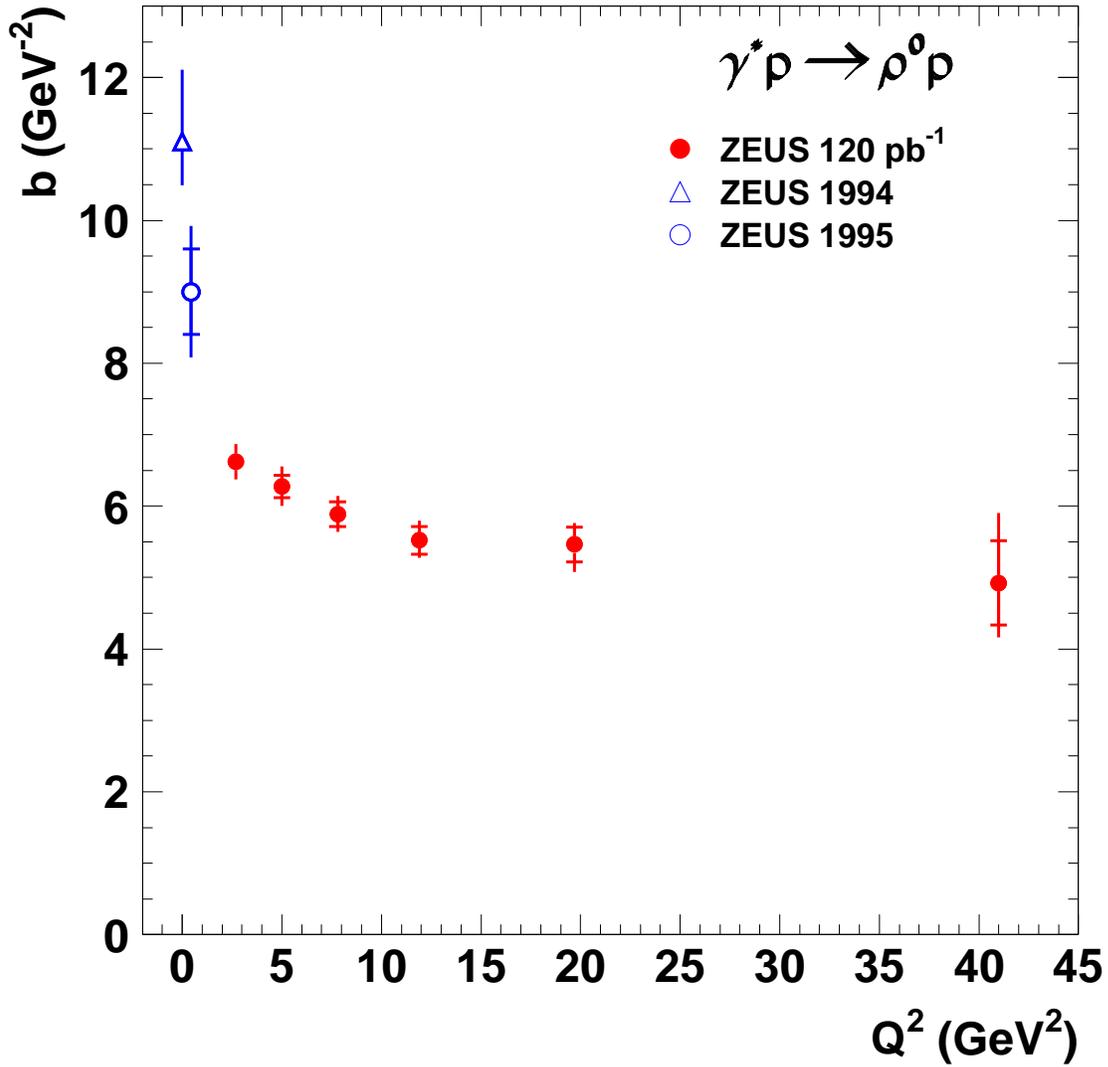}
\end{center}
\caption{\it
The value of the slope $b$ from a fit of the form $d\sigma/d|t| \propto
e^{-b|t|}$ for exclusive $\rho^0$ electroproduction, as a function
of $Q^2$. Also shown are values of $b$ obtained previously at lower
$Q^2$ values~\protect\cite{z-disrho,photob}. The
inner error bars indicate the statistical uncertainty, the outer error
bars represent the statistical and systematic uncertainty added in
quadrature.  }
\label{fig:b-q2}
\vfill
\end{figure}

The determination of $\sigma(\gamma^* p \to \rho^0 p)$ as a function
of $t$ for $W$ = 90 GeV was performed by averaging over $ 40 < W <
140$ GeV.  The differential cross-section $d\sigma/dt (\gamma^* p \to
\rho^0 p)$ is shown in Fig.~\ref{fig:tdep} and listed in
Table~\ref{tab:dsdt}, for different ranges of $Q^2$. An exponential
form proportional to $e^{-b|t|}$ was fitted to the data in each range
of $Q^2$; the results are shown in Fig.~\ref{fig:b-q2}.  The
exponent $b$, listed in Table~\ref{tab:slope}, decreases as a function
of $Q^2$. After including the previous results at lower
$Q^2$~\cite{photob,z-disrho}, a sharp decrease of $b$ is observed at low
$Q^2$; the value of $b$ then levels off at about 5 GeV$^{-2}$.

\begin{figure}[h]
\vfill
\begin{center}
\includegraphics[width=\hsize]{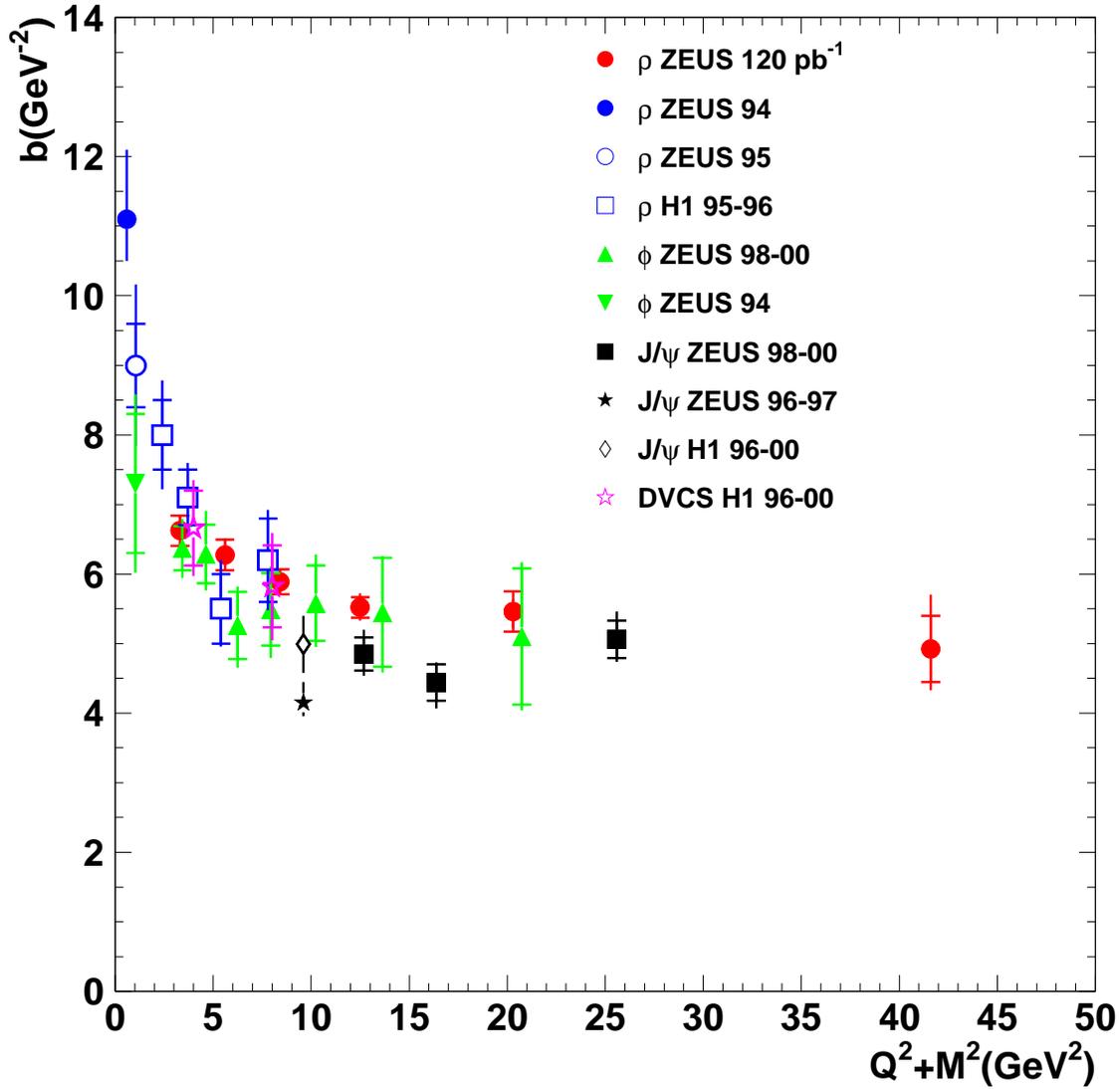}
\end{center}
\caption{\it
A compilation of the value of the slope $b$ from a fit of the form
$d\sigma/d|t| \propto
e^{-b|t|}$ for exclusive vector-meson electroproduction, as a function
of $Q^2+M^2$. Also included is the DVCS result. The
inner error bars indicate the statistical uncertainty, the outer error
bars represent the statistical and systematic uncertainty added in
quadrature.}
\label{fig:b-q2m2}
\vfill
\end{figure}

A compilation of the value of the slope $b$ for exclusive VM
electroproduction, as a function of $Q^2+M^2$, is shown in
Fig.~\ref{fig:b-q2m2}. Here $M$ is the mass of the corresponding final
state. It also includes the exclusive production of a real photon, the
deeply virtual Compton scattering (DVCS) measurement~\cite{h1dvcs}.
When $b$ is plotted as a function of $Q^2+M^2$, the trend of $b$
decreasing with increasing scale to an asymptotic value of $5\ {\rm
GeV^{-2}}$, seems to be a universal property of exclusive processes,
as expected in perturbative QCD~\cite{strikfurtHA}.


\subsection{$Q^2$ dependence of  $\sigma(\gamma^* p \to \rho^0 p)$}
\label{sec:q2}

\begin{figure}[h]
\vfill
\begin{center}
\includegraphics[width=\hsize]{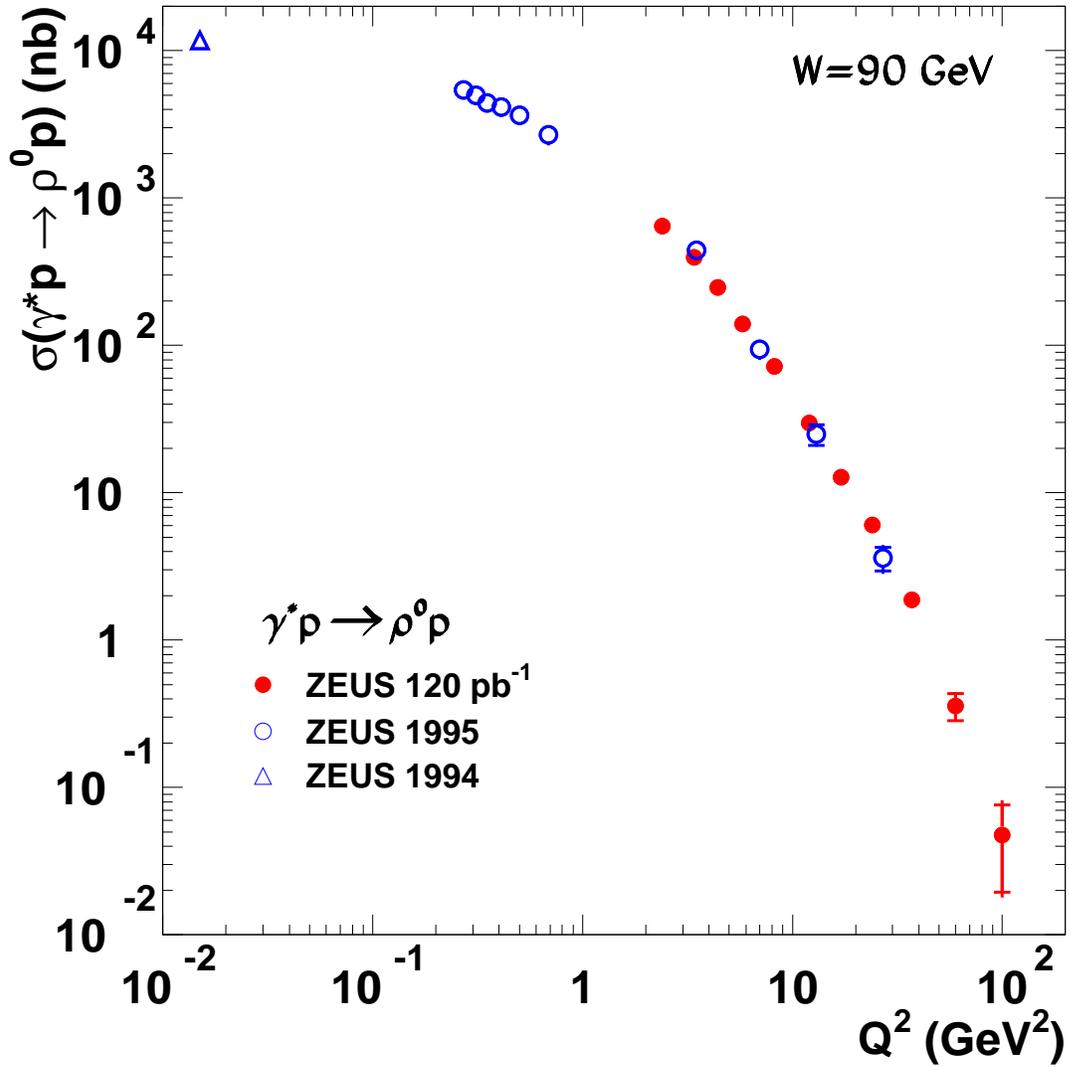}
\end{center}
\caption{\it
The $Q^2$ dependence of the cross section for exclusive $\rho^0$
electroproduction, at a $\gamma^* p$ centre-of-mass energy $W$=90
GeV. The ZEUS 1994~\protect\cite{photob} and the ZEUS
1995~\protect\cite{z-disrho} data points have been extrapolated to $W
= 90$ GeV using the parameterisations reported in the respective
publications. The
inner error bars indicate the statistical uncertainty, the outer error
bars represent the statistical and systematic uncertainty added in
quadrature.}
\label{fig:q2dep}
\vfill
\end{figure}

The determination of $\sigma(\gamma^* p \to \rho^0 p)$ as a function
of $Q^2$ for $W$ = 90 GeV was performed by averaging over $ 40 < W <
140$ GeV. The results are shown in Fig.~\ref{fig:q2dep} with
corresponding values given in Table~\ref{tab:q2dep}. As expected,
a steep decrease of the cross section with $Q^2$ is observed. The
photoproduction and the low-$Q^2$ ($<$ 1 GeV$^2$) measurements are also
shown in the figure. An attempt to fit the $Q^2$ dependence with a
simple propagator term
\begin{equation}
\sigma(\gamma^* p \to \rho^0 p)\sim (Q^2 + m_\rho^2)^{-n}, \nonumber
\label{eq:prop-q2}
\end{equation}
with the normalisation and $n$ as free parameters, failed to produce
results with an acceptable $\chi^2$. The data appear to favour an $n$
value which increases with $Q^2$.

\subsection{$W$ dependence of  $\sigma(\gamma^* p \to \rho^0 p)$}
\label{sec:W}

\begin{figure}[h]
\vfill
\begin{center}
\includegraphics[width=\hsize]{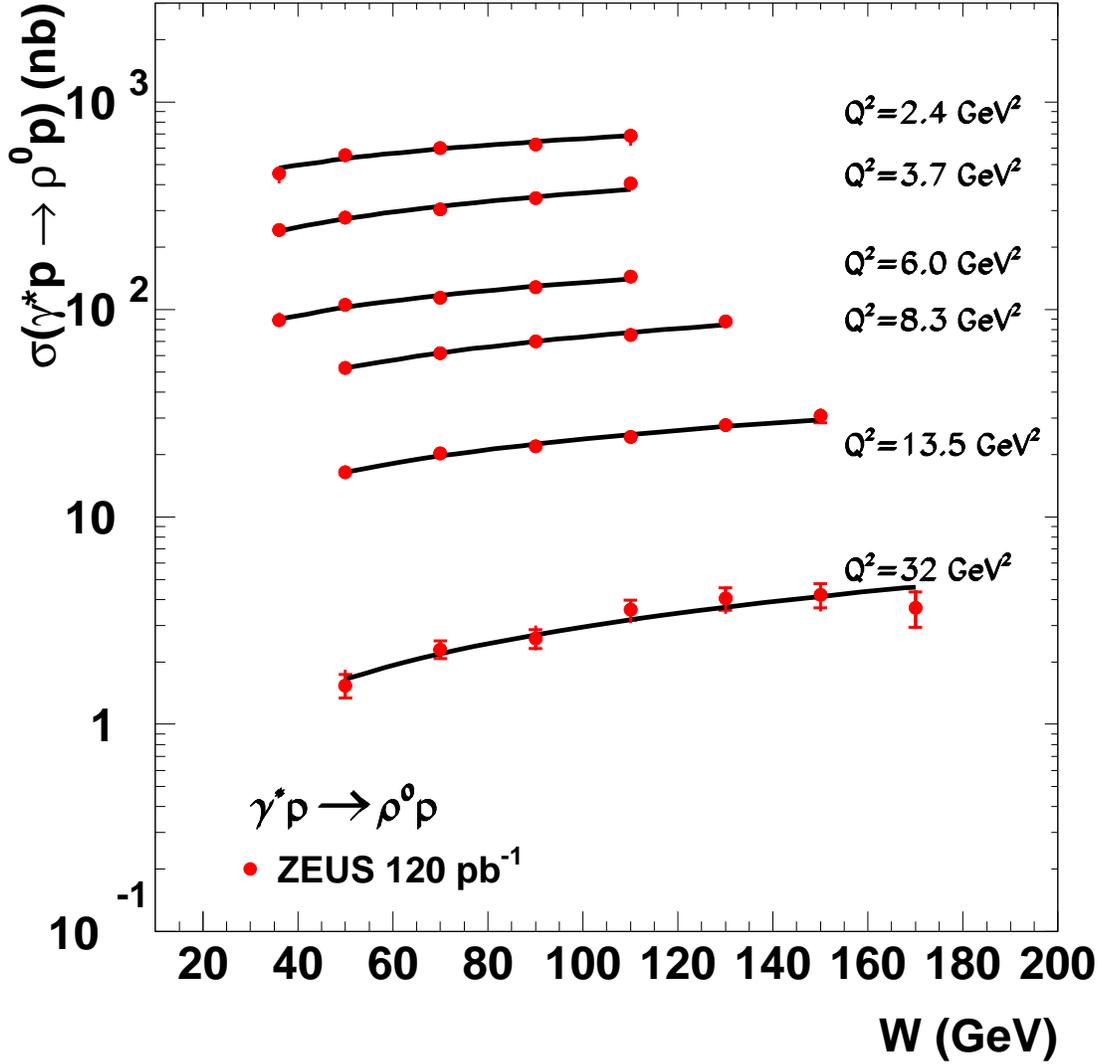}
\end{center}
\caption{\it
The $W$ dependence of the cross section for exclusive $\rho^0$
electroproduction, for different $Q^2$ values, as indicated in the
figure. The
inner error bars indicate the statistical uncertainty, the outer error
bars represent the statistical and systematic uncertainty added in
quadrature. The lines are the result of a fit of the form $ W^\delta$
to the data.}
\label{fig:wdep}
\vfill
\end{figure}

The values of the cross section $\sigma(\gamma^* p \to \rho^0 p)$ as a
function of $W$, for fixed values of $Q^2$, are plotted in
Fig.~\ref{fig:wdep} and given in Table~\ref{tab:wdep}. The cross
sections increase with increasing $W$, with the rate of increase growing
with increasing $Q^2$.

\begin{figure}[h]
\vfill
\begin{center}
\includegraphics[width=\hsize]{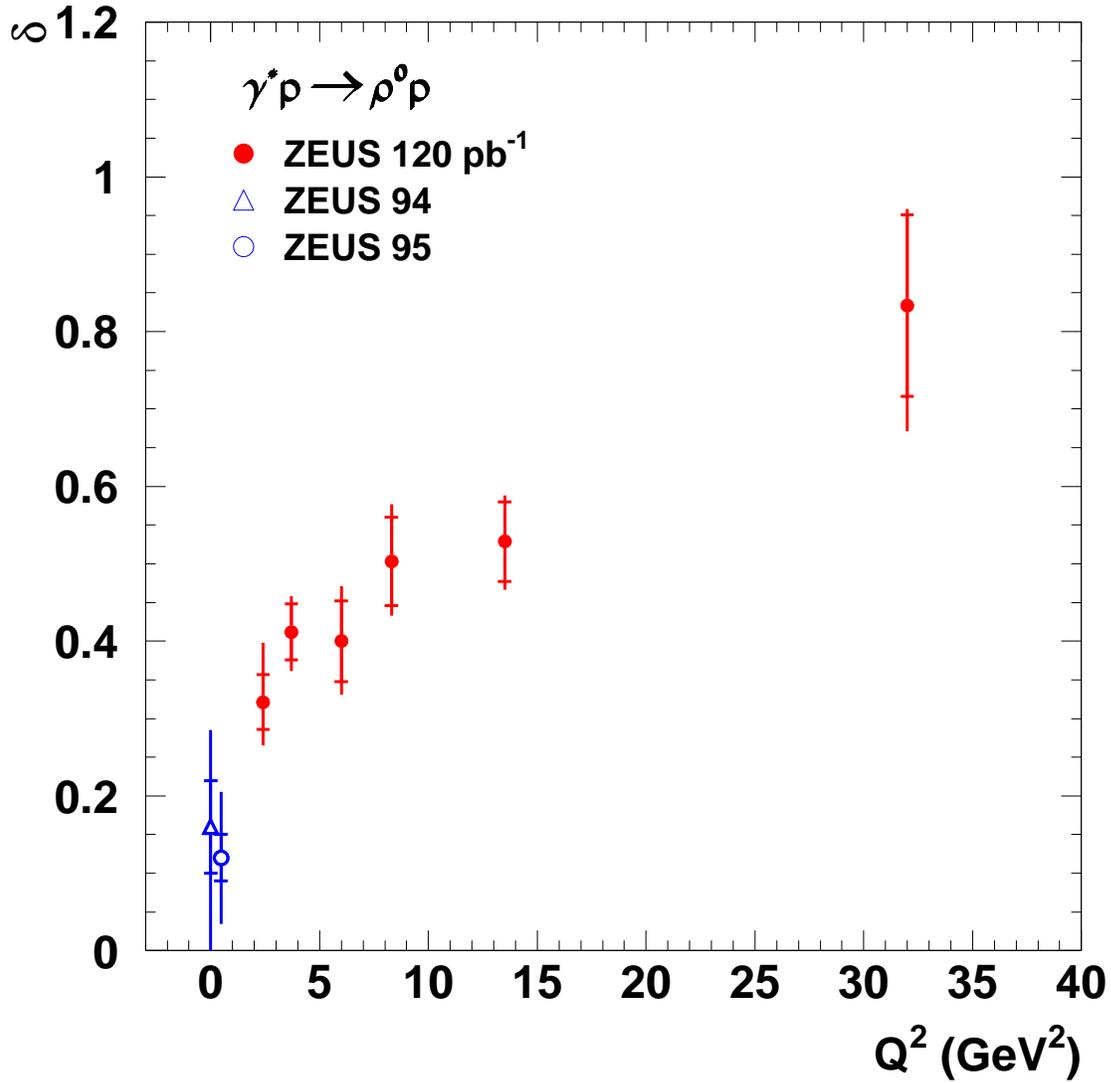}
\end{center}
\caption{\it
The value of $\delta$ from a fit of the form $W^\delta$ for exclusive
$\rho^0$ electroproduction , as a function of $Q^2$. Also shown are
values of $\delta$ obtained previously at lower $Q^2$
values~\protect\cite{z-disrho,photob}. The
inner error bars indicate the statistical uncertainty, the outer error
bars represent the statistical and systematic uncertainty added in
quadrature.}
\label{fig:delta}
\vfill
\end{figure}

In order to quantify the rate of growth and its significance, the $W$
dependence for each $Q^2$ value was fitted to the functional form
\begin{equation}
\label{eq:wdep}
\sigma \sim W^\delta. \nonumber
\end{equation}
The resulting $\delta$ values are presented as a function of $Q^2$ in
Fig.~\ref{fig:delta} and listed in Table~\ref{tab:delta}. For
completeness, the $\delta$ values from lower $Q^2$ are also
included. A clear increase of $\delta$ with $Q^2$ is observed. Such an
increase is expected in pQCD, and reflects the change of the low-$x$
gluon distribution of the proton with $Q^2$.

\begin{figure}[h]
\vfill
\begin{center}
\includegraphics[width=\hsize]{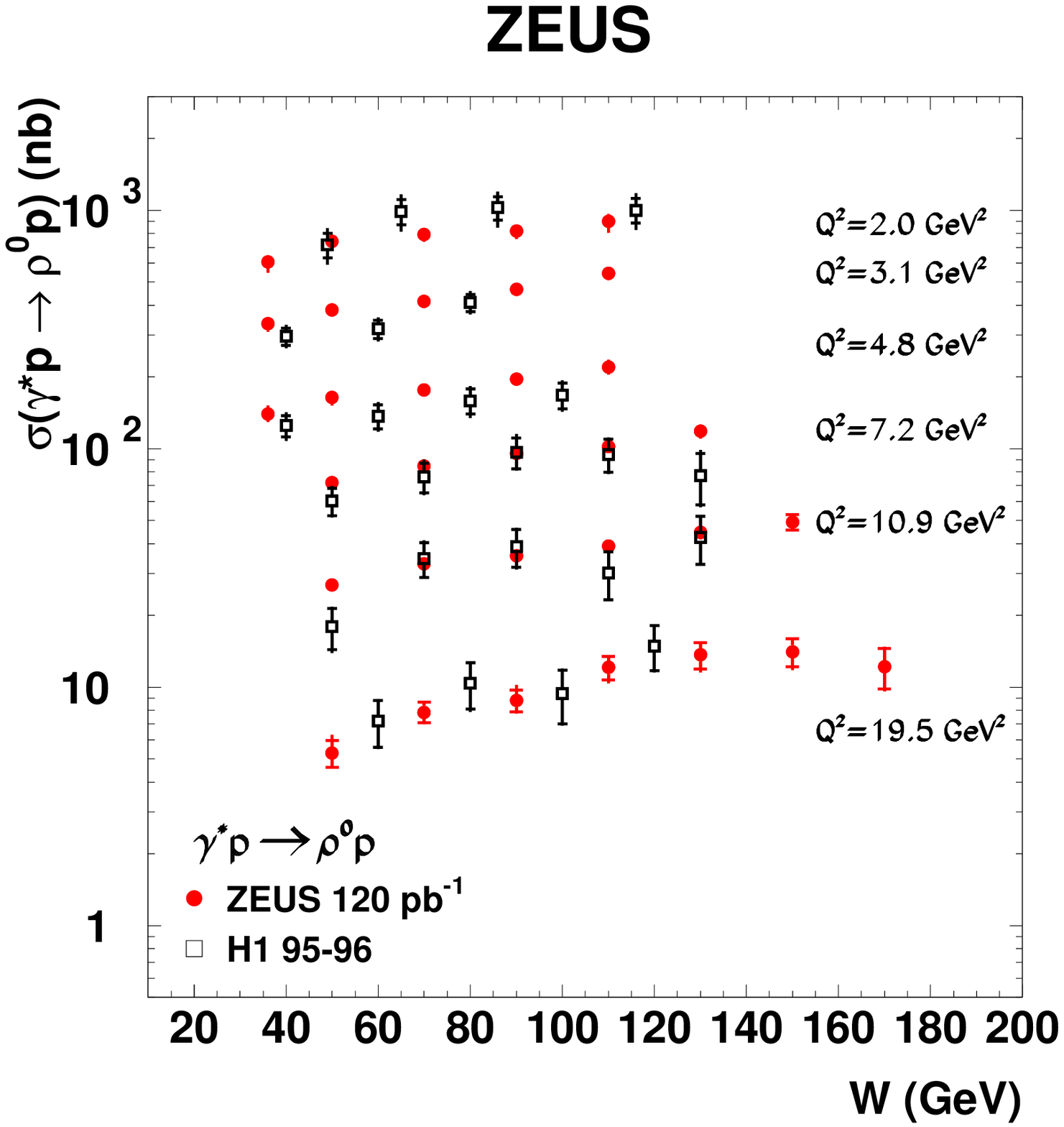}
\end{center}
\caption{\it
Comparison of the H1 (squares) and ZEUS (dots) measurements of the $W$
dependence of $\sigma^{\gamma^* p \to \rho^0 p}$, for different $Q^2$
values, as indicated in the figure. The
inner error bars indicate the statistical uncertainty, the outer error
bars represent the statistical and systematic uncertainty added in
quadrature. }
\label{fig:wdep-z-h1}
\vfill
\end{figure}

To facilitate the comparison, the ZEUS cross-section data as a
function of $W$ have been replotted in the $Q^2$ bins used by
H1~\cite{h1-disrho}.  The results are shown in
Fig.~\ref{fig:wdep-z-h1}. The agreement between the two measurements
is reasonable. However, in some $Q^2$ bins the shape of the $W$
dependence is somewhat different.

\begin{figure}[h]
\vfill
\begin{center}
\includegraphics[width=\hsize]{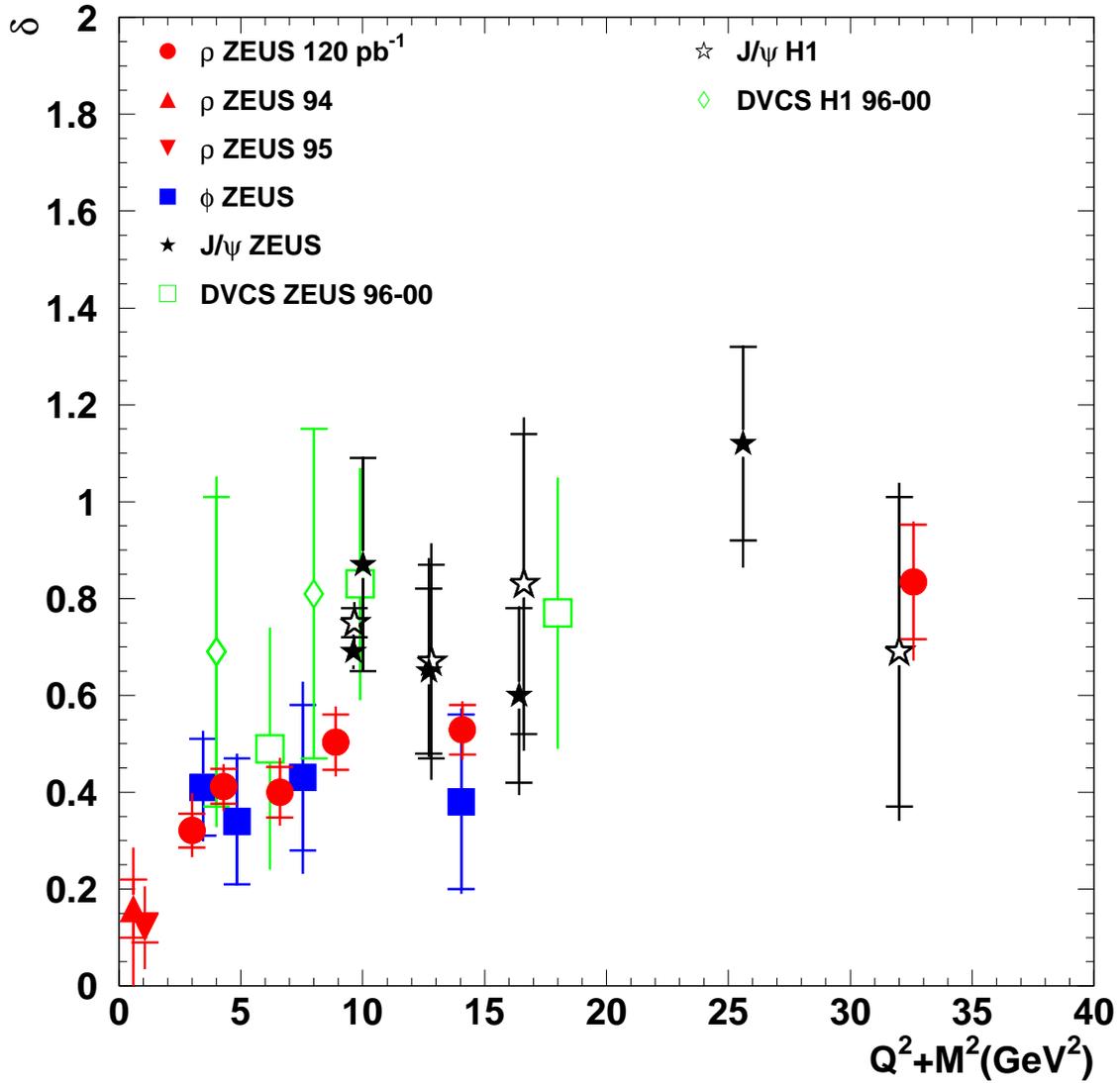}
\end{center}
\caption{\it
A compilation of the value of $\delta$ from a fit of the form
$W^\delta$ for exclusive vector-meson electroproduction, as a function
of $Q^2+M^2$. It includes also the DVCS results. The
inner error bars indicate the statistical uncertainty, the outer error
bars represent the statistical and systematic uncertainty added in
quadrature.}
\label{fig:delta07-pub}
\vfill
\end{figure}

A compilation of the value of the slope $\delta$ for exclusive VM
electroproduction, as a function of $Q^2+M^2$, is shown in
Fig.~\ref{fig:delta07-pub}. It also includes the DVCS
result~\cite{h1dvcs}. When plotted as a function of $Q^2+M^2$, the value
of $\delta$ and its increase with the scale are similar for
all the exclusive processes, as expected in perturbative
QCD~\cite{strikfurtHA}.


\section{$R = \sigma_L/\sigma_T$ and $r^{04}_{00}$}
\label{sec:R}

The SCHC hypothesis implies that $r_{1-1}^1=-{\rm
Im}\{r_{1-1}^2\}$ and ${\rm Re}\{r_{10}^5\}=-{\rm Im}\{r_{10}^6\}$.
In this case, the ratio $R=\sigma_L/\sigma_T$ can be related to the
$r^{04}_{00}$ matrix element,
\begin{equation}
R = \frac{1}{\epsilon}  \frac{r^{04}_{00}}{1 - r^{04}_{00}},
\label{eq:Rr}
\end{equation}
and thus can be extracted from the $\theta_h$ distribution alone.

If the SCHC requirement is relaxed, then the relation between $R$ and
$r^{04}_{00}$ is modified,
\begin{equation}\label{eq:R_ratio_no_schc}
R = \frac{1}{\epsilon} \frac{r^{04}_{00}-\Delta^2}{1 - (r^{04}_{00}-\Delta^2)}, \nonumber
\end{equation}
\noindent
with
\begin{equation}\label{eq:delta}
\Delta \simeq \frac{r^5_{00}}{\sqrt{2r^{04}_{00}}}. \nonumber
\end{equation}
\noindent
In the kinematic range of the measurements presented in this paper,
the non-zero value of $\Delta$ implies a correction of $\sim$3\% on $R$ up
to the highest $Q^2$ value, where it is $\sim$10\%, and is
neglected.

Under the assumption that Eq.~(\ref{eq:Rr}) is valid and for values of
$\epsilon$ studied in this paper, $<\epsilon>$=0.996, the matrix
element $r^{04}_{00}$ may be interpreted as
\begin{equation}
 r^{04}_{00} = \sigma_L/\sigma_\mathrm{tot},  \nonumber
\end{equation}
where $\sigma_{\rm tot} = \sigma_L + \sigma_T$. When the value of
$r^{04}_{00}$ is close to one, as is the case for this analysis, the
error on $R$ becomes large and highly asymmetrical. It is then
advantageous to study the properties of $r^{04}_{00}$ itself which
carries the same information, rather than $R$.

\begin{figure}[h]
\vfill
\begin{center}
\includegraphics[width=\hsize]{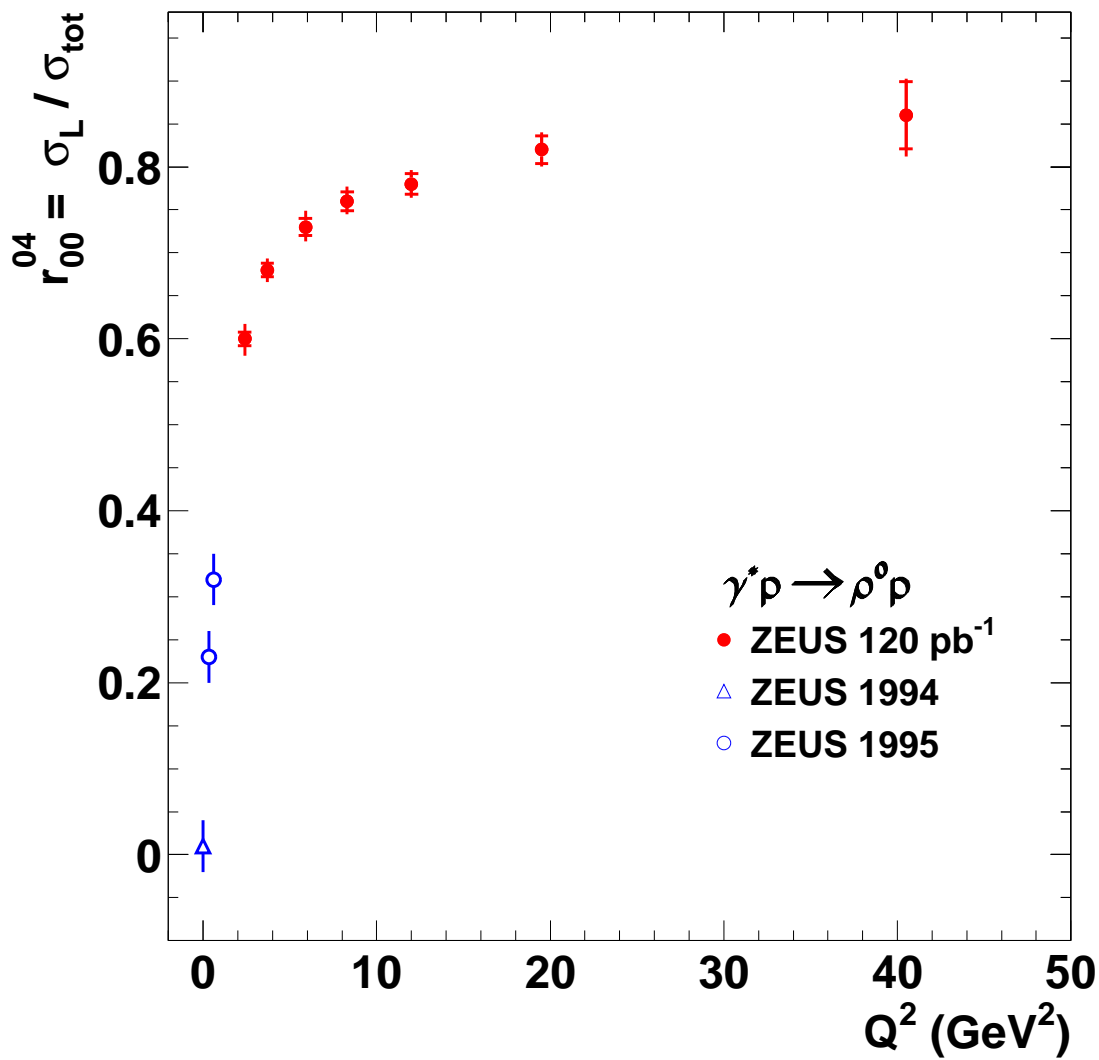}
\end{center}
\caption{\it
The ratio $r^{04}_{00}$ as a function of $Q^2$ for $W =$ 90 GeV.  Also
included are values of $r^{04}_{00}$ from previous measurements at
lower $Q^2$ values~\protect\cite{z-disrho,photob}. The
inner error bars indicate the statistical uncertainty, the outer error
bars represent the statistical and systematic uncertainty added in
quadrature.}
\label{fig:r04vsq2}
\vfill
\end{figure}

The $Q^2$ dependence of $r^{04}_{00}$ for $W$ = 90 GeV, averaged over
the range $40 < W <140$ GeV, is shown in Fig.~\ref{fig:r04vsq2} and
listed in Table~\ref{tab:R-q2} together with the corresponding $R$
values.  The figure includes three data points at lower $Q^2$ from
previous studies~\cite{photob,z-disrho}.  An initial steep rise of
$r^{04}_{00}$ with $Q^2$ is observed and above $Q^2\simeq 10$~GeV$^2$, 
the rise with $Q^2$ becomes milder.  At $Q^2$ = 40 GeV$^2$, $\sigma_L$
constitutes about 90\% of the total $\gamma^*p$ cross section.

\begin{figure}[h]
\vfill
\begin{center}
\includegraphics[width=\hsize]{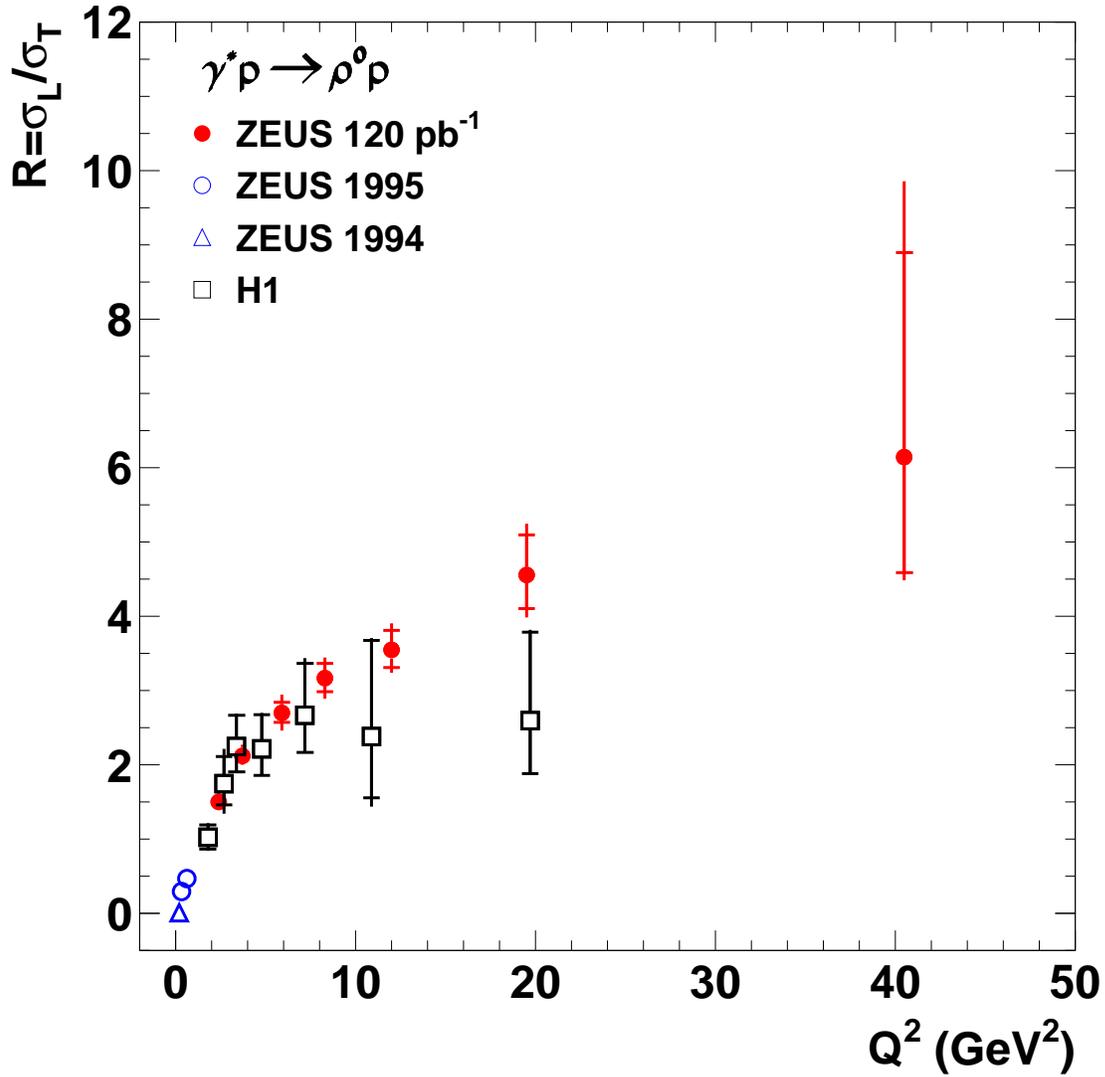}
\end{center}
\caption{\it
Comparison of the H1 (squares) and ZEUS (dots)measurements of $R$ as a
function of $Q^2$. The H1 data are at $W$ = 75 GeV and those of ZEUS
at $W$ = 90 GeV. Also included are measurements performed previously
at lower $Q^2$ values~\protect\cite{z-disrho,photob}. The
inner error bars indicate the statistical uncertainty, the outer error
bars represent the statistical and systematic uncertainty added in
quadrature.}
\label{fig:R-Q2-tot+H1}
\vfill
\end{figure}

The comparison of the H1 and ZEUS results is presented in
Fig.~\ref{fig:R-Q2-tot+H1} in terms of the ratio $R$. The H1
measurements are at $W$ = 75 GeV and those of ZEUS at $W$ = 90 GeV.
Given the fact that $R$ seems to be independent of $W$ (see
below), both data sets can be directly compared.  The two measurements
are in good agreement.

\begin{figure}[h]
\vfill
\begin{center}
\includegraphics[width=\hsize]{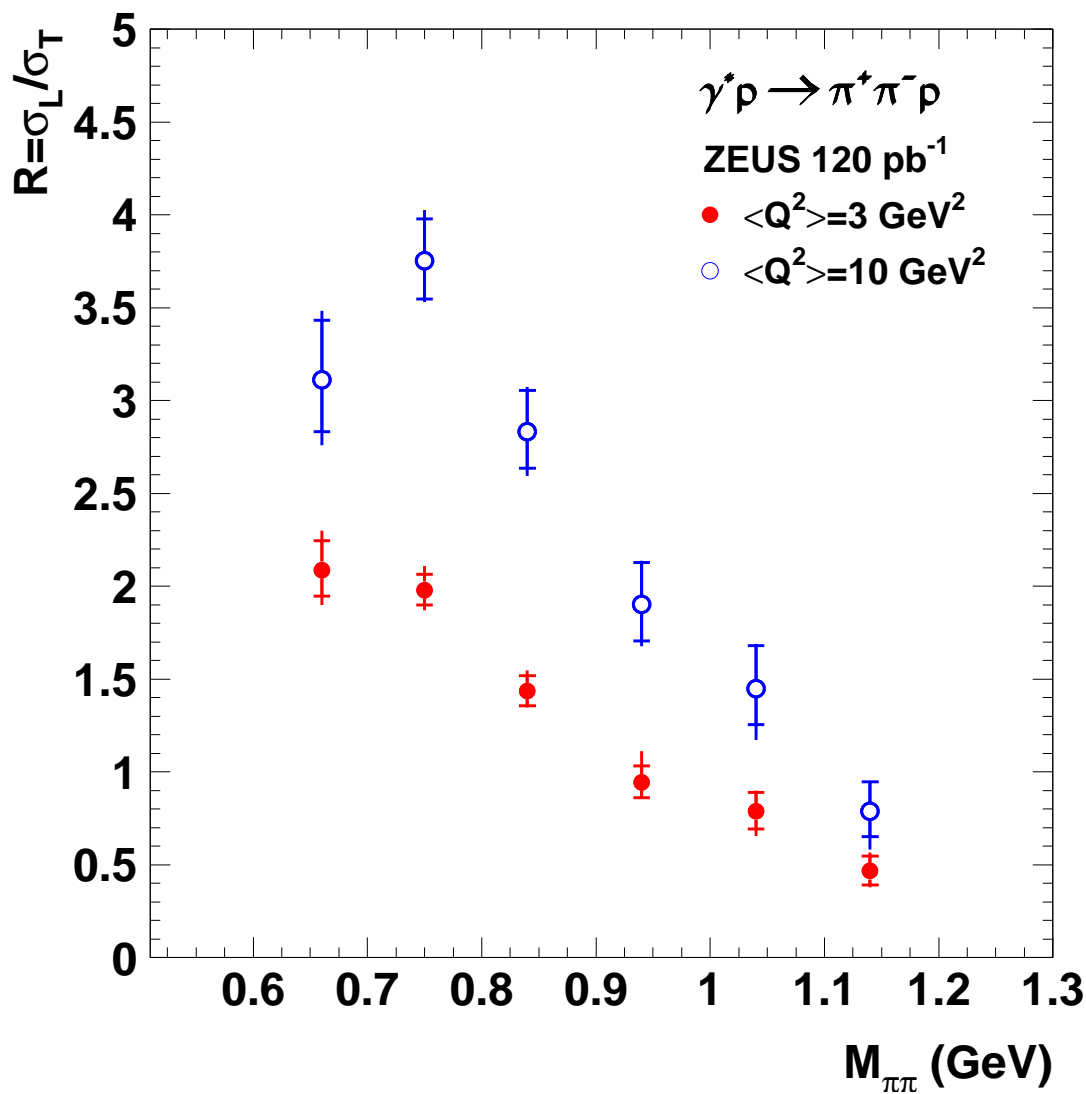}
\end{center}
\caption{\it
The ratio $R$ as a function of $M_{\pi\pi}$, for $W$ = 80 GeV, and for
two values of $Q^2$, as indicated in the figure. The
inner error bars indicate the statistical uncertainty, the outer error
bars represent the statistical and systematic uncertainty added in
quadrature.
}
\label{fig:R-mpipi}
\vfill
\end{figure}

The dependence of $R$ on $M_{\pi\pi}$ is presented in
Fig.~\ref{fig:R-mpipi} for two $Q^2$ intervals. The value of $R$ falls
rapidly with $M_{\pi\pi}$ above the central $\rho^0$ mass value.
Although a change of $R$ with $M_{\pi\pi}$ was anticipated to be
$\sim$ 10\%~\cite{shabelsky}, the effect seen in the data is much
stronger. The effect remains strong also at higher $Q^2$, contrary to
expectations~\cite{shabelsky}.  Once averaged over the $\rho^0$ mass
region, the main contribution to $R$ comes from the central $\rho^0$
mass value.

\begin{figure}[h]
\vfill
\begin{center}
\includegraphics[width=\hsize]{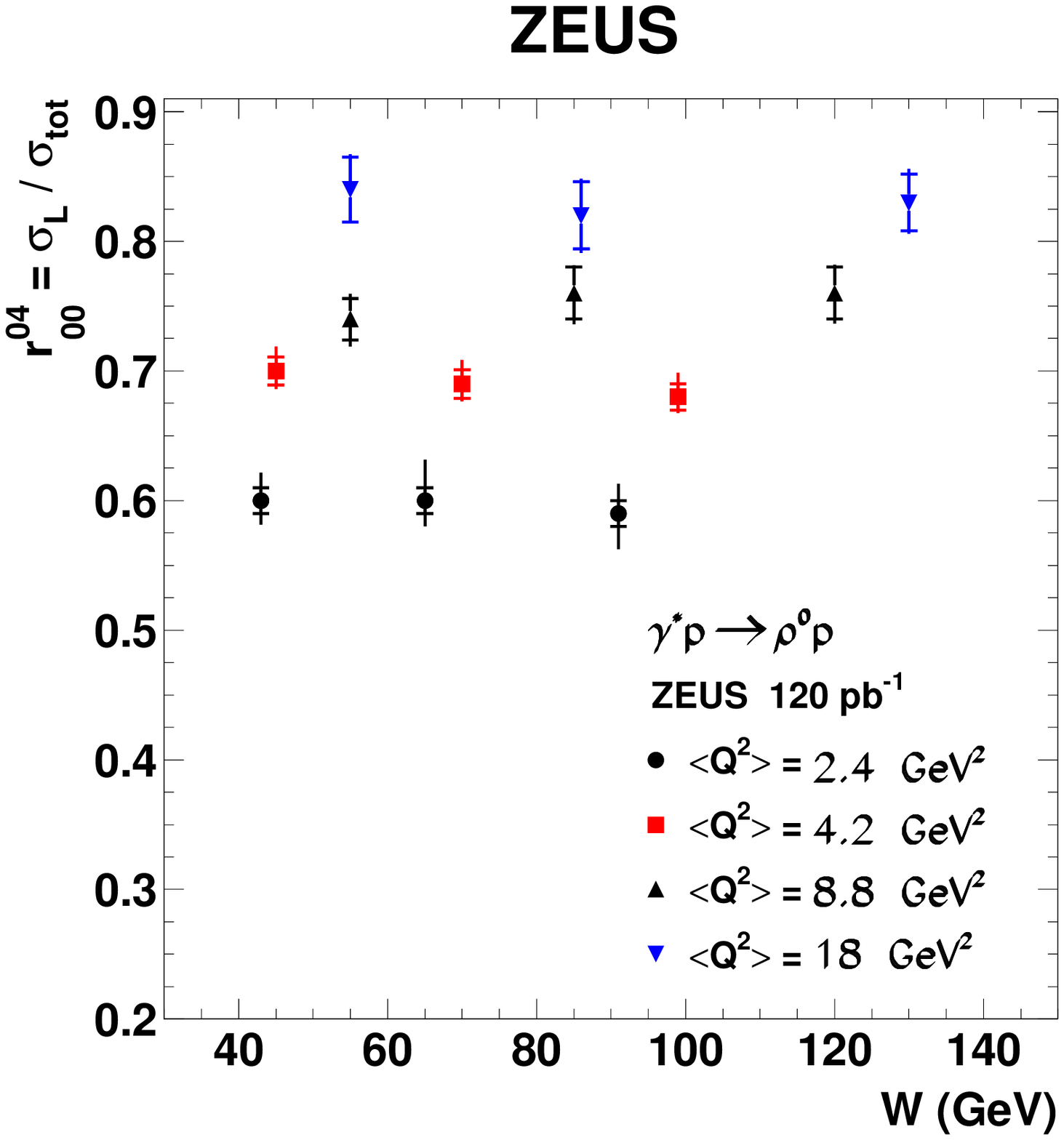}
\end{center}
\caption{\it
The ratio $r^{04}_{00}$ as a function of $W$ for different values of $Q^2$, as
indicated in the figure. The
inner error bars indicate the statistical uncertainty, the outer error
bars represent the statistical and systematic uncertainty added in
quadrature.}
\label{fig:R-W-all}
\vfill
\end{figure}

The $W$ dependence of $r^{04}_{00}$, for different values of $Q^2$, is
shown in Fig.~\ref{fig:R-W-all} and listed in Table~\ref{tab:R-w}.
Within the measurement uncertainties, $r^{04}_{00}$ is independent of
$W$, for all $Q^2$ values. This implies that the $W$ behaviour of
$\sigma_L$ is the same as that of $\sigma_T$, a result which is
somewhat surprising. The $q\bar{q}$ configurations in the wave
function of $\gamma^*_L$ have typically a small transverse size, while
the configurations contributing to $\gamma^*_T$ may have large
transverse size.  The contribution to $\sigma_T$ of large-size
$q\bar{q}$ configurations, which are more hadron-like, is expected to
lead to a shallower $W$ dependence than in case of $\sigma_L$.  Thus,
the result presented in Fig.~\ref{fig:R-W-all} suggests that the
large-size configurations of the transversely polarised photon are
suppressed.

\begin{figure}[h]
\vfill
\begin{center}
\includegraphics[width=\hsize]{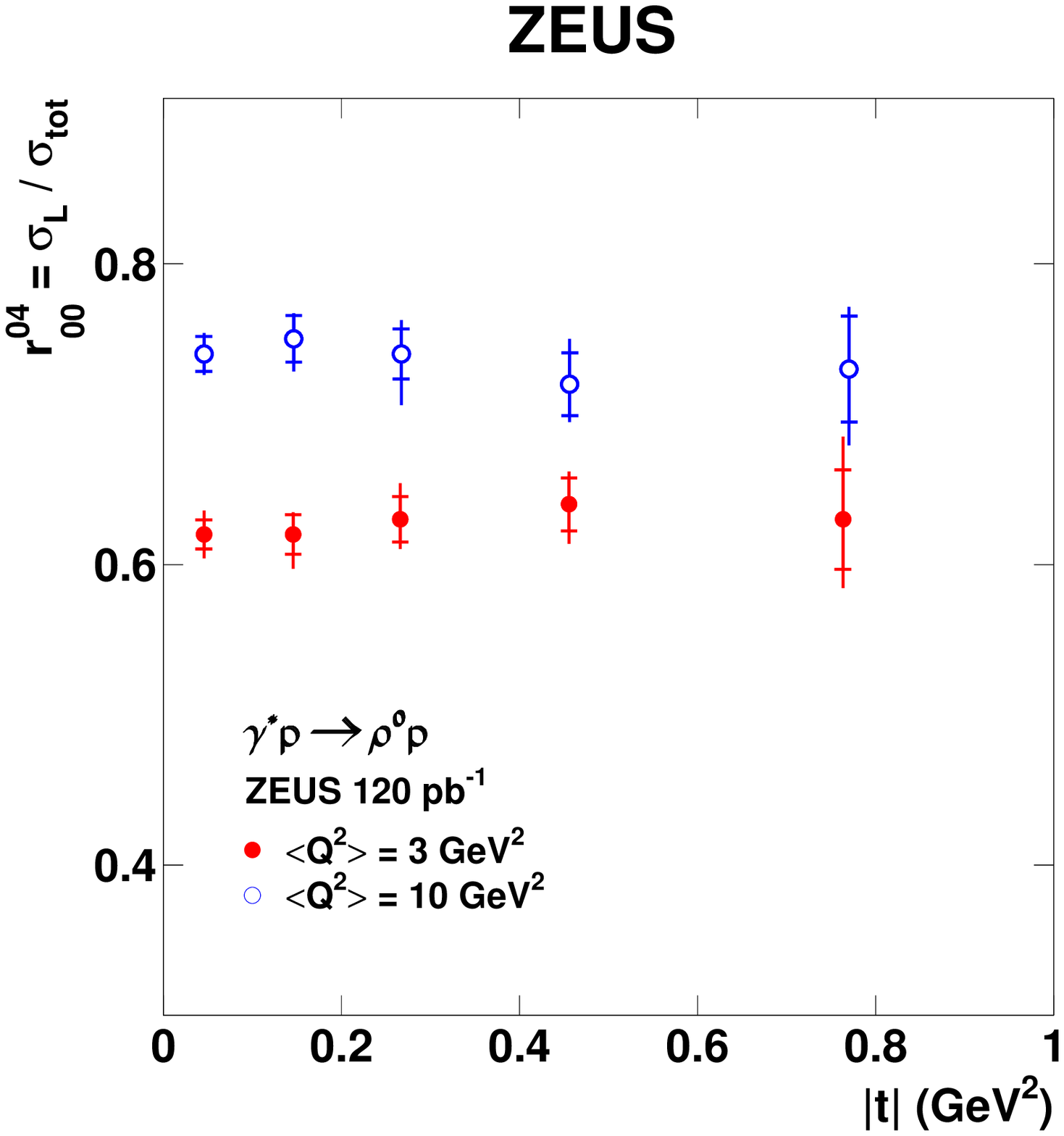}
\end{center}
\caption{\it
The ratio $r^{04}_{00}$ as a function of $|t|$ for different values of $Q^2$, as
indicated in the figure. The
inner error bars indicate the statistical uncertainty, the outer error
bars represent the statistical and systematic uncertainty added in
quadrature.}
\label{fig:R-t}
\vfill
\end{figure}

The above conclusion can also explain the behaviour of $r^{04}_{00}$
as a function of $t$, shown in Fig.~\ref{fig:R-t} and presented in
Table~\ref{tab:R-t} for two $Q^2$ values. Different sizes of
interacting objects imply different $t$ distributions, in particular a
steeper $d\sigma_T/dt$ compared to $d\sigma_L/dt$. This turns out not
to be the case.  In both $Q^2$ ranges, $r^{04}_{00}$ is independent of
$t$, reinforcing the earlier conclusion about the suppression of the
large-size configurations in the transversely polarised photon.


\section{Effective Pomeron trajectory}
\label{sec:alpha}

An effective Pomeron trajectory can be determined from exclusive
$\rho^0$ electroproduction by using Eq.~(\ref{eq:apom}). Since the $W$
dependence of the proton-dissociative contribution was established to
be the same as the exclusive $\rho^0$ sample, no subtraction for
proton-dissociative events was performed.

\begin{figure}[h]
\vfill
\begin{center}
\includegraphics[width=\hsize]{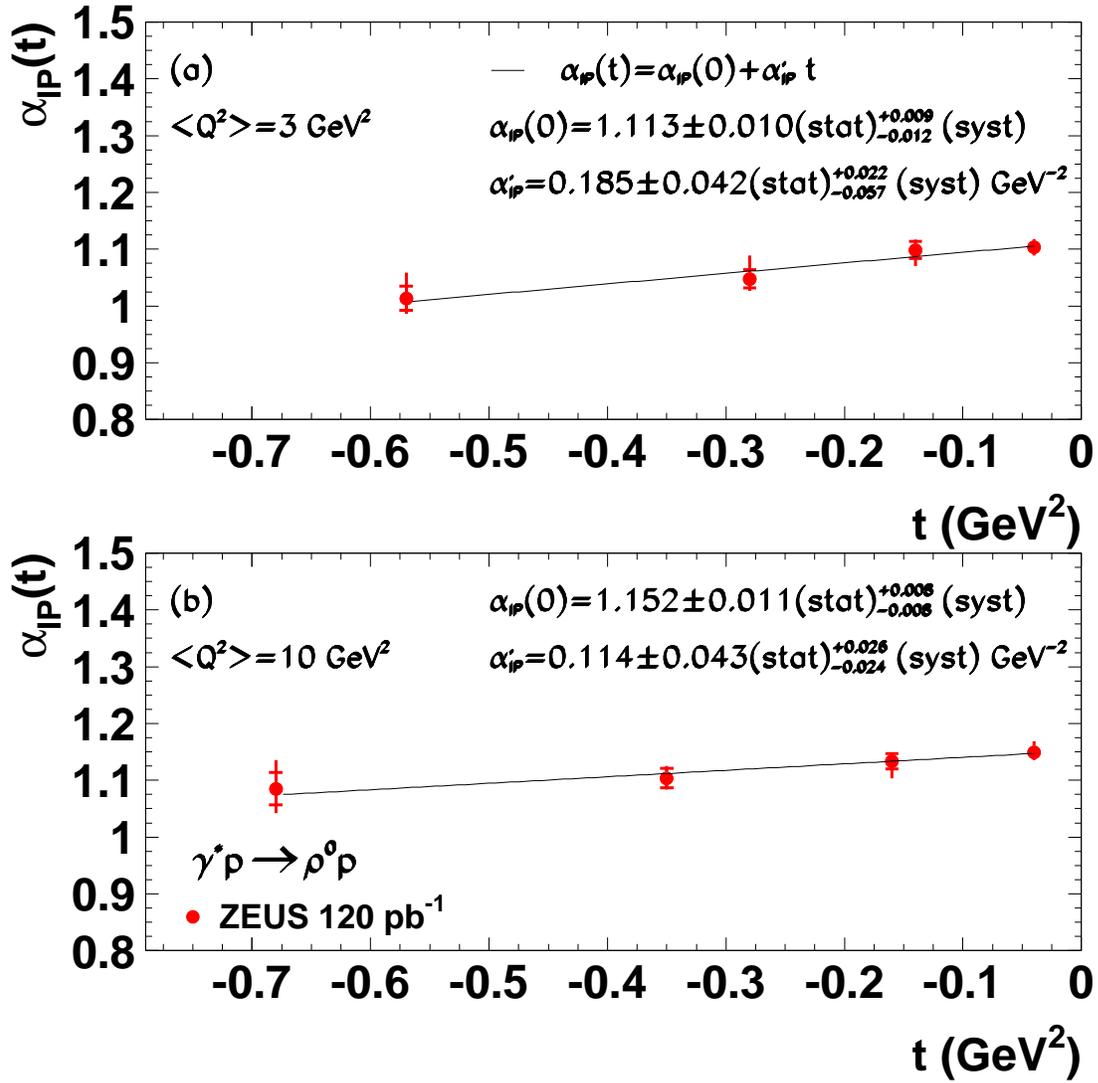}
\end{center}
\caption{\it
The effective Pomeron trajectory $\apom(t)$ as a function of $t$, for two values
of $Q^2$, with average values indicated in the figure. The
inner error bars indicate the statistical uncertainty, the outer error
bars represent the statistical and systematic uncertainty added in
quadrature.}
\label{fig:trajectory}
\vfill
\end{figure}

\begin{figure}[h]
\vfill
\begin{center}
\includegraphics[width=\hsize]{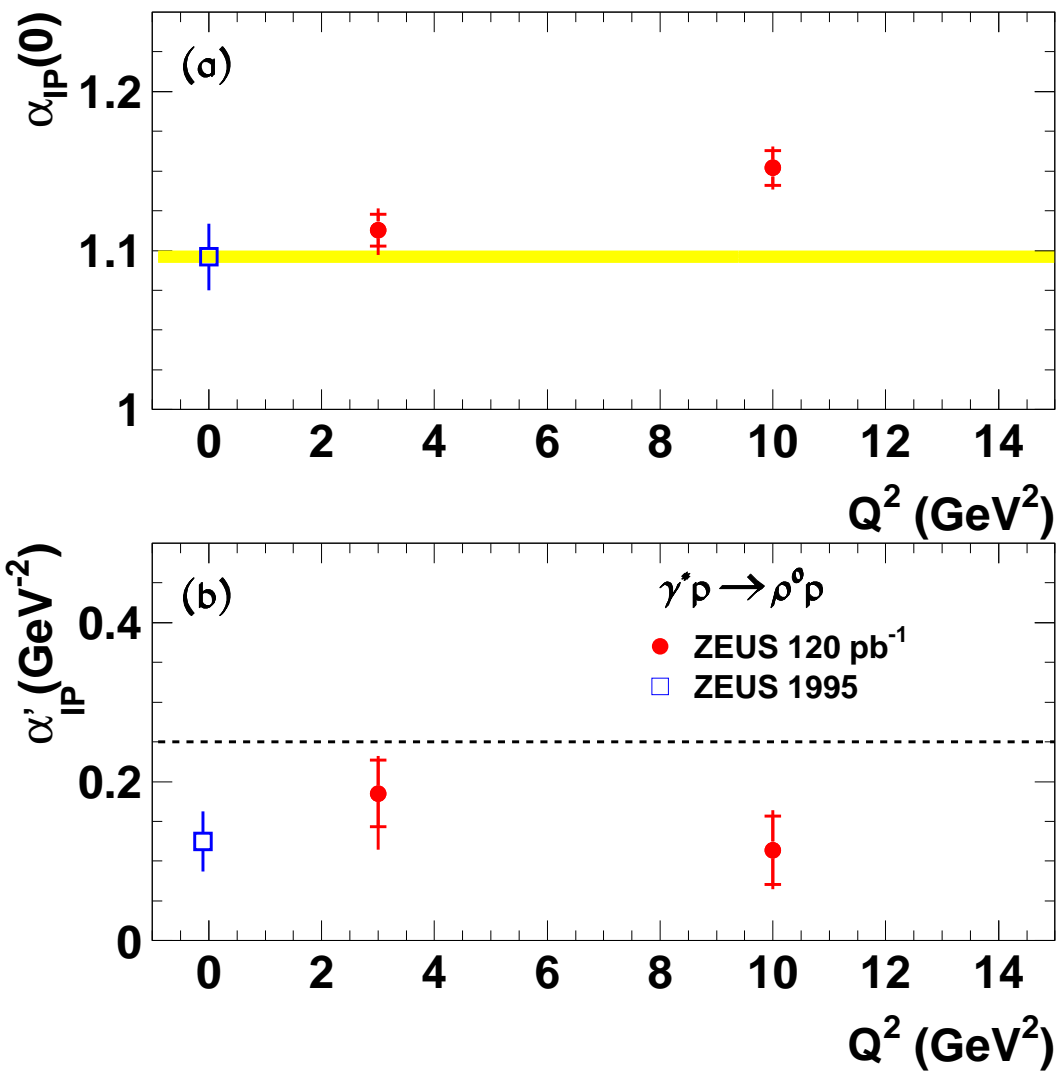}
\end{center}
\caption{\it
The parameters of the effective Pomeron trajectory in exclusive
$\rho^0$ electroproduction, (a) $\apom(0)$ and (b) $\aprime$, as a
function of $Q^2$. The inner error bars indicate the statistical
uncertainty, the outer error bars represent the statistical and
systematic uncertainty added in quadrature.  The band in (a) and the
dashed line in (b) are at the values of the parameters of the soft
Pomeron~\protect\cite{cudell,landshoff-aprime}.}
\label{fig:alpha}
\vfill
\end{figure}

A study of the $W$ dependence of the differential $d\sigma/dt$ cross
section at fixed $t$ results in values of $\apom(t)$, listed in
Table~\ref{tab:trajectory} and displayed in Fig.~\ref{fig:trajectory},
for $Q^2$ = 3 GeV$^2$ (upper plot) and 10 GeV$^2$ (lower plot). A
linear fit of the form of Eq.~(\ref{eq:trajectory}), shown in the figures,
yields values of $\apom(0)$ and $\aprime$ shown in
Fig.~\ref{fig:alpha}, and listed in Table~\ref{tab:alpha}. The value
of $\apom(0)$ increases slightly with $Q^2$, while the value of
$\aprime$ is $Q^2$ independent, within the measurement
uncertainties. Its value tends to be lower than that of the soft
Pomeron~\cite{dl}.

\begin{figure}[h]
\vfill
\begin{center}
\includegraphics[width=\hsize]{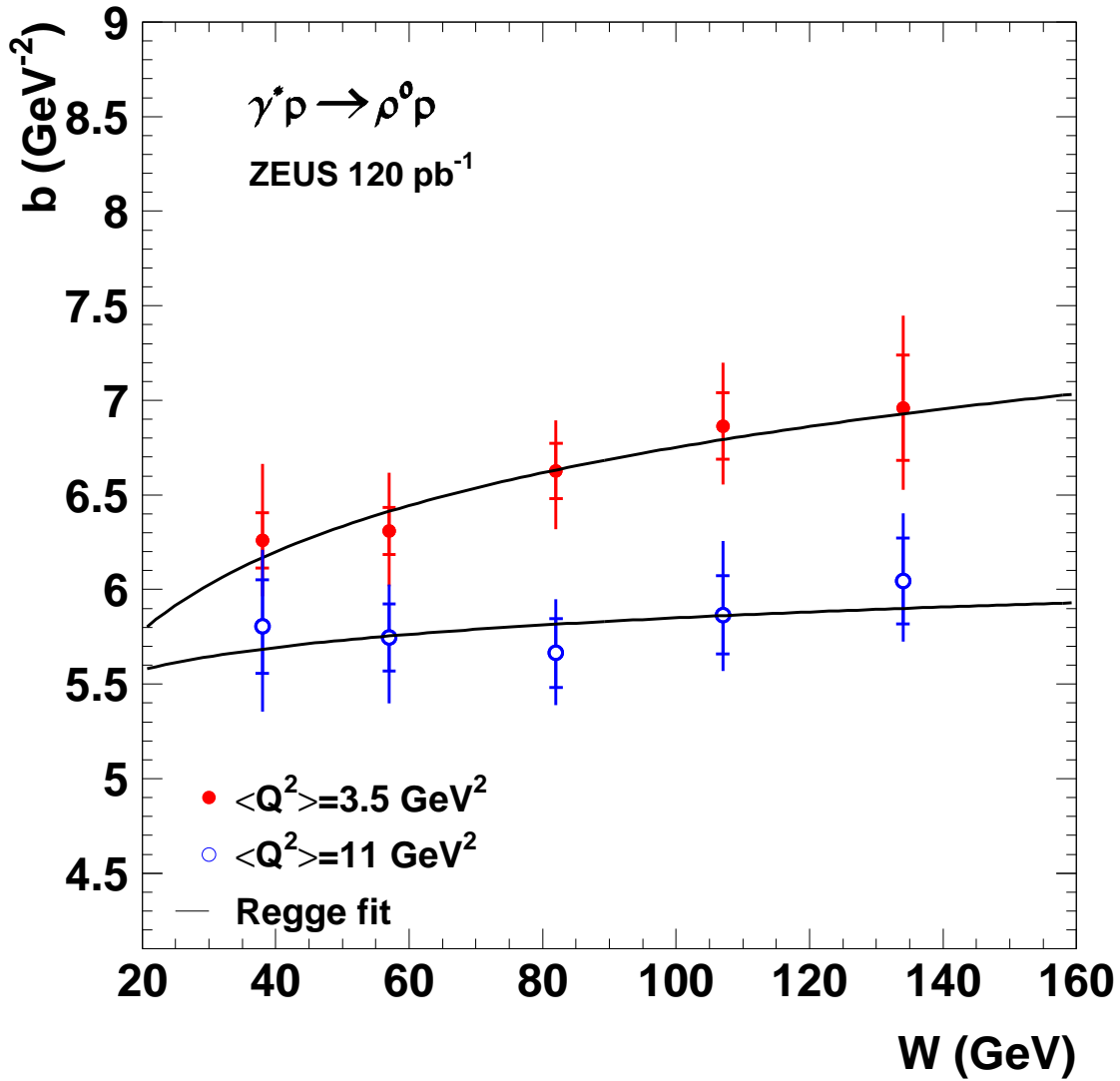}
\end{center}
\caption{\it
The $b$ slope as a function of $W$ for two ranges of $Q^2$, with
average values as indicated in the figure. The
inner error bars indicate the statistical uncertainty, the outer error
bars represent the statistical and systematic uncertainty added in
quadrature. The lines are the
results of fitting Eq.~(\ref{eq:apom}) to the data.}
\label{fig:b-w}
\vfill
\end{figure}

An alternative way of measuring the slope of the Pomeron trajectory is
to study the $W$ dependence of the $b$ slope, for fixed $Q^2$
values. Figure~\ref{fig:b-w} displays the values of $b$ as a function
of $W$ for two $Q^2$ intervals (see also Table~\ref{tab:slopew}). The
curves are a result of fitting the data to the expression $b = b_0 +
4\aprime \ln (W/W_0)$. The resulting slopes of the trajectory are
$\aprime$~=~$0.15~\pm~0.04~(stat.)~^{+0.04}_{-0.06}~(syst.)$ for
$<Q^2>$ = 3.5 GeV$^2$ and $\aprime$~=~$0.04~\pm~0.06~(stat.)~^{+0.07}_{-0.02}~(syst.)$
for $<Q^2>$ = 11 GeV$^2$. These results are consistent with those
presented in Table~\ref{tab:alpha}.


\section{Comparison to models}
\label{sec:comp}

In this section, predictions from several pQCD-inspired models are
compared to the measurements.

\subsection{The models}
All models are based on the dipole representation of the virtual
photon, in which the photon first fluctuates into a $q\bar{q}$ pair
(the colour dipole), which then interacts with the proton to produce
the $\rho^0$. The ingredients necessary in such calculations are the
virtual-photon wave-function, the dipole-proton cross section, and the
$\rho^0$ wave-function. The photon wave-function is known from QED.
The models differ in the treatment of the dipole-proton cross section
and the assumed $\rho^0$ wave-function.

The models of Frankfurt, Koepf and Strikman (FKS)~\cite{fks1,fks2} and of
Martin, Ryskin and Teubner (MRT)~\cite{mrt1,mrt2} are based on two-gluon
exchange as the dominant mechanism for the dipole-proton interaction.
The gluon distributions are derived from inclusive measurements of the
proton structure function. In the FKS model, a three-dimensional
Gaussian is assumed for the $\rho^0$ wave-function, while MRT use
parton-hadron duality and normalise the calculations to the data. For
the comparison with the present measurements the MRST99~\cite{mrst99}
and CTEQ6.5M~\cite{cteq65m} parameterisations for the gluon density
were used.

Kowalski, Motyka and Watt (KMW)~\cite{kmw} use an improved version of
the saturation model~\cite{gbw1,gbw2}, with an explicit dependence on
the impact parameter and DGLAP~\cite{dglap1,dglap2,dglap3,dglap4}
evolution in $Q^2$, introduced through the unintegrated gluon
distribution~\cite{un-gl}. Forshaw, Sandapen and Shaw (FSS)~\cite{fss}
model the dipole-proton interaction through the exchange of a
soft~\cite{dl} and a hard~\cite{dl-hard} Pomeron, with (Sat) and
without (Nosat) saturation, and use the DGKP and Gaussian $\rho^0$
wave-functions. In the model of Dosch and Ferreira (DF)~\cite{df}, the
dipole cross section is calculated using Wilson loops, making use of
the stochastic vacuum model for the non-perturbative QCD contribution.

While the calculations based on two-gluon exchange are limited to
relatively high-$Q^2$ values (typically $\sim$ 4 GeV$^2$), those
based on modelling the dipole cross section incorporate both the
perturbative and non-perturbative aspects of $\rho^0$ production.

\subsection{Comparison with data}

\begin{figure}[h]
\vfill
\begin{center}
\includegraphics[width=\hsize]{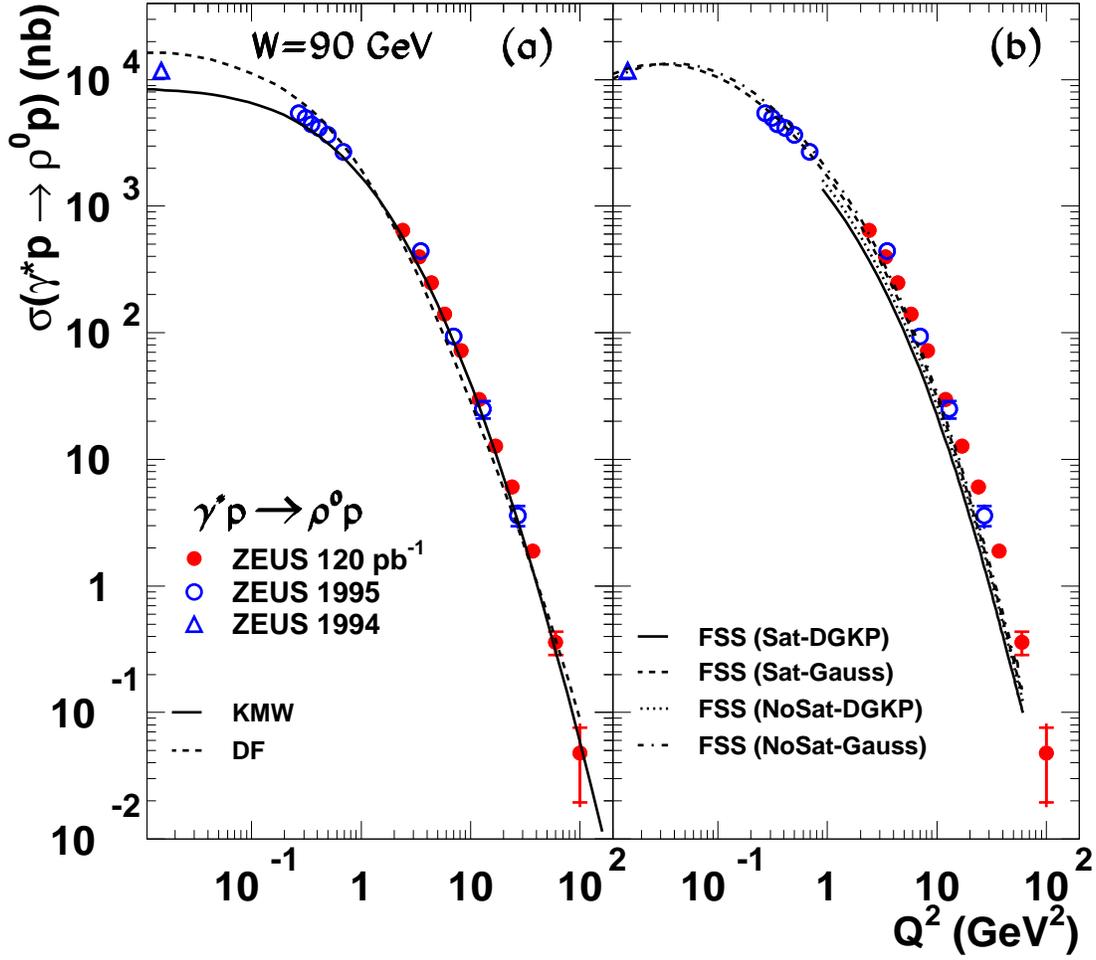}
\end{center}
\caption{\it
The $Q^2$ dependence of the $\gamma^* p \to \rho^0 p$ cross section at
$W$=90 GeV.  The same data are plotted in (a) and (b), compared to
different models, as described in the text. The predictions are
plotted in the range as provided by the authors. }
\label{fig:q2dep-models}
\vfill
\end{figure}

The different predictions discussed above are compared to the $Q^2$
dependence of the cross section in Fig.~\ref{fig:q2dep-models}.  None
of the models gives a good description of the data over the full
kinematic range of the measurement. The FSS model with the
three-dimensional Gaussian $\rho^0$ wave-function describes the
low-$Q^2$ data very well, while the KMW and DF models describe
the $Q^2>1$ GeV$^2$ region well.

\begin{figure}[h]
\vfill
\begin{center}
\includegraphics[width=\hsize]{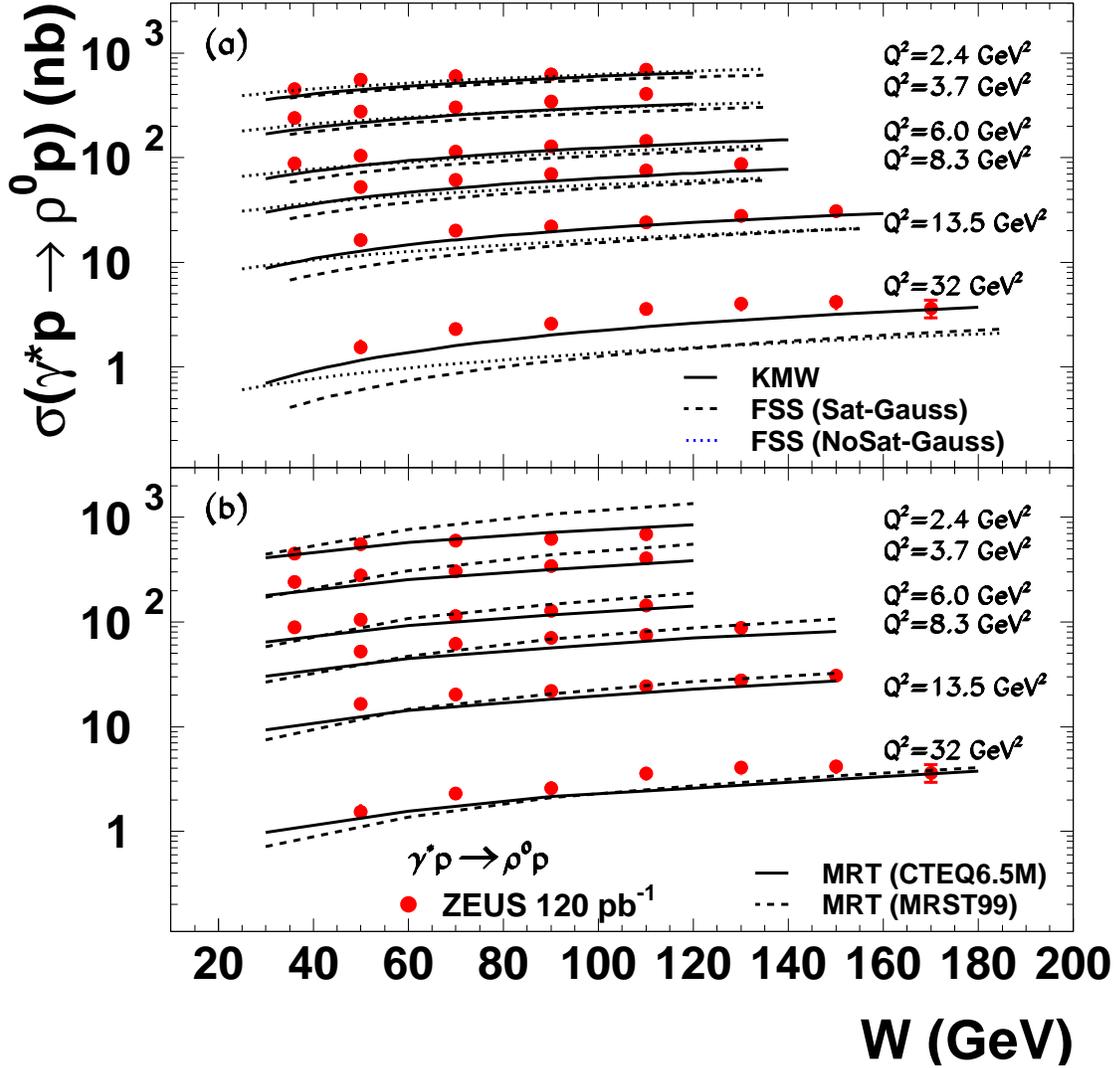}
\end{center}
\caption{\it
The $W$ dependence of the $\gamma^* p \to \rho^0 p$ cross section for
different values of $Q^2$, as indicated in the figure.  The same data
are plotted in (a) and (b), compared to different models, as described
in the text. The predictions are plotted in the range as
provided by the authors. }
\label{fig:wdep-models}
\vfill
\end{figure}

\begin{figure}[h]
\vfill
\begin{center}
\includegraphics[width=\hsize]{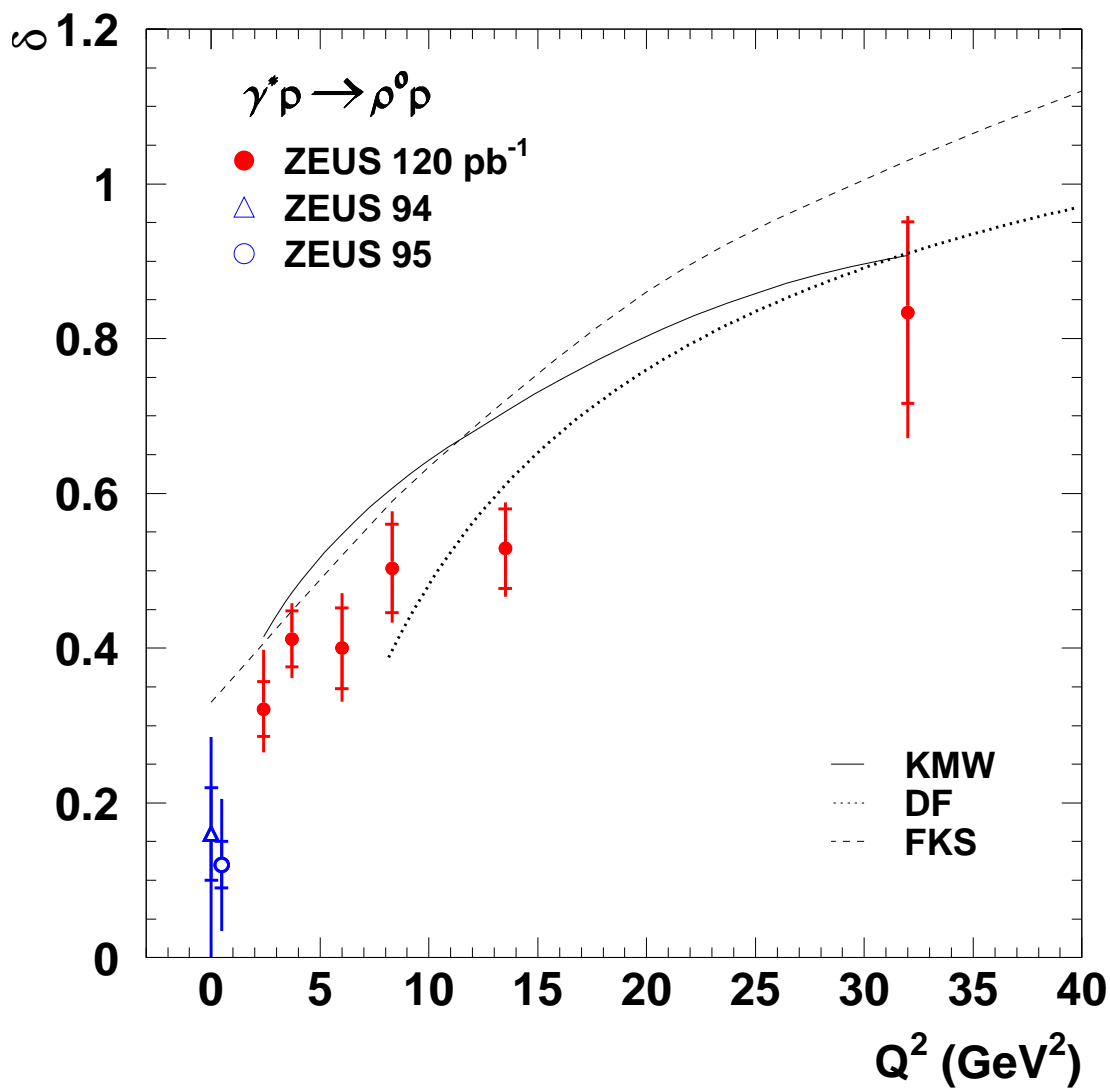}
\end{center}
\caption{
The value of $\delta$ from a fit of the form $\sigma \sim W^\delta$
for the reaction $\gamma^* p \to \rho^0 p$, as a function of $Q^2$.
The lines are the predictions of models as denoted in the figure
(see text).}
\label{fig:delta-models}
\vfill
\end{figure}

The various predictions are also compared with the $W$ dependence of
the cross section, for different $Q^2$ values, in
Fig.~\ref{fig:wdep-models}.  Here again, none of the models reproduces
the magnitude of the cross section measurements. The closest to the
data, in shape and magnitude, are the MRT model with the CTEQ6.5M
parametrisation of the gluon distribution in the proton and the KMW
model. The KMW model gives a good description of the $Q^2$ dependence
of $\delta$, as shown in Fig.~\ref{fig:delta-models}.

\begin{figure}[h]
\vfill
\begin{center}
\includegraphics[width=\hsize]{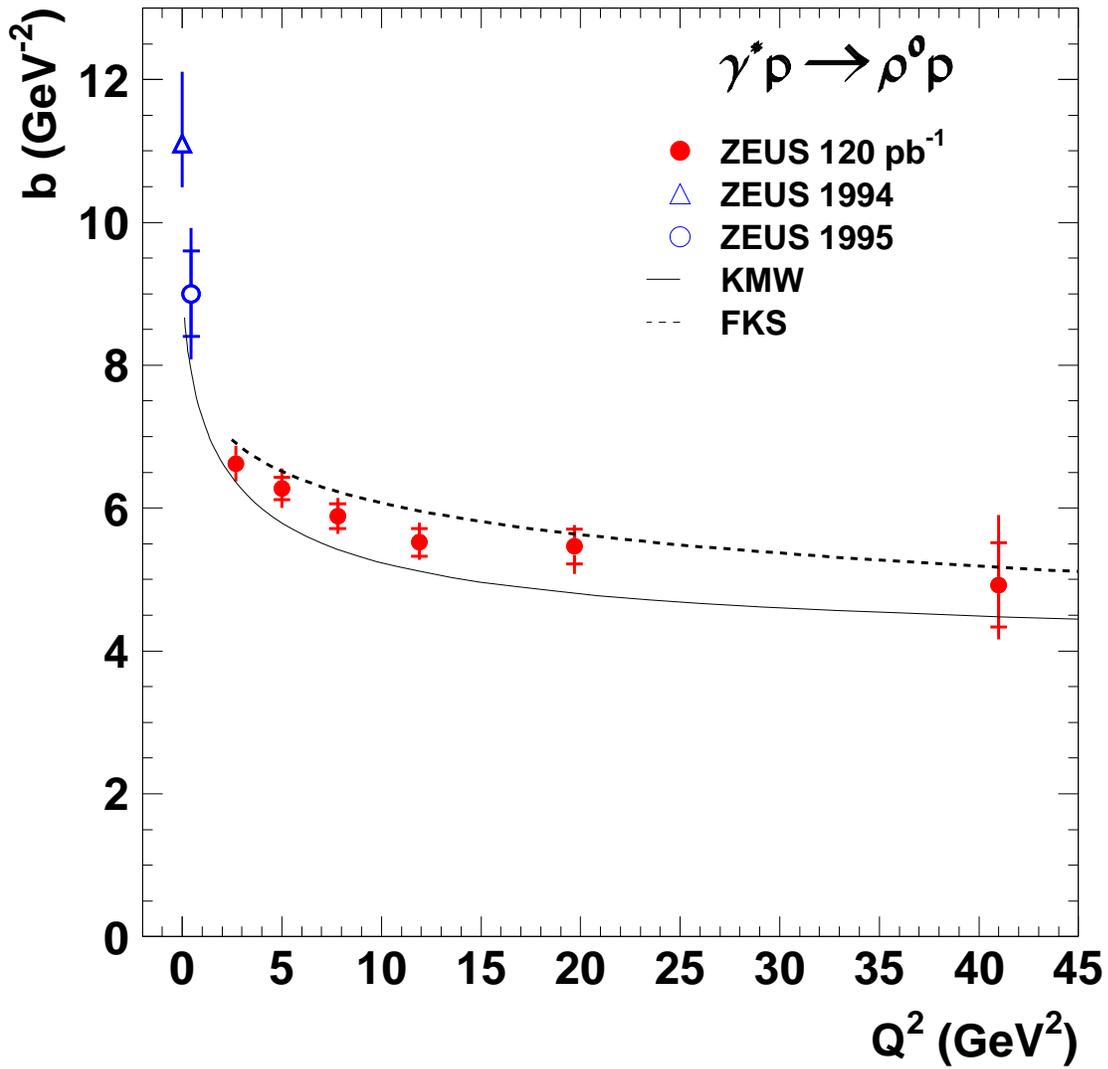}
\end{center}
\caption{
The value of the slope $b$ from a fit of the form $d\sigma/d|t| \sim
e^{-b|t|}$ for the reaction $\gamma^* p \to \rho^0 p$, as a function
of $Q^2$.  The lines are the predictions of models as denoted in the
figure (see text).}
\label{fig:b-models}
\vfill
\end{figure}

The dependence of $b$ on $Q^2$ is given only in the FKS and the KMW
models as shown in Fig.~\ref{fig:b-models}. The FKS expectations are
somewhat closer to the data.

\begin{figure}[h]
\vfill
\begin{center}
\includegraphics[width=\hsize]{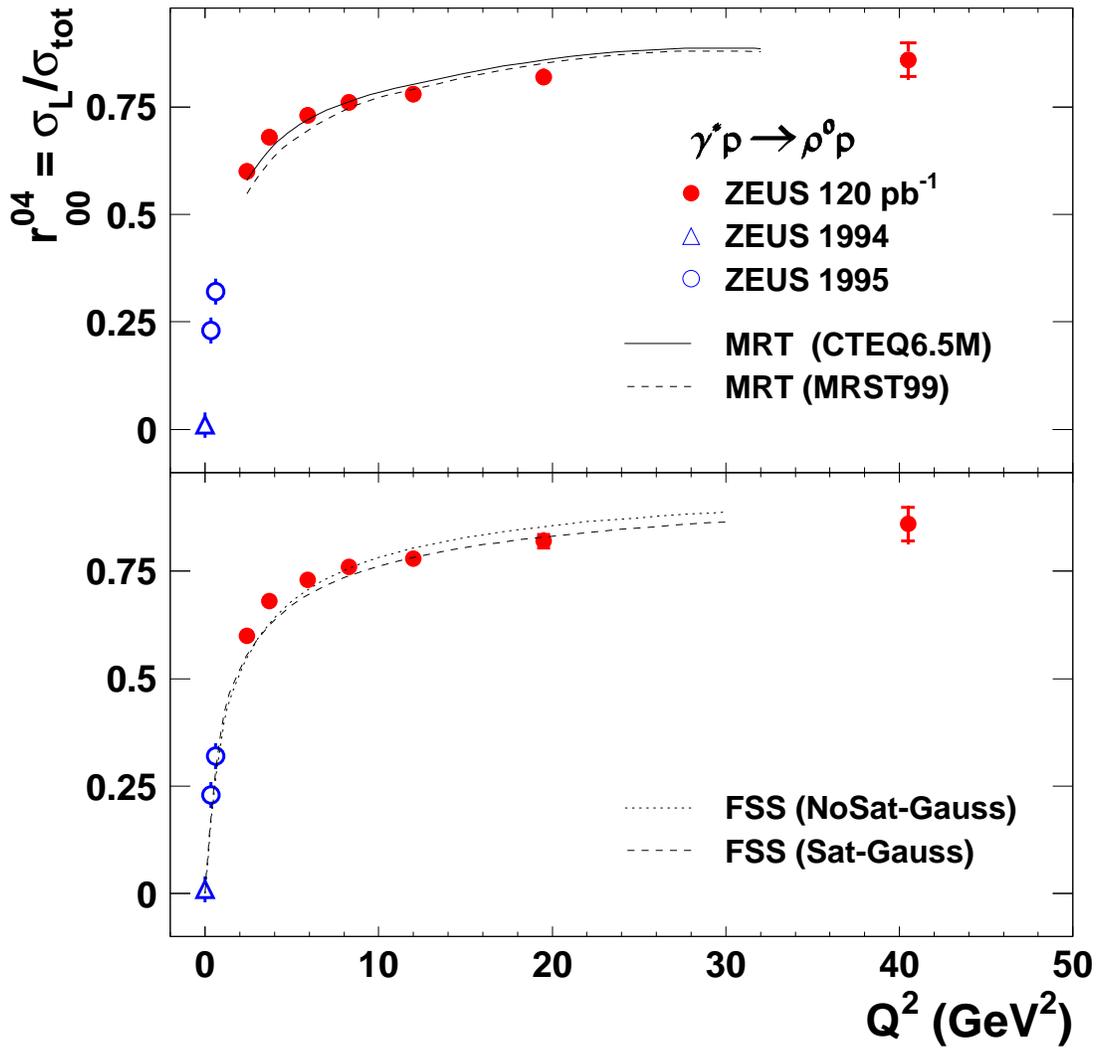}
\end{center}
\caption{
The ratio $r^{04}_{00}$ as a function of $Q^2$ compared to the
predictions of models as denoted in the figure (see text).}
\label{fig:RQ2-models}
\vfill
\end{figure}

\begin{figure}[h]
\vfill
\begin{center}
\includegraphics[width=\hsize]{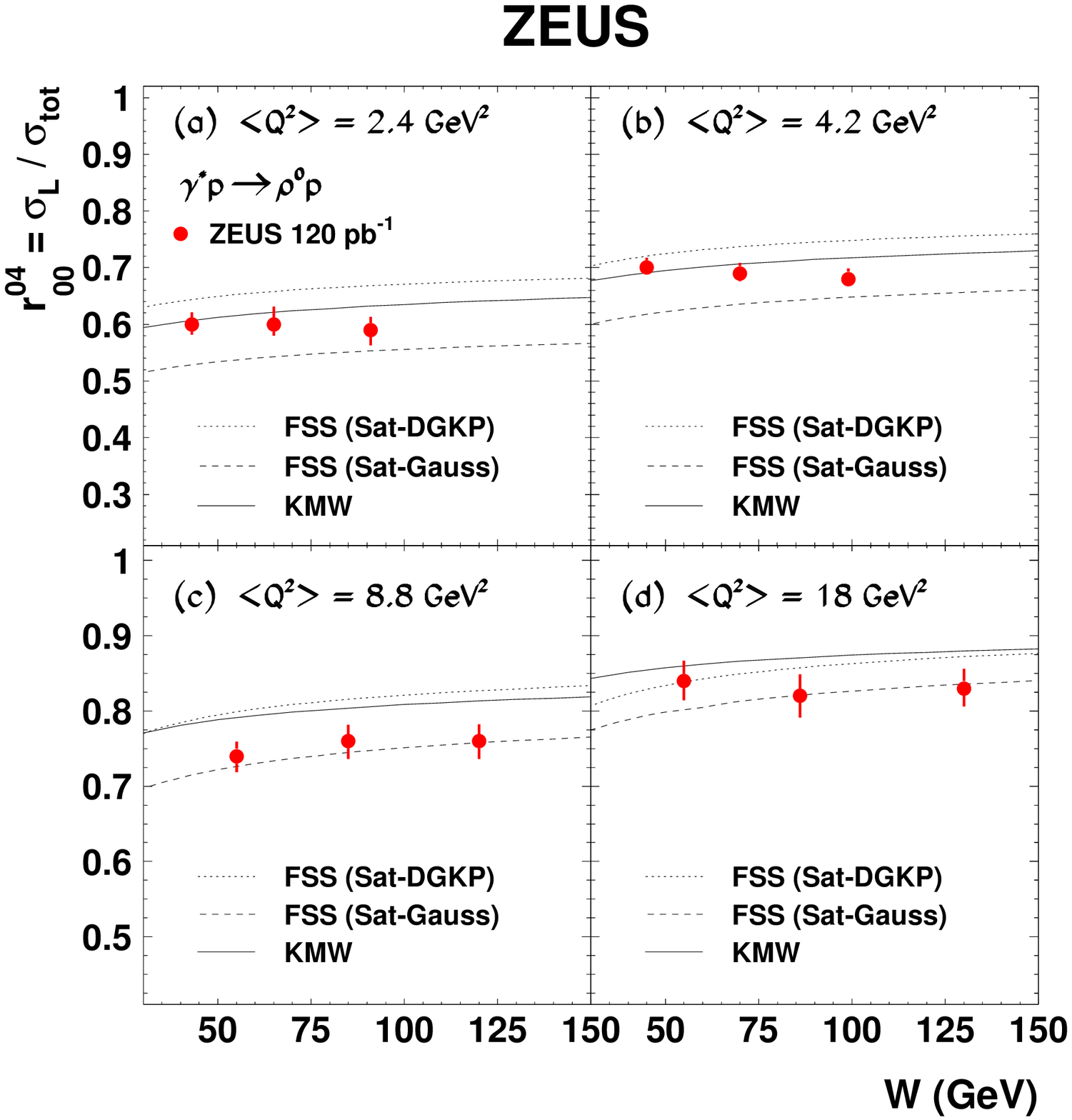}
\end{center}
\caption{
The ratio $r^{04}_{00}$ as a function of $W$ for different values of
$Q^2$ compared to the predictions of models as indicated in the figure
(see text).}
\label{fig:RW-models2}
\vfill
\end{figure}

\begin{figure}[h]
\vfill
\begin{center}
\includegraphics[width=\hsize]{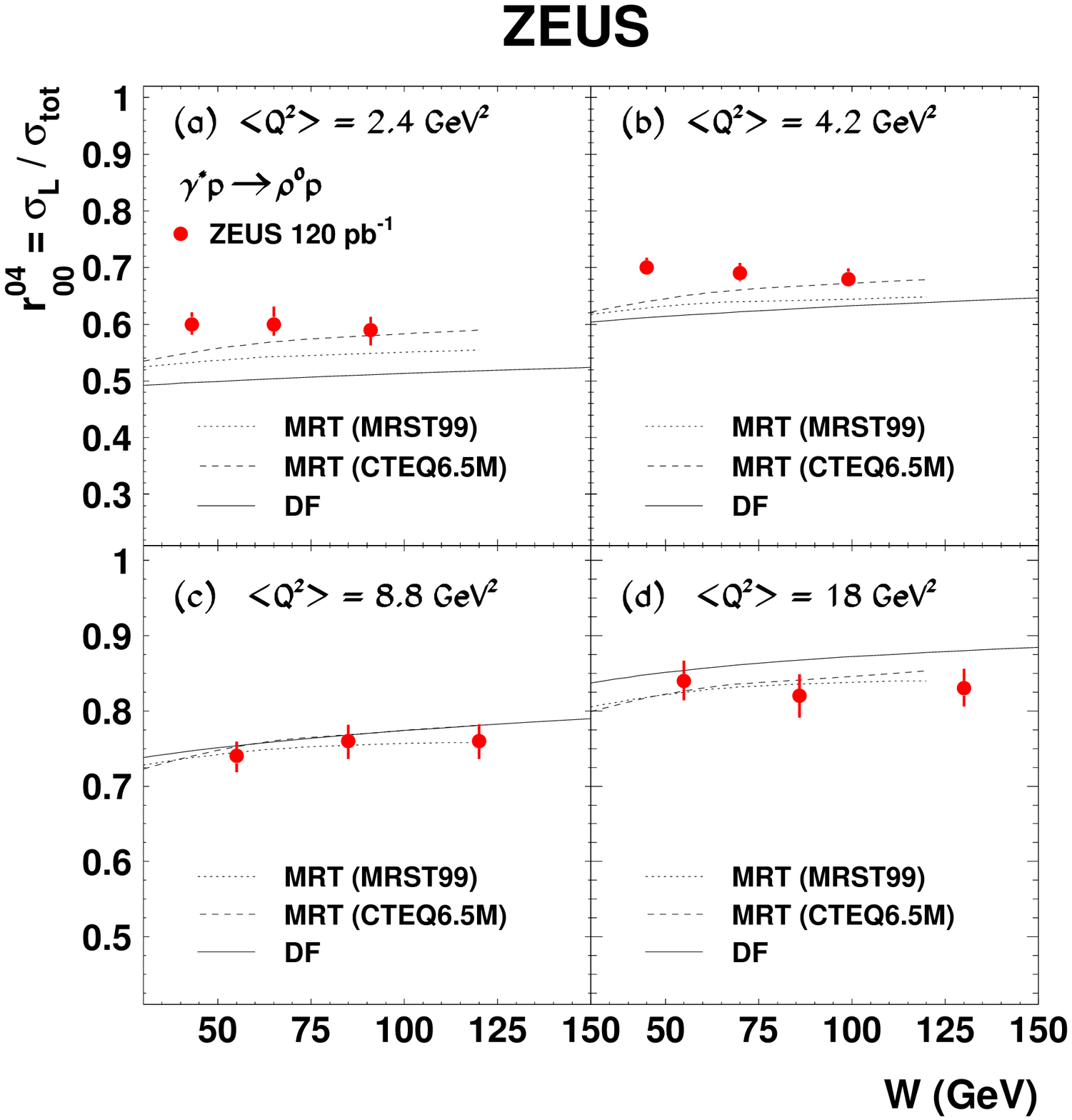}
\end{center}
\caption{
The ratio $r^{04}_{00}$ as a function of $W$ for different values of
$Q^2$ compared to the predictions of models as indicated in the figure
(see text).}
\label{fig:rvswmod2}
\vfill
\end{figure}

The expected $Q^2$ dependence of $r^{04}_{00}$ is compared to the
measurements in Fig.~\ref{fig:RQ2-models}. The MRT prediction, using
the CTEQ6.5M gluon density, is the only prediction which describes the data
in the whole $Q^2$ range. While all the models exhibit a mild
dependence of $r^{04}_{00}$ on $W$, consistent with the data as shown
in Figs.~\ref{fig:RW-models2} and~\ref{fig:rvswmod2}, none of them
reproduces correctly the magnitude of $r^{04}_{00}$ in all the $Q^2$
bins.

In summary, none of the models considered above is able to describe
all the features of the data presented in this paper. The high
precision of the measurements can be used to refine models for
exclusive $\rho^0$ electroproduction.


\section{Summary and Conclusions}
\label{sec:end}
Exclusive $\rho^0$ electroproduction has been studied by ZEUS at HERA
in the range $2<Q^2<160~\mathrm{GeV}^2$ and $32<W<180~\mathrm{GeV}$
with a high statistics sample. The $Q^2$ dependence of the $\gamma^* p
\rightarrow \rho^0 p$ cross section is a steeply falling function of $Q^2$.
The cross section rises with $W$ and its logarithmic derivative in $W$
increases with increasing $Q^2$.  The exponential slope of the $t$
distribution decreases with increasing $Q^2$ and levels off at about
$b=5~\mathrm{GeV}^{-2}$. The decay angular distributions of the
$\rho^0$ indicate $s$-channel helicity breaking. The ratio of cross
sections induced by longitudinally and transversely polarised virtual
photons increases with $Q^2$, but is independent of $W$ and of $|t|$,
suggesting suppression of large-size configurations of the
transversely polarised photon. The effective Pomeron trajectory,
averaged over the full $Q^2$ range, has a larger intercept and a
smaller slope than those extracted from soft interactions. All these
features are compatible with expectations of perturbative
QCD. However, none of the available models which have been compared to
the measurements is able to reproduce all the features of the data.

\section*{Acknowledgments}

It is a pleasure to thank the DESY Directorate for their strong
support and encouragement. The remarkable achievements of the HERA
machine group were essential for the successful completion of this
work and are greatly appreciated. The design, construction and
installation of the ZEUS detector has been made possible by the
efforts of many people who are not listed as authors.  We thank
E. Ferreira, J. Forshaw, M. Strikman, T. Teubner and G. Watt, for
providing the results of their calculations.

\vfill\eject



\clearpage
\begin{sidewaystable}

\begin{center}
\begin{tabular}{|c|r@{$\pm$}l|r@{$\pm$}l|r@{$\pm$}l|r@{$\pm$}l|r@{$\pm$}l|}
\hline
Element & \multicolumn{2}{c|}{$2<Q^2<3$ GeV$^2$}  &\multicolumn{2}{c|} {$3<Q^2<4$ GeV$^2$}  &\multicolumn{2}{c|} {$4<Q^2<6$ GeV$^2$}  &\multicolumn{2}{c|} {$6<Q^2<10$ GeV$^2$}  & 
\multicolumn{2}{c|} {$10<Q^2<100$ GeV$^2$}   \\
\hline
$r^{04}_{00}$&     
    0.590&    0.006$^{+    0.012}_{-    0.010}$&
    0.659&    0.008$^{+    0.009}_{-    0.015}$&
    0.725&    0.008$^{+    0.014}_{-    0.008}$&
    0.752&    0.008$^{+    0.011}_{-    0.008}$&
    0.814&    0.010$^{+    0.008}_{-    0.019}$\\
\hline
Re($r^{04}_{10}$)& 
    0.024&    0.005$^{+    0.003}_{-    0.009}$&
    0.025&    0.007$^{+    0.008}_{-    0.009}$&
    0.007&    0.007$^{+    0.004}_{-    0.017}$&
    0.014&    0.007$^{+    0.005}_{-    0.010}$&
    0.014&    0.009$^{+    0.016}_{-    0.007}$\\
\hline
$r^{04}_{1-1}  $&  
   -0.009&    0.007$^{+    0.008}_{-    0.012}$&
   -0.010&    0.008$^{+    0.006}_{-    0.016}$&
    0.000&    0.007$^{+    0.015}_{-    0.006}$&
   -0.016&    0.007$^{+    0.018}_{-    0.004}$&
   -0.001&    0.010$^{+    0.021}_{-    0.006}$\\
\hline
$r^1_{11}    $&   
   -0.008&    0.007$^{+    0.006}_{-    0.019}$&
   -0.023&    0.008$^{+    0.008}_{-    0.016}$&
   -0.015&    0.008$^{+    0.010}_{-    0.019}$&
   -0.032&    0.008$^{+    0.017}_{-    0.001}$&
   -0.002&    0.011$^{+    0.008}_{-    0.020}$\\
\hline
$r^1_{00}    $&  
   -0.037&    0.019$^{+    0.047}_{-    0.014}$&
   -0.014&    0.026$^{+    0.046}_{-    0.015}$&
    0.020&    0.028$^{+    0.072}_{-    0.013}$&
    0.019&    0.030$^{+    0.008}_{-    0.060}$&
   -0.018&    0.042$^{+    0.053}_{-    0.034}$\\
\hline
Re($r^1_{10}$)& 
   -0.032&    0.007$^{+    0.018}_{-    0.004}$&
   -0.023&    0.010$^{+    0.008}_{-    0.024}$&
   -0.016&    0.009$^{+    0.018}_{-    0.013}$&
   -0.006&    0.011$^{+    0.003}_{-    0.030}$&
   -0.042&    0.016$^{+    0.029}_{-    0.009}$\\
\hline
$r^1_{1-1}   $&  
    0.195&    0.009$^{+    0.012}_{-    0.019}$&
    0.151&    0.011$^{+    0.014}_{-    0.011}$&
    0.121&    0.011$^{+    0.016}_{-    0.011}$&
    0.095&    0.011$^{+    0.006}_{-    0.029}$&
    0.100&    0.016$^{+    0.023}_{-    0.032}$\\
\hline
Im($r^2_{10}  $)&  
    0.040&    0.007$^{+    0.010}_{-    0.020}$&
    0.024&    0.009$^{+    0.005}_{-    0.020}$&
    0.029&    0.009$^{+    0.012}_{-    0.011}$&
    0.031&    0.009$^{+    0.016}_{-    0.012}$&
    0.026&    0.015$^{+    0.028}_{-    0.005}$\\
\hline
Im($r^2_{1-1} $)&  
   -0.186&    0.009$^{+    0.009}_{-    0.024}$&
   -0.148&    0.011$^{+    0.019}_{-    0.015}$&
   -0.124&    0.012$^{+    0.029}_{-    0.013}$&
   -0.107&    0.011$^{+    0.004}_{-    0.027}$&
   -0.052&    0.016$^{+    0.039}_{-    0.012}$\\
\hline
$r^5_{11}    $& 
    0.018&    0.003$^{+    0.004}_{-    0.005}$&
    0.018&    0.004$^{+    0.006}_{-    0.004}$&
    0.007&    0.003$^{+    0.005}_{-    0.007}$&
    0.018&    0.004$^{+    0.005}_{-    0.002}$&
    0.004&    0.005$^{+    0.007}_{-    0.003}$\\
\hline
$r^5_{00}    $&   
    0.085&    0.009$^{+    0.007}_{-    0.015}$&
    0.089&    0.013$^{+    0.019}_{-    0.016}$&
    0.106&    0.013$^{+    0.010}_{-    0.016}$&
    0.093&    0.013$^{+    0.013}_{-    0.010}$&
    0.168&    0.018$^{+    0.011}_{-    0.020}$\\
\hline
Re($r^5_{10}$)  & 
    0.167&    0.003$^{+    0.007}_{-    0.003}$&
    0.164&    0.004$^{+    0.005}_{-    0.006}$&
    0.143&    0.005$^{+    0.004}_{-    0.013}$&
    0.132&    0.005$^{+    0.004}_{-    0.003}$&
    0.110&    0.007$^{+    0.011}_{-    0.008}$\\
\hline
$r^5_{1-1}   $&  
    0.000&    0.005$^{+    0.006}_{-    0.008}$&
   -0.006&    0.006$^{+    0.009}_{-    0.006}$&
    0.001&    0.005$^{+    0.009}_{-    0.003}$&
    0.000&    0.006$^{+    0.018}_{-    0.003}$&
    0.001&    0.007$^{+    0.011}_{-    0.002}$\\
\hline
Im($r^6_{10}  $)& 
   -0.157&    0.003$^{+    0.006}_{-    0.004}$&
   -0.147&    0.004$^{+    0.004}_{-    0.007}$&
   -0.145&    0.004$^{+    0.003}_{-    0.009}$&
   -0.135&    0.004$^{+    0.007}_{-    0.003}$&
   -0.125&    0.006$^{+    0.012}_{-    0.002}$\\
\hline
Im($r^6_{1-1} $)& 
    0.010&    0.005$^{+    0.004}_{-    0.013}$&
   -0.005&    0.005$^{+    0.008}_{-    0.005}$&
   -0.001&    0.005$^{+    0.005}_{-    0.017}$&
    0.008&    0.005$^{+    0.003}_{-    0.006}$&
   -0.002&    0.007$^{+    0.005}_{-    0.007}$\\
\hline 
\end{tabular}
\vspace{1cm}
\caption{\it
Spin density matrix elements for electroproduction of $\rho^0$, for
different intervals of $Q^2$. The first uncertainty is statistical,
the second systematic.
  }
\label{tab:15dme}
\end{center}

\end{sidewaystable}
%
%
%

\clearpage
\begin{table}
\begin{center}
\begin{tabular}{|r@{$-$}l|c|c|lcr|}
\hline
\multicolumn{2}{|c|} {$Q^2$ bin} &$Q^2$ & $ |t| $   
 &\multicolumn{3}{c|}{$d\sigma/dt $}
 \\ \multicolumn{2}{|c|} {$ (\rm{GeV^2}) $} & {$ (\rm{GeV^2}) $} & {$ (\rm{GeV^2}) $} 
  & $(\rm{nb/GeV^2})$ & stat. & syst. \\ \hline
2&4&$2.7$&$     0.05$&$  2636.4$&$\pm    49.5$&$^{+   117.3}_{-   155.3}$\\
2&4&$2.7$&$     0.15$&$  1284.2$&$\pm    32.8$&$^{+    65.4}_{-    87.7}$\\
2&4&$2.7$&$     0.29$&$  450.7$&$\pm    13.5$&$^{+    30.8}_{-    39.1}$\\
2&4&$2.7$&$     0.53$&$  127.5$&$\pm     6.2$&$^{+    17.2}_{-    17.0}$\\
2&4&$2.7$&$     0.83$&$  28.1$&$\pm     3.3$&$^{+    10.3}_{-     5.1}$\\
4&6.5&$5.0$&$     0.05$&$ 842.7$&$\pm    23.7$&$^{+    33.3}_{-    40.5}$\\
4&6.5&$5.0$&$     0.15$&$ 415.8$&$\pm    15.4$&$^{+    18.9}_{-    26.1}$\\
4&6.5&$5.0$&$     0.29$&$ 159.8$&$\pm     7.0$&$^{+    10.6}_{-    13.8}$\\
4&6.5&$5.0$&$     0.53$&$ 43.7$&$\pm     3.2$&$^{+     5.7}_{-     5.8}$\\
4&6.5&$5.0$&$     0.83$&$ 12.5$&$\pm     1.8$&$^{+     2.2}_{-     2.2}$\\
6.5&10&$7.8$&$      0.05$&$  338.4$&$\pm    10.8$&$^{+    15.4}_{-    15.0}$\\
6.5&10&$7.8$&$      0.15$&$  156.2$&$\pm     7.4$&$^{+     5.3}_{-    13.3}$\\
6.5&10&$7.8$&$      0.29$&$  67.3$&$\pm     3.3$&$^{+     4.9}_{-     4.7}$\\
6.5&10&$7.8$&$      0.53$&$  22.1$&$\pm     1.6$&$^{+     2.3}_{-     3.1}$\\
6.5&10&$7.8$&$      0.83$&$   5.03$&$\pm     0.94$&$^{+     1.48}_{-     0.92}$\\
10&15&$11.9$&$     0.05$&$  118.0$&$\pm     5.0$&$^{+     5.5}_{-     5.7}$\\
10&15&$11.9$&$     0.15$&$  70.2$&$\pm     3.9$&$^{+     5.2}_{-     3.6}$\\
10&15&$11.9$&$     0.29$&$  26.8$&$\pm     1.7$&$^{+     1.7}_{-     2.6}$\\
10&15&$11.9$&$     0.53$&$  8.40$&$\pm     0.76$&$^{+     0.97}_{-     1.36}$\\
10&15&$11.9$&$     0.83$&$  2.67$&$\pm     0.51$&$^{+     0.48}_{-     0.52}$\\
15&30&$19.7$&$     0.05$&$  39.6$&$\pm     2.2$&$^{+     1.7}_{-     3.3}$\\
15&30&$19.7$&$     0.15$&$  20.4$&$\pm     1.5$&$^{+     1.9}_{-     1.4}$\\
15&30&$19.7$&$     0.29$&$  9.12$&$\pm     0.71$&$^{+     0.59}_{-     0.94}$\\
15&30&$19.7$&$     0.53$&$  2.73$&$\pm     0.31$&$^{+     0.39}_{-     0.38}$\\
15&30&$19.7$&$     0.83$&$  0.84$&$\pm     0.19$&$^{+     0.19}_{-     0.30}$\\
30&80&$41.0$&$      0.05$&$ 5.44$&$\pm     0.83$&$^{+     0.76}_{-     0.80}$\\
30&80&$41.0$&$      0.15$&$ 2.28$&$\pm     0.50$&$^{+     0.37}_{-     0.54}$\\
30&80&$41.0$&$      0.29$&$ 1.40$&$\pm     0.26$&$^{+     0.26}_{-     0.35}$\\
30&80&$41.0$&$      0.53$&$  0.42$&$\pm     0.11$&$^{+     0.07}_{-     0.11}$\\
30&80&$41.0$&$      0.83$&$ 0.15$&$\pm     0.07$&$^{+     0.06}_{-     0.07}$\\
\hline
\end{tabular}
\caption{\it 
The differential cross-section $d\sigma/dt$ for the reaction $\gamma^*
p \to \rho^0 p$ for different $Q^2$ intervals.  The first column gives
the $Q^2$ bin, while the second column gives the $Q^2$ value at which the
cross section is quoted.  The normalisation uncertainty due to
luminosity ($\pm$2\%) and proton-dissociative background ($\pm$4\%),
is not included.  }
\label{tab:dsdt}
\end{center}
\end{table}

\clearpage
\begin{table}
\begin{center}
\begin{tabular}{|r@{$-$}l|r@{.}l|c|}
\hline
\multicolumn{2}{|c|}{$ Q^2$ bin (GeV$^2)$} & \multicolumn{2}{c|}{$Q^2$ (GeV$^2$)} & $b$ (GeV$^{-2}$)  \\
\hline
2&4&    2&7 &$    6.6\pm 0.1^{+  0.2}_{- 0.2}$\\
4&6.5&  5&0 &$    6.3\pm 0.2^{+  0.2}_{- 0.2}$\\
6.5&10& 7&8 &$    5.9\pm 0.2^{+  0.2}_{- 0.2}$\\
10&15&  11&9&$    5.5\pm 0.2^{+  0.2}_{- 0.2}$\\
15&30&  19&7&$    5.5\pm 0.3^{+  0.2}_{- 0.3}$\\
30&80&  41&0&$    4.9\pm 0.6^{+  0.8}_{- 0.5}$\\
\hline
\end{tabular}

\vspace{1cm}
\caption{\it
The slope $b$ resulting from a fit to the differential cross-section
$d\sigma/dt$ to an exponential form for the reaction $\gamma^* p
\to \rho^0 p$, for different $Q^2$ intervals.  
The first column gives the $Q^2$ bin, while the second column gives the
$Q^2$ value at which the differential cross sections are quoted. The
first uncertainty is statistical, the second systematic.  }
\label{tab:slope}
\end{center}
\end{table}

\begin{table}
\begin{center}
\begin{tabular}{|r@{$-$}l|c|c|c|lcr|} 
\hline
 \multicolumn{2}{|c|}{$Q^2$ bin} &$W$ bin & $ Q^2 $ & $ W $ &
 &\multicolumn{2}{c|}{$\sigma({\gamma^{*}p \rightarrow \rho^{0} p}) $}
 \\ \multicolumn{2}{|c|}{ $ (\rm{GeV^2}) $} & $ (\rm{GeV}) $ & $ (\rm{GeV^2}) $ & $
 (\rm{GeV})$ & $(\rm nb)$ & stat. & syst. \\
\hline
2&3&$    40-100$&  $  2.4 $&$   90$&$   647.1$&$\pm   8.7$&$^{+  28.4}_{-    41.7}$\\
3&4&$    40-100$&  $  3.4 $&$   90$&$   396.7$&$\pm   6.7$&$^{+  14.6}_{-    19.4}$\\
4&5&$    40-100$&  $  4.4 $&$   90$&$   247.8$&$\pm   5.8$&$^{+  8.9 }_{-    12.6}$\\
5&7&$    40-120$&  $  5.8 $&$   90$&$   140.3$&$\pm   2.6$&$^{+  3.9 }_{-    5.9}$\\
7&10&$   40-140$&  $  8.2 $&$   90$&$    71.9$&$\pm   1.4$&$^{+  1.7 }_{-    2.9}$\\
10&15&$  40-140$&  $  12 $&$   90$&$    29.73$&$\pm   0.68$&$^{+  0.75 }_{-    1.14}$\\
15&20&$  40-140$&  $  17 $&$   90$&$    12.77$&$\pm   0.50$&$^{+  0.27 }_{-    0.42}$\\
20&30&$  40-140$&  $  24 $&$   90$&$     6.03$&$\pm   0.31$&$^{+  0.37 }_{-    0.13}$\\
30&50&$  40-140$&  $  37 $&$   90$&$     1.88$&$\pm   0.16$&$^{+  0.07 }_{-    0.15}$\\
50&80&$  40-140$&  $  60 $&$   90$&$     0.36$&$\pm   0.07$&$^{+  0.04 }_{-    0.03}$\\
80&160&$ 40-140$&  $ 100 $&$   90$&$     0.05$&$\pm   0.03$&$^{+  0.02 }_{-    0.01}$\\
\hline
\end{tabular}

\caption{\it  Cross-section  measurements at $Q^2$ and $W = 90~\rm{GeV}$ 
averaged over the $Q^2$ and $W$ intervals given in the table.
The normalisation
uncertainty due to luminosity ($\pm$2\%) and proton-dissociative
background ($\pm$4\%) is not included. 
}
\label{tab:q2dep}
\end{center}
\end{table}

\clearpage
\begin{table}
\vspace{-1cm}
\begin{center}
\begin{tabular}{|r@{$-$}l|r@{$-$}l|r@{$.$}l|r@{$.$}l|lcr|} 
\hline
\multicolumn{2}{|c|}{ $Q^2$ bin} &\multicolumn{2}{c|}{$W$ bin}  &\multicolumn{2}{c|}{$ Q^2 $} & \multicolumn{2}{c|}{$  W $}  &
&\multicolumn{2}{c|}{$\sigma({\gamma^{*}p \rightarrow \rho^{0} p}) $} \\ 
\multicolumn{2}{|c|}{  $ (\rm{GeV^2}) $} &\multicolumn{2}{c|}{ $ (\rm{GeV}) $} &\multicolumn{2}{c|}{ $ (\rm{GeV^2}) $}
&\multicolumn{2}{c|}{ $  (\rm{GeV})$} &  $(\rm nb)$ & stat. & syst. \\ 
\hline
2&3 & 32&40  &      2&4&     36&0&$    451.9$&$\pm     15.1$&$^{+     25.5}_{-     43.6}$\\
2&3 & 40&60  &      2&4&     50&0&$    554.1$&$\pm     11.5$&$^{+     31.6}_{-     39.2}$\\
2&3 & 60&80  &      2&4&     70&0&$    599.9$&$\pm     13.9$&$^{+     28.5}_{-     38.5}$\\
2&3 & 80&100  &      2&4&     90&0&$    622.5$&$\pm     17.3$&$^{+     33.8}_{-     43.2}$\\
2&3 & 100&120  &      2&4&    110&0&$    690.1$&$\pm     30.3$&$^{+     40.9}_{-     66.9}$\\
\hline
3&5 & 32&40  &     3&7&     36&0&$    240.8$&$\pm      8.0$&$^{+      9.5}_{-     15.5}$\\
3&5 & 40&60  &      3&7&     50&0&$    277.5$&$\pm      5.9$&$^{+     12.2}_{-     15.3}$\\
3&5 & 60&80  &      3&7&     70&0&$    303.7$&$\pm      7.3$&$^{+     11.1}_{-     14.4}$\\
3&5 & 80&100  &      3&7&     90&0&$    344.6$&$\pm      9.4$&$^{+     10.4}_{-     17.2}$\\
3&5 & 100&120  &      3&7&    110&0&$    404.7$&$\pm     15.5$&$^{+     15.2}_{-     22.5}$\\
\hline
5&7 & 32&40  &      6&0&     36&0&$     88.5$&$\pm      5.1$&$^{+      6.0}_{-      4.1}$\\
5&7 & 40&60  &      6&0&     50&0&$    104.9$&$\pm      3.6$&$^{+      3.6}_{-      6.9}$\\
5&7 & 60&80  &      6&0&     70&0&$    113.6$&$\pm      4.1$&$^{+      6.0}_{-      3.9}$\\
5&7 & 80&100  &      6&0&     90&0&$    127.6$&$\pm      4.9$&$^{+      4.0}_{-      5.8}$\\
5&7 & 100&120  &      6&0&    110&0&$    144.0$&$\pm      6.1$&$^{+      8.6}_{-      8.4}$\\
\hline
7&10 & 40&60  &      8&3&     50&0&$     52.3$&$\pm      1.9$&$^{+      1.7}_{-      2.7}$\\
7&10 & 60&80  &      8&3&     70&0&$     61.7$&$\pm      2.4$&$^{+      2.1}_{-      2.9}$\\
7&10 & 80&100  &      8&3&     90&0&$     70.1$&$\pm      2.9$&$^{+      2.0}_{-      3.3}$\\
7&10 & 100&120  &      8&3&    110&0&$     75.2$&$\pm      3.4$&$^{+      3.1}_{-      3.0}$\\
7&10 & 120&140  &      8&3&    130&0&$     87.5$&$\pm      4.7$&$^{+      2.5}_{-      4.1}$\\
\hline
10&22 & 40&60  &     13&5&     50&0&$     16.4$&$\pm      0.6$&$^{+      0.6}_{-      0.7}$\\
10&22 & 60&80  &     13&5&     70&0&$     20.2$&$\pm      0.8$&$^{+      0.8}_{-      0.7}$\\
10&22 & 80&100  &     13&5&     90&0&$     21.9$&$\pm      0.9$&$^{+      0.7}_{-      0.9}$\\
10&22 & 100&120  &     13&5&    110&0&$     24.3$&$\pm      1.1$&$^{+      0.9}_{-      1.2}$\\
10&22 & 120&140  &     13&5&    130&0&$     27.7$&$\pm      1.4$&$^{+      0.9}_{-      1.0}$\\
10&22 & 140&160  &     13&5&    150&0&$     30.7$&$\pm      2.3$&$^{+      1.2}_{-      1.1}$\\
\hline
22&80 & 40&60  &     32&0&     50&0&$      1.5$&$\pm      0.2$&$^{+      0.2}_{-      0.1}$\\
22&80 & 60&80  &     32&0&     70&0&$      2.3$&$\pm      0.2$&$^{+      0.1}_{-      0.1}$\\
22&80 & 80&100  &     32&0&     90&0&$      2.6$&$\pm      0.3$&$^{+      0.3}_{-      0.2}$\\
22&80 & 100&120  &     32&0&    110&0&$      3.6$&$\pm      0.4$&$^{+      0.1}_{-      0.3}$\\
22&80 & 120&140  &     32&0&    130&0&$      4.0$&$\pm      0.5$&$^{+      0.2}_{-      0.4}$\\
22&80 & 140&160  &     32&0&    150&0&$      4.2$&$\pm      0.6$&$^{+      0.2}_{-      0.4}$\\
22&80 & 160&180  &     32&0&    170&0&$      3.6$&$\pm      0.7$&$^{+      0.3}_{-      0.3}$\\
\hline 
\end{tabular}

\caption{\it  Cross-sections values obtained at $Q^2$ and $W$ as a 
result of averaging over bins of the $Q^2$ and $W$ intervals given in
the table.  The normalisation uncertainty due to luminosity ($\pm$2\%)
and proton-dissociative background ($\pm$4\%), are not included.  }
\label{tab:wdep}
\end{center}
\end{table}

\begin{table}[ht]
\begin{center}
\begin{tabular}{|r@{$-$}l|r@{$.$}l|lcr|} 
\hline 
\multicolumn{2}{|c|}{ $Q^2$ bin}  &\multicolumn{2}{c|}{ $ Q^2 $} 
&\multicolumn{3}{c|}{} \\ 
\multicolumn{2}{|c|}{$ (\rm{GeV^2}) $} & \multicolumn{2}{c|}{$ (\rm{GeV^2}) $} & $\delta$ & stat. & syst. \\ 
\hline  
\hline  
2&3   &     2&4&$  0.321$&$\pm0.035$&$^{+  0.068}_{-  0.043}$\\
3&5   &     3&7&$  0.412$&$\pm0.036$&$^{+  0.029}_{-  0.035}$\\
5&7   &     6&0&$  0.400$&$\pm0.052$&$^{+  0.048}_{-  0.045}$\\ 
7&10  &     8&3&$  0.503$&$\pm0.057$&$^{+  0.047}_{-  0.041}$\\ 
10&22 &     13&5&$ 0.529$&$\pm0.051$&$^{+  0.030}_{-  0.035}$\\ 
22&80 &     32&0&$ 0.834$&$\pm0.118$&$^{+  0.043}_{-  0.112}$\\
\hline
\end{tabular}
\end{center}
\caption{\it The value of $\delta$ obtained from fitting 
$\sigma^{\gamma^{*}p \rightarrow \rho^{0} p} \propto W^{\delta}$.
The first column gives
the $Q^2$ bin, while the second column gives the $Q^2$ value at which the
cross section was quoted.
}
\label{tab:delta}
\end{table}

\begin{table}
  \begin{center}
    \begin{tabular}{|r@{$-$}l|r@{$.$}l|c|c|r|} 
\hline 
     \multicolumn{2}{|c|}{ $Q^2$ bin (GeV$^2$)} &\multicolumn{2}{c|} {$Q^2$ (GeV$^2$)} & W bin (GeV) & $r_{00}^{04}$ &
      $R=\sigma_L/\sigma_T$
\\ \hline 
      2&3 & 2&4 & $32-120$ & $0.60\pm 0.01 ^{+0.03}_{-0.03}$ &
      $1.50^{+0.05}_{-0.05}$ $ ^{+0.20}_{-0.15}$
\\[5pt]

      3&5 & 3&7 & $32-120$ & $0.68\pm 0.01 ^{+0.02}_{-0.02}$ &
      $2.10^{+0.08}_{-0.08}$ $^{+0.18}_{-0.14}$
\\[5pt] 
      5&7 & 5&9 & $40-140$ & $0.73\pm 0.01 ^{+0.01}_{-0.02}$ &
      $2.70^{+0.14}_{-0.13}$ $^{+0.26}_{-0.28}$
\\[5pt] 
      7&10 & 8&3 & $40-140$ & $0.76\pm 0.01 ^{+0.01}_{-0.02}$ &
      $3.20^{+0.20}_{-0.18}$ $^{+0.25}_{-0.27}$
\\[5pt]
     10&15 & 12&0 & $40-140$ & $0.78\pm 0.01 ^{+0.01}_{-0.01}$ &
     $3.50^{+0.26}_{-0.24}$ $^{+0.30}_{-0.26}$
\\[5pt] 
     15&30 & 19&5 & $40-140$ & $0.82\pm 0.02 ^{+0.01}_{-0.02}$ &
     $4.60^{+0.54}_{-0.45}$ $^{+0.48}_{-0.44}$
\\[5pt]
     30&100 & 40&5 & $40-160$ & $0.86\pm 0.04 ^{+0.03}_{-0.02}$ &
     $6.10^{+2.75}_{-1.56}$ $^{+2.15}_{-0.85}$
     
\\   \hline
    \end{tabular} 
\caption{\it The spin matrix element $r_{00}^{04}$ and
    the ratio of cross sections for longitudinally and transversely
    polarised photons, $R=\sigma_L/\sigma_T$, as a function of $Q^2$,
    averaged over the $Q^2$ and $W$ bins given in the table.  The
    first uncertainty is statistical, the second systematic. }
    \label{tab:R-q2}
\end{center}
\end{table}

\clearpage

\begin{table}
  \begin{center}
    \begin{tabular}{|r@{$-$}l|r@{$.$}l|r@{$-$}l|c|c|r|} 
\hline 
 \multicolumn{2}{|c|}{$Q^2$ bin (GeV$^2$)} & \multicolumn{2}{c|}{$Q^2$
 (GeV$^2$)} & \multicolumn{2}{c|}{$W$ bin (GeV)} & $W$ (GeV) &
 $r_{00}^{04}$ & $R=\sigma_L/\sigma_T$
\\ \hline 

2&3 & 2&4 & 32&55 & 43 &$0.60\pm 0.01 ^{+0.03}_{-0.02}$&$1.50^{+0.06}_{-0.06}$ $^{+0.21}_{-0.15}$
\\[5pt]
2&3 & 2&4 & 55&75 & 65 &$0.60\pm0.01 ^{+0.05}_{-0.03}$&$1.50^{+0.06}_{-0.06}$ $^{+0.35}_{-0.17}$
\\[5pt]
2&3 & 2&4 & 75&110& 91 &$0.59\pm 0.01 ^{+0.04}_{-0.04}$&$1.43^{+0.06}_{-0.06}$ $^{+0.23}_{-0.23}$
\\[5pt]  
\hline
3&7 & 4&2 & 32&60 & 45 &$0.70\pm 0.01 ^{+0.01}_{-0.01}$&$2.33^{+0.09}_{-0.09}$ $^{+0.13}_{-0.09}$
\\[5pt]
3&7 & 4&2 & 60&80 & 70 &$0.69\pm 0.01 ^{+0.02}_{-0.01}$&$2.23^{+0.12}_{-0.11}$ $^{+0.24}_{-0.10}$
\\[5pt]
3&7 & 4&2 & 80&120& 99 &$0.69\pm 0.01 ^{+0.01}_{-0.01}$&$2.23^{+0.10}_{-0.09}$ $^{+0.14}_{-0.09}$
\\[5pt] 
\hline 
7&12 & 8&8 & 40&70& 55 &$0.74\pm 0.01 ^{+0.01}_{-0.02}$&$2.85^{+0.25}_{-0.22}$ $^{+0.23}_{-0.26}$
\\[5pt]
7&12 & 8&8 & 70&100&85 &$0.76\pm 0.02 ^{+0.01}_{-0.02}$&$3.17^{+0.38}_{-0.32}$ $^{+0.19}_{-0.28}$
\\[5pt]
7&12 & 8&8 &100&140&120&$0.76\pm 0.02 ^{+0.01}_{-0.02}$&$3.17^{+0.38}_{-0.32}$ $^{+0.23}_{-0.26}$
\\[5pt]
 \hline
12&50 & 18&0 &40&70& 55&$0.84\pm 0.03 ^{+0.01}_{-0.01}$&$5.25^{+1.16}_{-0.84}$ $^{+0.54}_{-0.34}$
\\[5pt]
12&50 & 18&0 &70&100&85&$0.82\pm 0.03 ^{+0.01}_{-0.02}$&$4.55^{+0.94}_{-0.70}$ $^{+0.47}_{-0.43}$
\\[5pt]
12&50 &18&0&100&160&130&$0.83\pm 0.02 ^{+0.02}_{-0.01}$&$4.88^{+0.87}_{-0.67}$ $^{+0.64}_{-0.39}$
\\   \hline
    \end{tabular} 
    \caption{\it The spin matrix element $r_{00}^{04}$ and the ratio
    of cross sections for longitudinally and transversely polarised
    photons, $R=\sigma_L/\sigma_T$, as a function of $W$ for different
    values of $Q^2$, averaged over the $Q^2$ and $W$ bins given in the
    table.  The first uncertainty is statistical, the second
    systematic. }
\label{tab:R-w} 
\end{center}
\end{table}

\begin{table}
  \begin{center}
    \begin{tabular}{|r@{$-$}l|r@{$.$}l|c|c|c|r|} 
\hline 
\multicolumn{2}{|c|}{ $Q^2$ bin (GeV$^2$)}  &\multicolumn{2}{c|}{ $Q^2$ (GeV$^2$)} & W bin (GeV)   & $ |t|$ (GeV$^2$) &  $r_{00}^{04}$       &  $R=\sigma_L/\sigma_T$
\\ \hline 
2&5 & 3&0 &$32-120$ & 0.04 & $0.62\pm 0.01^{+0.02}_{-0.02} $   & $1.63 ^{+0.07}_{-0.06}$  $^{+0.15}_{-0.13}$
\\ [5pt]
2&5  &3&0& $32-120$ & 0.14 & $0.62\pm 0.01^{+0.01}_{-0.03} $   & $1.63 ^{+0.09}_{-0.09}$  $^{+0.10}_{-0.19}$
\\ [5pt]
2&5 &  3&0 & $32-120$ & 0.27 & $0.63\pm 0.01^{+0.04}_{-0.02} $   & $1.70 ^{+0.11}_{-0.11}$  $^{+0.24}_{-0.14}$
\\  [5pt]
2&5  & 3&0 &$32-120$ & 0.45 & $0.64\pm 0.02^{+0.02}_{-0.03} $   & $1.78 ^{+0.14}_{-0.13}$  $^{+0.16}_{-0.21}$
\\ [5pt]
2&5  & 3&0 & $32-120$ & 0.76 & $0.63\pm 0.03^{+0.07}_{-0.05} $   & $1.70 ^{+0.26}_{-0.22}$  $^{+0.63}_{-0.32}$
\\ \hline       
5&50  & 10&0 &$40-160$  & 0.04 & $0.74\pm 0.01^{+0.01}_{-0.01} $ & $2.84 ^{+0.18}_{-0.17}$  $^{+0.16}_{-0.15}$
\\ [5pt]
5&50  & 10&0 & $40-160$ & 0.15 & $0.75\pm 0.01^{+0.01}_{-0.02} $ & $3.00 ^{+0.26}_{-0.23}$  $^{+0.17}_{-0.30}$
\\ [5pt]
5&50 & 10&0 & $40-160$  & 0.27 & $0.74\pm 0.02^{+0.02}_{-0.04} $ & $2.84 ^{+0.26}_{-0.24}$  $^{+0.32}_{-0.51}$
\\  [5pt]
5&50 & 10&0 &$40-160$ & 0.45 & $0.72\pm 0.02^{+0.03}_{-0.02} $ & $2.57 ^{+0.29}_{-0.25}$  $^{+0.41}_{-0.22}$
\\ [5pt]
5&50 & 10&0 &$40-160$ & 0.76 & $0.73\pm 0.04^{+0.03}_{-0.05} $ & $2.70 ^{+0.56}_{-0.43}$  $^{+0.45}_{-0.57}$
\\ \hline  
\end{tabular}
    \caption{\it The spin matrix element $r_{00}^{04}$ and the ratio
    of cross sections for longitudinally and transversely polarised
    photons, $R=\sigma_L/\sigma_T$, as a function of $|t|$ for two
    values of $Q^2$, averaged over the $Q^2$ and $W$ bins given in the
    table.  The first uncertainty is statistical, the second
    systematic. }
\label{tab:R-t} 
\end{center}
\end{table}

\newpage

%
%
%

\begin{table}
\begin{center}
\begin{tabular}{|l|c|c|c|}
\hline
$Q^2$ bin (GeV$^2$) & $Q^2$ (GeV$^2$) & $|t|$ (GeV$^2$) & $\apom(t)$ \\
\hline
$2-5$  & $3$  & $0.04$ & $1.104\pm 0.011^{+ 0.010}_{- 0.010}$\\
$2-5$  & $3$  & $0.14$ & $1.099\pm 0.014^{+ 0.011}_{- 0.025}$\\
$2-5$  & $3$  & $0.28$ & $1.048\pm 0.016^{+ 0.038}_{- 0.014}$\\
$2-5$  & $3$  & $0.57$ & $1.013\pm 0.021^{+ 0.041}_{- 0.017}$\\
$5-50$ & $10$  & $0.04$ & $1.149\pm 0.012^{+ 0.015}_{- 0.006}$\\
$5-50$ & $10$  & $0.16$ & $1.134\pm 0.014^{+ 0.005}_{- 0.027}$\\
$5-50$ & $10$  & $0.35$ & $1.104\pm 0.017^{+ 0.012}_{- 0.011}$\\
$5-50$ & $10$  & $0.68$ & $1.085\pm 0.028^{+ 0.042}_{- 0.031}$\\

\hline
\end{tabular}
\vspace{1cm}
\caption{\it
The values of the effective Pomeron trajectory $\apom(t)$ as a function of
$|t|$, for two $Q^2$ values. The first uncertainty is statistical, 
the second systematic.
 }
\label{tab:trajectory}
\end{center}
\end{table}

\clearpage
\begin{table}
\begin{center}
\begin{tabular}{|l|c|c|c|}
\hline
$Q^2$ bin (GeV$^2$) & $Q^2$ (GeV$^2$) &  $\apom(0)$ &  $\apom^\prime$(GeV$^{-2}$) \\
\hline
$2-5$  & $3$ & $1.113\pm 0.010^{+ 0.009}_{- 0.012}$ &  $0.185\pm 0.042^{+ 0.022}_{- 0.057}$  \\
$5-50$ & $10$ & $1.152\pm 0.011^{+ 0.006}_{- 0.006}$ &  $0.114\pm 0.043^{+ 0.026}_{- 0.024}$  \\

\hline
\end{tabular}
\vspace{1cm}
\caption{\it
The values of the effective Pomeron trajectory intercept $\apom(0)$ and slope
$\apom^\prime$, for two $Q^2$ values. The first uncertainty
is statistical, the second systematic.  }
\label{tab:alpha}
\end{center}
\end{table}

\begin{table}
\begin{center}
\begin{tabular}{|c|c|c|}
\hline
$ Q^2$ (GeV$^2)$ & $W$ (GeV) & $b$ (GeV$^{-2}$)  \\
\hline
$3.5$&  $38 $&$    6.3\pm  0.2 ^{+  0.4}_{- 0.3}$\\
$3.5$&  $57 $&$    6.3\pm  0.1^{+  0.3}_{- 0.3}$\\
$3.5$&  $82 $&$    6.6 \pm 0.2^{+  0.2}_{- 0.3}$\\
$3.5$&  $107$&$    6.9\pm  0.2^{+  0.3}_{- 0.3}$\\
$3.5$&  $134$&$    7.0\pm  0.3^{+  0.4}_{- 0.3}$\\
$11$&   $38$&$     5.8\pm  0.3^{+  0.3}_{- 0.4}$\\
$11$&   $57$&$     5.8 \pm 0.2^{+  0.2}_{- 0.3}$\\
$11$&   $82$&$     5.7 \pm 0.2^{+  0.2}_{- 0.2}$\\
$11$&   $107$&$    5.9 \pm 0.2^{+  0.3}_{- 0.2}$\\
$11$&   $134$&$    6.1\pm  0.2^{+  0.3}_{- 0.2}$\\
\hline
\end{tabular}

\vspace{1cm}
\caption{\it
The slope $b$ resulting from a fit of the
differential cross section $d\sigma/dt$ for the reaction $\gamma^* p
\to \rho^0 p$ to an exponential form, for different $W$ values, for two $Q^2$ 
values.  The first uncertainty is statistical, the second
systematic. }
\label{tab:slopew}
\end{center}
\end{table}


 \clearpage

 \clearpage

 \clearpage

 \clearpage

 \clearpage

 \clearpage

 \clearpage

 \clearpage

 \clearpage

 \clearpage

 \clearpage

 \clearpage

 \clearpage

 \clearpage

 \clearpage

 \clearpage

 \clearpage

 \clearpage

 \clearpage

 \clearpage

 \clearpage

 \clearpage

 \clearpage

 \clearpage

 \clearpage

 \clearpage

 \clearpage

 \clearpage

 \clearpage

 \clearpage

 \clearpage


%
%
\end{document}